\documentclass{aa}
\bibpunct{(}{)}{;}{a}{}{,}
\usepackage{graphicx,natbib,url,twoopt}
\usepackage[T1]{fontenc}
\usepackage[labelfont=bf,font=footnotesize]{caption}
\usepackage{subcaption}
\usepackage{longtable}
\usepackage[]{overpic}
\usepackage{makecell}
\usepackage{colortbl}
\usepackage[table,xcdraw]{xcolor}
\usepackage{amsmath} 
\usepackage{booktabs}
\usepackage{float}
\usepackage{soul}
\usepackage{txfonts}
\usepackage{siunitx}
\usepackage{pdflscape}
\usepackage{comment}
\usepackage{soul}
\usepackage{array}
\usepackage{hyperref}
\def\ms{\hbox{\,m\,s$^{-1}$}}         
\def\m2s2{\hbox{\,m$^{2}$\,s$^{-2}$}} 
\def\Msun{\hbox{$\mathrm{M}_{\odot}$}}             

\def\Mjup{\hbox{$\mathrm{M}_{\rm Jup}$}}

\newcommand*{\mono}{\fontfamily{lmtt}\selectfont}
\newcommand\scalemath[2]{\scalebox{#1}{\mbox{\ensuremath{\displaystyle #2}}}}
\begin{document}


\title{Exploring the Brown Dwarf Desert with Precision Radial Velocities and Gaia DR3 Astrometric Orbits}

\author{
N.~Unger \inst{1,2}
\and D.~Ségransan   \inst{1,2}
\and D.~Barbato  \inst{1,2}
\and J.-B.~Delisle \inst{1,2}
\and J.~Sahlmann \inst{3}
\and B.~Holl  \inst{1,2}
\and S.~Udry \inst{1,2}
}

\institute{D\'epartement d'Astronomie, Universit\'e de Gen\`eve, Chemin Pegasi 51, CH-1290 Versoix, Suisse \\ \email{nicolas.unger@unige.ch}
\and
National Center of Competence in Research, PlanetS, Gesellschaftsstrasse 6, 3012 Bern, Switzerland
\and
RHEA Group for the European Space Agency (ESA), European Space Astronomy Centre (ESAC), Camino Bajo del Castillo s/n, E-28692 Villanueva de la Cañada, Madrid, Spain
}

\date{Received date /
Accepted date }

 
\abstract
{The observed scarcity of brown dwarfs in close orbits (within 10 au) around solar-type stars has posed significant questions about the origins of these substellar companions. These questions not only pertain to brown dwarfs but also impact our broader understanding of planetary formation processes. However, to resolve these formation mechanisms, accurate observational constraints are essential. Notably, most of the brown dwarfs have been discovered by radial velocity surveys, but this method introduces uncertainties due to its inability to determine the orbital inclination, leaving the true mass-and thus their true nature-unresolved. This highlights the crucial role of astrometric data, helping us distinguish between genuine brown dwarfs and stars.}
{This study aims to refine the mass estimates of massive companions to solar-type stars, mostly discovered through radial velocity measurements and subsequently validated using Gaia DR3 astrometry, to gain a clearer understanding of their true mass and occurrence rates.}
{We selected a sample of 31 sources with substellar companion candidates validated by Gaia DR3 and with available radial velocities. Using the Gaia DR3 solutions as prior information, we performed an MCMC fit with the available radial velocity measurements to integrate these two sources of data and thus obtain an estimate of their true mass.}
{Combining radial velocity measurements with Gaia DR3 data led to more precise mass estimations, leading us to reclassify several systems initially labeled as brown dwarfs as low-mass stars. Out of the 32 analyzed companions, 13 are determined to be stars, 17 are sub-stellar, and 2 have inconclusive results with the current data. Importantly, using these updated masses, we reevaluated the occurrence rate of brown dwarf companions (13-80 \Mjup) on close orbits (<10 au) in the CORALIE sample, determining a tentative occurrence rate of $0.8^{+0.3}_{-0.2}\%$.}
{}


\maketitle
%

\section{Introduction} \label{sec:intro}

Determining the true masses of planetary and brown dwarf (BD) companions is crucial, as it contributes to the development of accurate formation and evolution models. This is specially true for companions in and around the BD desert, which refers to the observed scarcity of BD companions in close orbits around solar type stars. The occurrence rate for these substellar companions, with masses ranging from 13 to 80 \Mjup, is estimated to be a mere 0.6\% \citep{2000A&A...355..581H, 2006ApJ...640.1051G, 2011A&A...525A..95S, 2017MNRAS.467.4264G, 2019A&A...631A.125K, 2023A&A...674A.114B}.

The occurrence rate of BDs follows a power law on both sides of the scarcity region \citep{2006ApJ...640.1051G}. Notably, the scarcity is most prominent within the 30-55 \Mjup\ range \citep{2014MNRAS.439.2781M, 2022A&A...661A.151H}, with approximately half of the BD population on each side. The two populations of BDs barely overlap, suggesting that they originate from different formation mechanisms \citep{2014prpl.conf..619C, 2014MNRAS.439.2781M, 2018haex.bookE..95W}. Therefore, gaining insight into the companion population at the boundaries of the BD mass limits plays a crucial role in understanding their formation origins.

The scarcity of BD companions on short orbits around solar-like stars can be attributed to several hypotheses. One proposes that the orbital migration of BD within evolving protoplanetary discs results in their merger with the host star \citep{2002MNRAS.330L..11A}. Another suggests ejections resulting from dynamical interactions between different companions \citep{2018haex.bookE..95W}. Additionally, there is the hypothesis that BD predominantly form at much larger separations, on the scale of hundreds or thousands of au \citep{2013ApJ...769....9J}.

For gas giants, there are two main formation scenarios, core accretion \citep{1972epcf.book.....S, 1996Icar..124...62P, 2012A&A...541A..97M} and gravitational instability \citep{1978M&P....18....5C, 1997Sci...276.1836B, 2016ARA&A..54..271K}. Core accretion is believed to be the most prominent, but gravitational instability is still the leading formation mechanisms for some gas giants \citep{2010Natur.468.1080M, 2018ApJ...860L..12T}. How frequently gravitational instability happens is still an active area of research \citep{2023A&A...669A..31S}, so better observational constraints of giant planets is of high value to better refine these formation mechanisms.



Astrometry will be of great value in obtaining better observational constraints on giant planets and BDs, as it can provide a direct measurement of the orbital inclination, which in combination with the radial velocities gives us the true mass of a companion. The Gaia space mission \citep{2016A&A...595A...1G} is the instrument with the best precision in absolute astrometry to date and is particularly sensitive to detect orbital periods in the range of $0.3 \lesssim P \lesssim 6$ yr in the nominal 5.5-year mission \citep{2008A&A...482..699C}, and is expected to discover tens of thousands of giant planets and brown dwarfs \citep{2014ApJ...797...14P, 2022A&A...661A.151H}.

Prior to the full release of Gaia DR3, the astrometric solution for 1.46 billion stars had already been made public with the Early Data Release 3 (EDR3, \citet{2021A&A...649A...1G}). Using a technique called proper motion anomalies (PMa), one can compare the proper motion measurements from the Hipparcos mission \citep{1997A&A...323L..49P, 2007A&A...474..653V} with those from Gaia EDR3 to identify differences that could indicate the presence of possible companions \citep{2021ApJS..254...42B, 2021AJ....162..186B, 2022A&A...657A...7K}. This method has been effective in determining masses of long-period giant planets, brown dwarfs, and binary systems discovered using radial velocities \citep[e.g.,][]{2021A&A...645A...7K, 2021AJ....162...12V, 2021AJ....162..266L, 2023A&A...674A.114B}. 

With Gaia Data Release 3 (DR3, \cite{2023A&A...674A...1G}), we have access to the first astrometric orbital solutions of sub stellar companions \citep{2023A&A...674A..34G, 2023A&A...674A..10H}. This has led to some studies that refine the orbital parameters of giant planets using the orbital solutions from Gaia DR3 and RV data \citep{2022AJ....164..196W, 2023AJ....165..266M}, which we continue and extend to brown dwarfs with the present study.

The release of Gaia Data Release 4 (DR4), anticipated not before the end of 2025, is expected to bring a significant advancement in the field. DR4 will be based on twice the observation timespan, feature 30\% more precise parallaxes compared to Gaia DR3, and enhance the precision of proper motions by a factor of 2.6 \citep{2021ARA&A..59...59B}. This is expected to result in the discovery of tens of thousands of giant planet and brown dwarf companions. Since the full astrometric epoch data will be released, DR4 will make it possible to combine individual astrometric and RV measurements \citep{2022A&A...667A.172D}. This will be analogous to the combination of Hipparcos Intermediate Astrometric Data and RV \citep{2000A&A...355..581H, 2001ApJ...562..549Z, 2011A&A...525A..95S, 2011A&A...527A.140R}.

In Sect. \ref{sec:sample_selection}, we provide an overview of the selected sample of stars analyzed in this study, including their stellar parameters and the RV data employed. Moving on to Sect. \ref{sec:method}, we introduce the methodology employed to integrate the orbital solution from Gaia DR3 with the available RV data, enabling a joint fit. Subsequently, in Sect. \ref{sec:results}, we delve into individual stars, presenting their companions and the outcomes derived from our combined fit. Finally, we examine the results in Sect. \ref{sec:discussion} and conclude the study in Sect. \ref{sec:conclusion}.


\section{Sample Selection and RV Data} \label{sec:sample_selection}

\subsection{Sample selection}

We are interested in targets with companion candidates that have a minimum mass in the substellar mass regime ($<80$ \Mjup) in the RV solution, and that have a validated orbital solution in Gaia DR3. These validated solutions can be found in the {\mono gaiadr3.nss\_two\_body\_orbit} table, with the 'OrbitalTargetedSearchValidated' tag in the {\mono nss\_solution\_type} column.

The 'OrbitalTargetedSearch[Validated]' tag means that these are stars come from a targeted list of sources that contained known hosts to exoplanets, nearby bright stars, and some cooler and fainter stars \citep{2023A&A...674A..10H}. The 'Validated' tag at the end means that the Gaia team was able to validate the orbits via external or internal means. For the planets and brown dwarfs, these validations came from radial velocity measurements by ground based high resolution spectrographs. For the binaries, they were validated either also by ground based spectrographs or by Gaia itself with its on board RV spectrograph \citep{2023A&A...674A..34G}.

We started the selection by taking stars with the 'OrbitalTargetedSearchValidated' solution type, for which the minimum mass of the companion was on the substellar mass regime (<80 \Mjup). The minimum mass was taken from published orbits and from our own RV fits for the few cases where no publication was available (HIP66074, HD40503, and HD68638). We also removed double lined spectroscopic binaries (SB2) with help of the GaiaDR3 solution type (HD47391, GJ4331) or by identifying multiple peaks in the cross-correlation function (CCF) of the HARPS spectra (GJ812A). This sample is further reduced by removing sources that were not adequate for a combined analysis of RV and astrometry. The exoplanet pipeline in Gaia was only run for models with one Keplerian; this means that the solution provided by Gaia will be incomplete for sources where there are more than one confirmed companion or one companion and a long term drift that affects the model. This is the case for GJ876 \citep{2001ApJ...556..296M}, HD111232 \citep{2004A&A...415..391M, 2022ApJS..262...21F}, HD142 (\citealp{2012ApJ...753..169W}; Raimbault et al. (in review)), and HD164604 \citep{2010ApJ...711.1229A}. In Unger et al. (in prep) we present the full RV analysis of HD111232, as we have obtained new RV measurements that provide further insight into this planetary system.

\cite{2023A&A...674A..10H} mention that the initial validation didn't go without pitfalls and that six known exoplanet hosts were missed from the 'Validated' tag (see Sect. 6.3.2 of \cite{2023A&A...674A..10H}). Four of these sources (from the 'OrbitalTargetedSearch' solution type) were thus added to our sample: HR 810, HD5433, HD91669, and BD-004475. The remaining two sources were ignored, specifically HD142, previously mentioned for having multiple companions, and KIC7917485 \citep{2016ApJ...827L..17M} because this companion was found through phase modulation of the stellar pulsations, and there are no RVs available.

This results in a sample of 32 sources with single-companion solutions that are suitable for a combined fit of RV and the Gaia DR3 orbital solution.

\subsection{Stellar parameters}
\label{sec:stellar_parameters}

The physical properties of the stars in the sample were characterized by fitting the Spectral Energy Distribution (SED) of each star using the IDL suite EXOFASTv2 \citep{2019arXiv190709480E} and MESA Isochrones and Stellar Tracks (MIST) method \citep{2016ApJS..222....8D, 2016ApJ...823..102C}. This method simultaneously constrains the stellar parameters, such as effective temperature, radius, and metallicity, by incorporating information from various available archival magnitudes and evolutionary tracks. Gaussian priors were imposed on effective temperature, metallicity and parallax based on values in the \cite{2019A&A...628A..94A} catalog and from Gaia DR3 \citep{2023A&A...674A...1G}. A more detailed explanation of the full method used is described in section 2 of \cite{2023A&A...674A.114B}. Table \ref{tab:stellar_parameters} presents the stellar parameters for all the stars included in our study's sample.

\subsection{RV data}

We use radial velocity data from several instruments including CORAVEL \citep{1979VA.....23..279B, 2018A&A...619A..81H}, ELODIE \citep{1996A&AS..119..373B}, CES \citep{2000A&A...353L..33K}, CORALIE \citep{2000A&A...356..590U}, HARPS \citep{2003Msngr.114...20M, 2020A&A...636A..74T}, HIRES \citep{2017AJ....153..208B}, MIKE \citep{2003SPIE.4841.1694B}, TULL \citep{2009AJ....137.3529W}, COUDE \citep{1993ASPC...36..267C}, SOPHIE \citep{2008SPIE.7014E..0JP}, HAMILTON \citep{1987PASP...99.1214V, 2014ApJS..210....5F}, UCLES \citep{1990SPIE.1235..562D, 2001ApJ...551..507T}. Considering that each additional instrument adds two parameters to the model (an offset and the jitter term), if there were less than 3 RV measurements available for a particular instrument, we ignored that instrument as it would not further constrain the model. Table \ref{tab:rv_data} provides comprehensive details of the RV data utilized in this study.

We also present new RV measurements from the CORALIE spectrograph for 12 stars of this work's sample. For some of them, this represents more than ten years of additional measurements from the ones originally published at the time of discovery of their companions. \footnote{The data will be available at the DACE platform: \url{dace.unige.ch}}

\section{Method}
\label{sec:method}

Our final objective is to perform a combined fit that integrates the Gaia astrometry data with the available radial velocity measurements. However, the epoch astrometry has not been released yet, we only have the orbital solutions presently at our disposal. As such, the term "combined fit" in this context denotes the process of fitting the radial velocities by utilizing the Gaia DR3 solutions as prior information.


\subsection{Astrometric model}

The orbital parameters of an astrometric orbit are the seven elements that define the orbit's size, shape, and orientation. They are the period ($P$), time of periapsis ($T_p$), eccentricity ($e$), semi-major axis of the photocenter ($a_0$), inclination ($i$), longitude of the ascending node ($\Omega$), and argument of periapsis ($\omega$). 

The Gaia mission identifies the apparent center of light (photocenter) from star systems. When a star has a stellar companion, then the photocenter is shifted and may not align with the actual center of the primary star. Notably, this deviation becomes significant for systems with a mass ratio $M_2/M_1>0.6$. For such systems, the barycentric semi-major axis can be underestimated by 5-9\% at most \citep{2023A&A...674A.114B}. 
In our sample, all systems exhibit a mass ratio below 0.6, and therefore we do not take this effect into account in our analysis. There are two cases (HD3277 and HD17289) where the RVs and astrometry are affected by the companion, and the implications are discussed in Sect. \ref{sec:challeging_cases}.

The parameters $a_0$, $i$, $\Omega$, and $\omega$ are the geometrical elements (also called the Campbell elements), which can be replaced by the Thiele Innes elements, defined by the following relations:

\begin{align}
    A &= a_0 (\cos \omega \cos \Omega - \sin \omega \sin \Omega \cos i)   \\
    B &= a_0 (\cos \omega \sin \Omega + \sin \omega \cos \Omega \cos i)   \\
    F &= -a_0 (\sin \omega \cos \Omega + \cos \omega \sin \Omega \cos i)   \\
    G &= -a_0 (\sin \omega \sin \Omega - \cos \omega \cos \Omega \cos i)   
\end{align}

This is the basis used by Gaia to fit astrometric orbits. From the Thiele Innes elements, we can derive the Campbell elements with the following relations \citep{2001icbs.book.....H}:

\begin{align} 
    a_0 &= \sqrt{p + \sqrt{p^2 - q^2}}  \label{eq:TI2campbell_a} \\
    i &= \arccos \left( \frac{q}{a_0^2} \right) \label{eq:TI2campbell_i} \\
    \omega &= \frac{1}{2} \left(\arctan \left(\frac{B-F}{A+G}\right) + \arctan \left(\frac{-B-F}{A-G}\right) \right)  \label{eq:TI2campbell_w} \\
    \Omega &= \frac{1}{2} \left(\arctan \left(\frac{B-F}{A+G}\right) - \arctan \left(\frac{-B-F}{A-G}\right) \right) \enspace , \label{eq:TI2campbell_W}
\end{align}

\noindent where

\begin{align}
    p &= \frac{1}{2} \left(A^2 + B^2 + F^2 + G^2 \right) \\
    q &= A G - B F \enspace .
\end{align}

The position angle of the line that links the actual orbital plane and the tangential plane of projection is known as $\Omega$, or the node. In particular, the ascending node is the node at which the orbital motion is directed away from the Sun. This means that $\Omega$ can only be determined by combining radial velocities and positional measurements, since the two ellipses that are mirror images of one another with regard to the projection plane produce identical projections. As a result, there exists an inherent degeneracy of $\Omega \pm \pi$ in astrometric solutions, which in turn also means that the same degeneracy exists in $\omega$ as this angle is measured from the ascending node. This degeneracy is resolved by comparing the $\omega$ values from both the RV and Gaia solutions. A more detailed explanation is given in Sect. \ref{sec:data_analysis_rv_only}.




\subsection{Radial velocity model}

To fit these parameters to the radial velocity measurements we have to convert them to the standard RV parameters which are the semi-amplitude $K$, the argument of periapsis $\omega$, the period $P$, the periastron epoch $T_p$, and the eccentricity $e$.

From the Gaia solution, we already have the values for $P$, $T_p$, and $e$. Then, we convert the Thiele Innes elements to the Campbell elements using equations \ref{eq:TI2campbell_a} to \ref{eq:TI2campbell_W}, which gives us $\omega$, $a_0$ and $i$. Lastly, using $a_0$, $\varpi$, $i$, $P$ and $e$, we can calculate $K$ as

\begin{equation} \label{eq:K_from_astrometry}
    K = 2 \pi \frac{(a_0/\varpi)\sin i}{P \sqrt{1-e^2}} \enspace .
\end{equation}

We then build the model with the standard radial velocity equation:

\begin{equation} \label{eq:keplerians}
    \upsilon(t) = \gamma_j + K \left[\cos \left(\nu(t) + \omega \right) + e \cos (\omega) \right] \enspace ,
\end{equation}

\noindent where $\nu$ is the true anomaly for which its calculation requires the orbital period $P$, the time of periastron $T_p$, and the eccentricity $e$. To do this calculation, one has to solve the Kepler equation, which is a transcendental equation.

The final log likelihood we used is described as follows:

\begin{multline} \label{eq:rv_likelihood}
    \ln \mathcal{L} (\pmb{\theta}) = -\frac{n_{\mathrm{obs}}}{2} \ln(2 \pi) - \frac{1}{2} \ln(|\det \Sigma_{\mathcal{L}}|) \\ 
    - \frac{1}{2} (\pmb{v} - \pmb{v}_{\mathrm{pred}}(\pmb{\theta}))^T \Sigma_{\mathcal{L}}^{-1}  (\pmb{v} - \pmb{v}_{\mathrm{pred}}(\pmb{\theta})) \enspace ,
\end{multline}
where $\pmb{\theta}$ is the vector of parameters, $n_{\mathrm{obs}}$ is the total number of observations, $\Sigma$ is the covariance matrix of the data, $\pmb{v}$ is the vector of measurements and $\pmb{v}_{\mathrm{pred}}$ is the predicted model.

For the covariance matrix, we assumed independent normally distributed uncertainties with an additional jitter term $\sigma_{J_i}$, one for each instrument $i$, to account for the remaining systematic errors that are not taken into account in the reported uncertainties $\sigma_k$.

\begin{equation}
    \Sigma_{\mathcal{L}} = 
    \begin{pmatrix}
    \frac{1}{\sigma_0^2 + \sigma_{J_i}^2} & 0 & 0 & \dots & 0 \\
    0 & \frac{1}{\sigma_1^2 + \sigma_{J_i}^2} & 0 & \dots & 0 \\
    0 & 0 & \frac{1}{\sigma_2^2 + \sigma_{J_i}^2} & \dots & 0 \\
    \vdots &  &  & \ddots & \vdots \\
    0 & 0 & 0 & \dots & \frac{1}{\sigma_n^2 + \sigma_{J_i}^2} \\
    \end{pmatrix}
\end{equation}


\subsection{Data Analysis}

\subsubsection{RV only} \label{sec:data_analysis_rv_only}

We start by performing a maximum likelihood estimation (MLE) of the RV model only with the RV data. We use the MLE of the parameters as the starting point to run a Markov Chain Monte Carlo (MCMC) algorithm to explore the posterior. We use the code {\mono samsam} \citep{2022ascl.soft07011D}
which stands for Scaled Adaptive Metropolis SAMpler. We run the MCMC for at least 500,000 iterations after the burn-in, ensuring that the number of iterations is at least 50 times the autocorrelation time and that the chains reached a stationary state. We use the posterior distributions to estimate the minimum mass of the companion.

Before continuing to the joint analysis, we compare the values in common between the RV and the Gaia solution, which are, $P$, $T_p$, $e$, and $\omega$ to check their agreement and if the $\omega \pm \pi$ correction to the Gaia solution has to be applied. As previously mentioned, this degeneracy is inherent in astrometry but is resolved through the inclusion of radial velocities. Hence, if the eccentricities between the RV and Gaia solutions exhibit a rough match, we proceed to verify if $\omega$ aligns as well. If not, it is likely offset by $\pi$. In such cases, we adjust $\omega$ and $\Omega$ from Gaia by $\pi$ to account for this discrepancy. However, if the difference is less than $\pi/2$, we refrain from modifying the values of $\omega$ and $\Omega$.


\subsubsection{Joint RV and astrometric fit}

As we don't have access yet to the astrometric time series from Gaia to actually fit both datasets together, we use the Gaia DR3 orbital solution as a prior on the joint model, but only using RV data.

In the {\mono gaiadr3.nss\_two\_body\_orbit} table, we have access to the astrometric orbital solution and the covariance matrix of the fitted parameters. The 'Orbital' type solutions have 12 free parameters that are fitted, these are: the position in right ascension and declination, the parallax ($\varpi$), the proper motions in right ascension and declination, the Thiele Innes elements ($A$, $B$, $F$, $G$), the period ($P$), the eccentricity ($e$) and the time of periapsis ($T_p$) relative to the Gaia epoch for DR3 (2016.0, Julian Date = $2457389.0$).

We use all parameters except the positions and proper motions, as these are not relevant to the orbit itself. We keep the seven astrometric orbit parameters plus the parallax, which we will need to transform the angular measurement of the semi-major axis in mas to au by dividing $a_0$ by the parallax (Eq. \ref{eq:TI2campbell_a}).

We extract the relevant elements from the provided covariance matrix, which we will use as a multivariate normal prior to these parameters. For ease of computation, we add to the diagonal of this matrix the corresponding normal priors for the RV offsets of each instrument available. We use a standard deviation of 20 \ms, centered on the RV offset derived from the RV only fit. This value is wide enough to accommodate all targets, which allows for a standardized analysis.

The mean $\boldsymbol{\mu}$ and the covariance matrix $\boldsymbol{\Sigma_{\mathcal{P}}}$ of the Gaia prior are then described by:

\begin{equation}
    \boldsymbol{\mu} = [\varpi, A, B, F, G, P, e, T_p, \gamma_i] \enspace ,
\end{equation}

\noindent and

\begin{equation}
\scalemath{0.54}{
    \boldsymbol{\Sigma_{\mathcal{P}}} = 
    \begin{pmatrix}
\sigma_{\varpi}^2&cov(\varpi,A)&cov(\varpi,B)&cov(\varpi,F)&cov(\varpi,G)&cov(\varpi,P)&cov(\varpi,e)&cov(\varpi,T_p)&0&\\
cov(A,\varpi)&\sigma_{A}^2&cov(A,B)&cov(A,F)&cov(A,G)&cov(A,P)&cov(A,e)&cov(A,T_p)&0&\\
cov(B,\varpi)&cov(B,A)&\sigma_{B}^2&cov(B,F)&cov(B,G)&cov(B,P)&cov(B,e)&cov(B,T_p)&0&\\
cov(F,\varpi)&cov(F,A)&cov(F,B)&\sigma_{F}^2&cov(F,G)&cov(F,P)&cov(F,e)&cov(F,T_p)&0&\\
cov(G,\varpi)&cov(G,A)&cov(G,B)&cov(G,F)&\sigma_{G}^2&cov(G,P)&cov(G,e)&cov(G,T_p)&0&\\
cov(P,\varpi)&cov(P,A)&cov(P,B)&cov(P,F)&cov(P,G)&\sigma_{P}^2&cov(P,e)&cov(P,T_p)&0&\\
cov(e,\varpi)&cov(e,A)&cov(e,B)&cov(e,F)&cov(e,G)&cov(e,P)&\sigma_{e}^2&cov(e,T_p)&0&\\
cov(T_p,\varpi)&cov(T_p,A)&cov(T_p,B)&cov(T_p,F)&cov(T_p,G)&cov(T_p,P)&cov(T_p,e)&\sigma_{T_p}^2&0&\\
0&0&0&0&0&0&0&0&\sigma_{\gamma_i}^2&\\
    \end{pmatrix}
}
\end{equation}

\noindent Then the natural $\log$ of the prior $\mathcal{P}$ is:

\begin{equation} \label{eq:logprior}
\ln \mathcal{P} = -\frac{8 + N_{inst}}{2}\ln(2\pi) -\frac{1}{2}\ln|\boldsymbol{\Sigma_{\mathcal{P}}}| -\frac{1}{2}(\boldsymbol{x}-\boldsymbol{\mu})^T\boldsymbol{\Sigma_{\mathcal{P}}}^{-1}(\boldsymbol{x}-\boldsymbol{\mu})
\end{equation}

\noindent where $N_{inst}$ is the number of RV instruments used, and $\boldsymbol{x}$ is a proposed solution to the fitted parameters $\boldsymbol{\mu}$.

In the combined analysis, the free parameters to be fitted include ${ \varpi, A, B, F, G, P, e, T_p, \gamma_i, \sigma_{J_i} }$. For the RV offsets ($\gamma_i$) and jitter terms ($\sigma_{J_i}$), there is one of each for each instrument available.

For the likelihood function we first convert from the Thiele-Innes elements to the Campbell elements (Equations \ref{eq:TI2campbell_a} to \ref{eq:TI2campbell_W}) taking into account the possible $\pi$ flip of $\Omega$ and $\omega$. We then calculate the RV model (Eq. \ref{eq:keplerians}) and the likelihood (Eq. \ref{eq:rv_likelihood}).

The final $\log$ probability of the full model is then calculated by summing Eq. \ref{eq:rv_likelihood} and Eq. \ref{eq:logprior}:

\begin{equation}
    \ln P = \ln \mathcal{P} + \ln \mathcal{L}
\end{equation}

We use the results obtained from the RV-only MCMC as the starting point for the joint RV-Gaia MCMC. In the post-processing, we sample points of the primary mass directly from a Gaussian distribution, with the mean and standard deviation from the values we got from the SED fit.

\section{Results}
\label{sec:results}

In this section, we present the outcomes of the combined RV and astrometric fit for the previously described sample. Although a highly similar analysis was conducted by \cite{2022AJ....164..196W} for the planetary companions, we aim to replicate their study on these systems to validate their findings, ensure completeness, and provide new RV measurements for some of the stars. Moreover, our analysis extends to brown dwarfs and, as we will demonstrate, certain binary systems as well.

Table \ref{tab:rvonlyMCMC} contains the results from the RV-only MCMC fits, whereas the results of the MCMC analysis for the joint fits of all stars are presented in Table \ref{tab:jointMCMC}. In \ref{fig:app-phasefolded} plots for all analyzed targets are shown with the raw RVs, the final joint solution shown on top, the residuals from this model, the phase folded RVs, and a Z-score statistic showing the deviations for the orbital parameters between the joint solution and the Gaia DR3 solution. For the Z-score the corrected $\omega$ and $\Omega$ values are used.

We present two sections of results: one showcasing solutions we deem to be robust, and another addressing challenging cases with less reliable outcomes. The targets are listed in ascending order of the final true mass of the companion. We limit the extent of this results section by not providing individual descriptions for all targets. However, we do present descriptions for all planetary companions, targets that have new data, and cases in which the companion undergoes a noteworthy classification change.

A thorough examination of the joint RV-Gaia solution for HD132406, HD81040, HD175167, and HD114762 has already been conducted by \cite{2022AJ....164..196W}, and our analysis yields similar results. For a detailed analysis, we refer the reader to their study. The only notable dissimilarity lies in the estimation of companion masses, as we employed our own estimation of primary masses. Table \ref{tab:jointMCMC} still presents our MCMC solutions for these targets.


\subsection{Robust Solutions}


\subsubsection{BD-170063}

BD-170063 is a K4 star at a distance of 34.5 pc with a Gaia magnitude of $G = 9.2$ and a mass of $0.778\pm0.037$ \Msun. A giant planet at a period of $655.6\pm0.6$ days and a minimum mass of $5.1\pm0.12$ \Mjup was discovered by \cite{2009A&A...496..513M} using the HARPS spectrograph and since the publication, an additional 5 radial velocities were taken with HARPS. Additionally, there are also 12 radial velocities taken with the HIRES instrument, and here we present 14 radial velocity measurements taken with the CORALIE spectrograph. All these additional measurements have never been used to reanalyze this system, which brings the total to 55 measurements on an observation baseline of 19 years.


\cite{2022AJ....164..196W} already performed a combined Gaia and RV fit, but only used the original HARPS data released in 2009. We repeat the analysis here for completeness and include the 10 additional years of data available to improve the precision of the orbital parameters.

The joint fit converged well, however, there are strong correlations between all the Thiele Innes coefficients. Specifically, there are positive correlations between AF and BG, and negative correlations between the rest of the combinations. This results in a negative correlation between the mean anomaly and the longitude of the ascending node, and between the inclination and the semi major axis. The orbit is close to edge on at $82.4^{+2.8}_{-2}$ degrees, which results in a true mass of the companion of $5.325 \pm 0.036$ \Mjup, which locks its status as a giant planet.

All orbital parameters from the combined fit MCMC can be found in Table \ref{tab:jointMCMC}. Indeed, with the additional RV data, we obtain a 2 to 4-fold improvement in the precision of the orbital parameters compared to those obtained by \cite{2022AJ....164..196W}.


\subsubsection{HD68638}

HD68638 is a late G type star at a distance of 32.5 pc and a mass of $1.00\pm0.12$ \Msun.
A companion to HD68638 was announced in \cite{2023A&A...674A..10H} based on the availability of RVs from the ELODIE spectrograph and their agreement with the Gaia astrometric solution for this system.

From the RV-only fit we obtain a good fit with an orbital period of $240.7\pm0.4$ days, an RV semi-amplitude of $325^{+20}_{-17}$ \ms, an eccentricity of $0.56\pm0.04$, and a minimum mass of $8.25\pm0.07$ \Mjup. The period, time of periastron and $\omega$ are all within 2-$\sigma$ of the Gaia only solution, however there is disagreement in the eccentricity. Gaia reports an eccentricity of $0.31\pm0.06$, which is more than 4-$\sigma$ away from the RV-only fit. Still, the system looks good for a combined Gaia-RV fit.

A joint Gaia-RV fit was performed with good convergence on all parameters. The eccentricity tension is settled at $0.487^{+0.035}_{-0.038}$, which meets in the middle, but stays closer to the eccentricity obtained in the RV-only fit. The semi-amplitude got reduced by 25\ms\ to $300\pm15$ and the inclination converged to $166.51 \pm 0.63$ degrees, resulting in a true mass of the companion of $35.1 \pm 1.4$. Removing the planetary status it had from the minimum mass and moving it inside the 30-55 \Mjup\ mass range.


\subsubsection{HD91669}

HD91669 is an early K type star at 71.8 pc. A brown dwarf candidate was discovered by \cite{2009AJ....137.3529W} using data from the Tull Spectrograph installed at the 2.7 m Harlan J. Smith Telescope at the McDonald Observatory. This brown dwarf candidate has a minimum mass of $30.6 \pm 2.1$ \Mjup, and sits on an eccentric orbit ($e = 0.448 \pm 0.002$) with a period of $497.5 \pm 0.6$ days. Later, \cite{2011A&A...525A..95S} tried to constrain this system using Hipparcos astrometry, but without success.

In the RV-only fit, we obtain the same results as reported in \cite{2009AJ....137.3529W}. There is good agreement with the Gaia solution in the period, time of periastron, and $\omega$, however, there is a slight disagreement in the eccentricity. The RV-only fit gives an eccentricity of $0.449\pm0.003$, while Gaia reports a value of $0.317\pm0.062$, a difference of just over 2-$\sigma$.

We performed the joint Gaia-RV fit and obtained good convergence on all parameters. The eccentricity didn't budge from the RV-only result, staying at $0.4485^{+0.0032}_{-0.0029}$. An inclination of $51.2^{+1.3}_{-1.1}$ degrees cemented HD91669b's position in the BD regime with a true mass of $38.09 \pm 0.64$ \Mjup.


\subsubsection{HD30246}

HD 30246 (HIP 22203) is a G1 star located at 51 pc from the Sun with our SED fit resulting in a mass of $0.956 \pm 0.081$\Msun. A BD candidate was found orbiting HD30346 by \cite{2012A&A...538A.113D} using data from the SOPHIE spectrograph with a minimum mass of $55^{+20.3}_{-8.2}$ \Mjup\ on a $990.7 \pm 5.6$ day orbit and an eccentricity of $e=0.838 \pm 0.081$. Since no constraints were found from the Hipparcos astrometric data at the time, a true mass estimation was not possible.

Since the publication of \cite{2012A&A...538A.113D}, 30 new RV measurements are available for HD30246 in the SOPHIE archive, spanning from August 2011 to November 2016. We performed an RV-only fit including this new data to derive updated orbital parameters for HD30246b. The posterior estimates for the orbital parameters from this RV-only fit can be found in Table \ref{tab:rvonlyMCMC}.
We obtain a period of $989.53\pm0.55$ days, an RV semi-amplitude of $1145.2 \pm 4.6$\ms, a lower eccentricity of $0.6588 \pm 0.0029$, and a lower minimum mass of $41.0 \pm 2.3$ \Mjup, which is partly because our measurement for the primary mass is lower than the one reported by \cite{2012A&A...538A.113D}.

The Gaia solution agrees well with the RV solution we obtained, all parameters being within 1-$\sigma$. Thus, a combined Gaia-RV fit was performed successfully, giving good constrains on the orbital inclination of the orbiting body. We recover a $990.08 \pm 0.58$-day orbit, with a semi-amplitude $K=1143.8 \pm 4.9$\ms, an eccentricity of $e=0.6605 \pm 0.0030$, and an orbital inclination of $85.0 \pm 1.2$ degrees, close to edge-on, which leaves the mass of the BD similar to its minimum mass at $42.18 \pm 0.23$ \Mjup. This means that this companion retains its status as a BD and within the 30-55 \Mjup\ mass range.


\subsubsection{BD-004475}
BD-004475 (HIP 114458, Gaia DR3 2651390587219807744) is a G type star located at 42 pc from the Sun. From our SED fits, we estimate its mass at $1.07 \pm 0.11$ \Msun. A BD candidate companion to this star was announced in \cite{2021A&A...651A..11D} on a $723.2 \pm 0.74$ day orbit, an eccentricity of $e=0.39 \pm 0.01$, and a minimum mass of $25.05 \pm 2.23$ \Mjup. Using the astrometric excess noise from Gaia DR1, they found a mass upper limit for BD-004475b of 125 \Mjup.

The joint solution converges to an orbit with a period of $723.71 \pm 0.83$ days, an eccentricity of $e=0.3803^{+0.0066}_{-0.0079}$, and an inclination of $139.63^{+0.63}_{-0.79}$ degrees which results in a true mass for BD-004475b of $50.93^{+0.47}_{-0.71}$ \Mjup. This is the last BD companion of this sample to fall inside the 30-55 \Mjup\ mass range.


\subsubsection{HD77065}

HD77065 is a G type star located at a distance of 33 pc and a mass of $0.791 \pm 0.042$ \Msun. A BD companion candidate was first found by \cite{2002AJ....124.1144L} using data from the SOPHIE spectrograph. Later, \cite{2016A&A...588A.144W} present additional data and improved precision to the orbital elements, obtaining an orbit with a period of 119 days, an eccentricity of $e=0.69$, and a minimum mass of $41\pm2$ \Mjup. Gaia finds this same companion, and a joint Gaia-RV fit provides the same basic orbital parameters with an inclination of $41.52 \pm 0.55$ degrees, which results in a true mass of $66.73^{+0.79}_{-0.73}$ \Mjup. This new mass estimate removes the BD from the 30-55 \Mjup\ mass range, but still clearly in the BD domain.


\subsubsection{CD-4610046}

CD-4610046 is a G star at a distance of 103 pc from the Sun and a mass of $0.860 \pm 0.049$ \Msun. A BD candidate companion to this star was announced in \cite{2023A&A...674A..10H} with Gaia and the support of 17 radial velocity measurements of the CORALIE spectrograph that have been taken between 2017 and 2018, which we present in this work.

We performed an RV only fit to obtain a robust orbital solution. We ran the MCMC with one keplerian in the model and obtained an orbit with a $242.48 \pm 0.31$ day period, an RV semi-amplitude of $1978 \pm 13$ \ms, an eccentricity of $e=0.4276^{+0.0042}_{-0.0048}$, and a minimum mass of $49.6 \pm 1.9$ \Mjup. This would put this companion right at the upper edge of the 30-55 \Mjup\ mass range. We have the Gaia astrometric solution available for this system, and a joint Gaia-RV fit is possible.


In the joint Gaia-RV fit we obtain good results for the complete orbit of CD-4610046b, with a period of $242.32 \pm 0.28$ days, an eccentricity close to the RV result of $e=0.4255^{+0.0045}_{-0.0041}$ and an inclination of $128.9 \pm 1.1$ degrees, which results in a true mass of the companion of $66.8^{+1.2}_{-1.1}$ \Mjup. This removes the companion from the BD driest mass region and is now a high mass BD.


\subsubsection{HD52756}

HD52756 is an early K type star at a distance of 32 pc. A $59.3\pm2.0$ \Mjup\ BD candidate companion was first discovered by \cite{2011A&A...525A..95S} on a 52.8 day orbit.

In this study, we introduce a new radial velocity measurement acquired using the CORALIE spectrograph. We performed both RV-only and combined Gaia-RV fits with this additional data point. However, our analysis indicates that the results are almost identical to the ones presented in the previous study by \cite{2011A&A...525A..95S} except for an updated true mass estimate of $66.94^{+0.91}_{-0.86}$ \Mjup.



\subsubsection{HD140913}

HD140913 is G type star at 49 pc, with an estimated mass of $0.987 \pm 0.087$ \Msun. A BD candidate companion was first published by \cite{1994AAS...184.4307S} with data from the CORAVEL spectrograph, and later analyzed by \cite{1998AcAau..42..593M} and \cite{2000A&A...355..581H}. This companion orbits at a period of $147.931^{+0.020}_{-0.022}$ days, with an eccentricity of $e=0.542^{+0.015}_{-0.018}$ and with a minimum mass of $40.0^{+3.2}_{-3.0}$ \Mjup. A combined Gaia-RV fit gives an orbital inclination of $30.3 \pm 1.3$ degrees, which results in a true mass of $93.3^{+1.7}_{-1.6}$ \Mjup. This is the first companion in this sample that turns out to be a very low mass star.


\subsubsection{HD148284}

HD148284 is a G type star at a distance of 122 pc from the Sun and an estimated mass of $1.39\pm0.2$ \Msun. \cite{2018AJ....156..213M} reported the existence of a BD candidate companion using data from the HIRES spectrograph. The companion was reported to have an orbital period of 339 days, an eccentricity of $e=0.39$, and a minimum mass of $33.7\pm5.5$ \Mjup. Our combined Gaia-RV fit converged well to similar orbital parameters, and an orbital inclination of $156.00^{+0.45}_{-0.56}$ degrees which results in a true mass of the companion of $103.6^{+2}_{-0.56}$ \Mjup, well within the stellar domain.


\subsubsection{HD48679}

HD48679 is a G0 type star at a distance of 67 pc from the Sun and has an estimated mass of $1.034 \pm 0.097$ \Msun. \cite{2019A&A...631A.125K} announced the presence of a $36.0 ± 1.3$ \Mjup\ BD candidate companion discovered with RV measurements from the SOPHIE spectrograph. This companion orbits at a period of $1111.61 \pm 0.30$ days and a semi-major axis of $2.145 \pm 0.037$ au. Using the astrometric excess noise from Gaia DR2, they predicted the inclination to be between 41 and 65 degrees. However, our combined Gaia-RV fit results in an inclination of $20.87 \pm 0.18$ degrees, which puts this companion in the very low mass star regime at $108.88^{+0.91}_{-1.50}$ \Mjup.





\subsubsection{HD112758}

HD112758 is an early K type star at a distance of 20 pc from the sun and has an estimated mass of $0.815 \pm 0.071$ \Msun. A 33 \Mjup\ BD candidate companion was first discovered by \cite{1997abos.conf..313M} using data from the CORAVEL spectrograph. It was later found to be a binary using astrometric data from Hipparcos \citep{2000A&A...355..581H, 2001ApJ...562..549Z}.

We present 9 new RV measurements taken with the CORALIE spectrograph taken in 2018. The results of the RV only analysis can be found in Table \ref{tab:rvonlyMCMC}, where we find a 5-fold improvement in the precision of the orbital parameters because of the much higher instrumental precision of CORALIE compared to CORAVEL. The combined Gaia-RV fit converged well without any significant tension in its results. We obtain an orbital inclination of $8.716 \pm 0.029$ degrees, which puts the true mass of the companion at $257.28 \pm 0.93$ \Mjup.


\subsubsection{HD164427}

HD164427 is an early G type star at a distance of 38 pc from the Sun. A BD candidate companion was first announced by \cite{2001ApJ...551..507T} using data from the UCLES spectrograph installed at the Anglo Australian Telescope. They found a $46.4 \pm 3.4$ \Mjup\ companion with an orbital period of $108.55 \pm 0.04$ days. Later, using Hipparcos data, \cite{2001ApJ...562..549Z} found its companion to be stellar with an estimated orbital inclination of $8.5$ degrees, reporting a true mass estimate of $367 \pm 84$ \Mjup. Lastly, \cite{2011A&A...525A..95S} analyzed the system again with Hipparcos data, but this time adding new data from the CORALIE spectrograph, and they found a slightly less inclined orbit with an inclination of 11.8 degrees and thus a lower mass of 272 \Mjup.

In this work, we present an additional 9 RV measurements taken with the CORALIE spectrograph from 2011 until 2017. In our analysis we combine the Doppler data from all available instruments (CORAVEL, UCLES, and CORALIE). The results from the RV-only fit can be found in Table \ref{tab:rvonlyMCMC}. Merging the supplementary data with all other datasets led to a significant improvement in the precision of orbital parameters, with an increase of nearly an order of magnitude compared to \cite{2011A&A...525A..95S}.

We performed the combined Gaia-RV fit with success and find an orbital period of $108.53855 \pm 3.3e-4$ days, an eccentricity of $e = 0.54944 \pm 7.3e-4$, and an inclination of $9.361^{+0.058}_{-0.050}$ degrees, which results in a true mass of the companion of $354.6 \pm 2.1$ \Mjup.


\subsubsection{HD162020}

HD162020 is an early K-type star at a distance of 31 pc from the Sun and has an estimated mass of $0.797 \pm 0.042$ \Msun. A companion to this star was first discovered by \cite{2002A&A...390..267U}, using RV data from the CORALIE spectrograph. They reported the existence of a hot brown dwarf orbiting with an orbital period of 8.4 days, an eccentricity of $e=0.277$, and a minimum mass of 14.4\Mjup. From calculations of the measured projected rotational velocity and circularization time, they estimated that the companion is probably a low mass brown dwarf. However, they couldn't exclude the possibility of it being a low mass star. Later, \cite{2011A&A...525A..95S} reanalyzed HD162020 using Hipparcos astrometry, but the sensitivity of Hipparcos was not high enough to pick up its companion.

Here, we present two decades of updated CORALIE data, encompassing 55 additional RV measurements. By conducting an RV-only analysis, we achieved notable enhancements in the accuracy of the orbital parameters. This includes a substantial 20-fold increase in the precision of the orbital period.

We proceeded with the combined Gaia-RV fit and obtained results that agree well with the independent RV analysis. We obtain an orbital period of $8.4282388^{+1.4e-6}_{-2.6e-6}$ days, an eccentricity of $e = 0.28126 \pm 5.7e-4$, a time of periastron of $T_p = -18.8742^{+0.0108}_{-0.0079}$ days, and a near face on inclination of $I = 177.273^{+0.030}_{-0.027}$ degrees. This results in a true mass of the companion of $410.8^{+5.8}_{-5.3}$ \Mjup, which puts it firmly in the stellar regime.



\subsection{Challenging cases}
\label{sec:challeging_cases}

In this subsection, we explore challenging cases encountered during our joint Gaia-RV fit. Discrepancies between the Gaia and RV solutions primarily contribute to the lack of a robust and reliable combined fit. We observe that these inconsistencies are mainly attributed to the unreliability of the Gaia data for these targets, while RV-only solutions remain mostly robust with the available Doppler data.


\subsubsection{HIP66074}

HIP66074 is a late K dwarf according to an effective temperature estimation of $4161^{+5}_{-3}$ K, from the {\mono teff\_gspspec} column in the {\mono gaiadr3.astrophysical\_parameters} table from Gaia DR3. From the SED fits, we obtained a primary mass of $0.73 \pm 0.03$ \Msun.

HIP66074b is one of only two astrometric planet discoveries by Gaia, presented by \cite{2023A&A...674A..10H} and \cite{2023A&A...674A..34G}. The cited works report the existence of a planetary companion at a period of 297.6 days.

Using the available HIRES RV data, the RV-only fit converges well to a period of $300.5^{+2.5}_{-3.1}$ days, K = $17.7^{+3.3}_{-1.8}$ \ms, eccentricity of $0.38 \pm 0.11$, and a minimum mass of $0.44 \pm 0.05$ \Mjup. However, because there are only 10 RV measurements available, the False Alarm Probability of this signal is at 33\%, much higher than the usual 1\% or 0.1\% used in radial velocity surveys to confirm the presence of a companion.

The joint fit MCMC had a lot of struggle to converge, needing 600,000 iterations to reach convergence, as a result of a discrepancy between the Gaia solution and the RV only solution. There is a big discrepancy between the estimated RV semi-amplitude from the Gaia solution $\left(297^{+82}_{-62}\ms\right)$ and the actual RV semi amplitude that was measured with the HIRES spectrograph $\left(17.7^{+3.3}_{-1.8} \ms\right)$. Additionally, the Gaia solution reports an inclination that is edge-on, but this is incompatible with the fact that the minimum mass of the RV solution is only 0.44 \Mjup, when the Gaia only solution would put the mass of the companion at 9 \Mjup.

In the joint solution, the inclination goes all the way down to $9.7^{+2.1}_{-1.4}$ degrees to reconcile the difference in semi amplitudes. This puts the estimation of the inclination in the joint solution more than 17$\sigma$ away from the Gaia solution. This shows that the tension in the model was much greater than the prior constraint to keep the inclination close to 90 degrees.

\cite{2022AJ....164..196W} shows that a non-zero flux ratio can partially explain this problem, however, they also show that the fitted flux ratio they obtain is incompatible with the expected planetary mass of the companion. As they already hypothesized, this big difference between the Gaia and the RV solution could be due to an unknown companion in the system. However, it's also possible that the astrometry of HIP66074 suffers from instrument systematics. More radial velocity measurements would be needed to confirm this signal at 300 days, and the release of Gaia DR4 will allow for a thorough joint analysis between the RVs and astrometry. Until then, the planetary nature of HIP66074b can not be confirmed and remains a candidate.


\subsubsection{HD40503}

HD40503 is an early K dwarf located at a distance of 39 pc. Based on our spectral energy distribution (SED) fit, we estimate the primary mass to be $0.797 \pm 0.044$ \Msun. This discovery represents the second independent candidate exoplanet detection from Gaia DR3 astrometry \citep{2023A&A...674A..10H, 2023A&A...674A..34G}. As previously noted by \cite{2023A&A...674A..10H}, \cite{2022AJ....164..196W}, and \cite{2023AJ....165..266M}, although Gaia's determined orbital period aligns with the best-fit Keplerians obtained from RV measurements, the available RV data remains insufficient to confidently constrain this system due to the presence of stellar activity or contamination. Therefore, it is currently unfeasible to perform a joint fit until additional RV data and/or the complete astrometry become available.



\subsubsection{HR810}

HR810 (Iota Horologium, HD 17051) is an F8 star at a distance of 17 pc. A giant planet was discovered around HR810 by \cite{2000A&A...353L..33K}. A good review about this systems' solution was already shown by \cite{2022AJ....164..196W}.
They report that the analysis of HR 810 using RV-only and Gaia data reveals significant discrepancies and uncertainties in orbital parameters, notably affected by systematic errors and a known issue with Gaia's two-body fitting code for low eccentricity orbits. Due to these inconsistencies, the joint analysis of both datasets is inconclusive, necessitating caution in its interpretation. Our joint fit analysis resulted in the same conclusion, which is why we restrain from showing the joint fit solution.




\subsubsection{HD5433}

HD5433 is a G-type star at a distance of 64 pc from the Sun and with a mass of $0.984 \pm 0.096$ \Msun. A highly eccentric BD candidate companion was found by \cite{2021A&A...651A..11D} with data from the SOPHIE spectrograph. They announced an orbital period of $576.6 \pm 1.59$ days, an eccentricity of $e=0.81 \pm 0.02$ and a minimum mass of $49.11 \pm 3.4$ \Mjup. A joint Gaia-RV fit was performed with success, resulting in an orbital inclination of $41.34 \pm 0.77$ degrees, which gives a true mass of $69.19^{+0.91}_{-0.83}$ retaining its position as a BD, but moving away from the desert.

There exists a significant disparity between the semi-major axis estimates obtained from Gaia and the joint fit. This discrepancy is further evident when comparing the expected RV semi-amplitudes. Gaia predicts a value of 295 \ms, which is six times smaller than the estimate derived from the RVs, measuring 1790 \ms. One possible explanation for this discrepancy is an inaccurate inclination estimation by Gaia, which reports an inclination of $I=12\pm39$ degrees. In contrast, our joint fit yielded an inclination four times higher at $I = 41.34 \pm 0.77$ degrees in order to reconcile the astrometry with the RVs, but still compatible within 1$\sigma$. 

Even though the exact cause of the semi-major axis discrepancy remains uncertain, and thus the provided solution should be taken with care, we anticipate gaining greater clarity upon the release of Gaia DR4.


\subsubsection{HD82460}

HD82460, is a G type star at a distance of 50 pc from the Sun. This star was observed with the SOPHIE spectrograph and \cite{2019A&A...631A.125K} announced a $73.2 \pm 3.0$ \Mjup\ BD candidate companion at a $590.90 \pm 0.24$ day orbit.

The results of the combined fit of the data from SOPHIE and Gaia exhibit suboptimal performance, as evidenced by a jitter of approximately 80 m/s. Additionally, the orbital parameters obtained from the combined fit are all approximately 1.5 standard deviations higher than the solutions from Gaia. However, the convergence of the combined fit is impressive, as it converges rapidly and the sampling is effective.

Our analysis suggests that the difference in the results may be attributed to the eccentricity estimation of Gaia, which obtains $e = 0.73\pm0.03$, in contrast to the eccentricity from the radial velocity data, which is $e = 0.95\pm0.02$. This is a tension of more than 7 $\sigma$. In the joint fit, it seems that the likelihood is minimized by converging to the lower Gaia eccentricity and consequently increasing the RV jitter, even though this is likely the wrong solution.

Additionally, the lower eccentricity also results in a lower RV semi-amplitude. According to the Gaia solution, the estimated RV semi-amplitude (Eq. \ref{eq:K_from_astrometry}) is $K=2046$\ms, which is less than half the semi-amplitude obtained from the RV-only analysis. The lower $K$ value from the combined fit leads to a contradictory smaller true mass than the minimum mass obtained from the RV data alone.

These discrepancies suggest potential issues with the Gaia solution, given our confidence in the RV solution. Further analysis of HD82460 is required when the full astrometric time series becomes available in Gaia DR4 in order to identify an explanation for this tension between the two solutions. We present the complete results of the combined fit in Table \ref{tab:jointMCMC}, but they should be interpreted with caution, as they may not be reliable.


\subsubsection{HD89707}

HD89707 is a G2 type star at a distance of 35 pc from the Sun with an estimated mass of $0.99\pm0.13$ \Msun. It was first started to be observed in 1982 with the CORAVEL spectrograph and was first identified as a spectroscopic binary by \cite{1991A&A...248..485D}. Later it was analyzed by \cite{2000A&A...355..581H} and \cite{2011A&A...525A..95S}, where they flag it as $53.6\pm7.4$ \Mjup\ BD candidate with an orbital period of 298 days and an eccentricity of $e=0.9$.

We present 13 new radial velocity measurements taken with the CORALIE spectrograph between 2010 and 2021, taken after \cite{2011A&A...525A..95S} did their analysis. Additionally, at the time of their publication, the 64 RV measurements from the CORAVEL spectrograph taken between 1982 and 1999 were not used in their analysis. In this work, we are presenting an updated analysis of this system, now with the complete dataset of both instruments. This represents 77 additional RV measurements for a total of 108, which means four times the observation baseline (close to 40 years) and triple the amount of RV measurements.

We did an RV-only analysis, and we found a period of $298.238^{+0.015}_{-0.014}$ days, a higher eccentricity of $e=0.9461^{+0.0034}_{-0.0038}$, an RV semi-amplitude of $5070 \pm 180$ \ms, and a minimum mass of $53.7 \pm 4.5$ \Mjup. Throughout all orbital parameters, we obtain almost an order of magnitude better precision to the ones reported by \cite{2011A&A...525A..95S}. The full orbital parameters from the RV fit can be seen in Table \ref{tab:rvonlyMCMC}.

After comparing the Gaia orbital solution to the RV solution, we find that the period and time of periastron match very well at less than $1\sigma$. However, Gaia finds a lower eccentricity of $0.67\pm0.20$, but still less than $2\sigma$ away from the RV eccentricity. Still, the agreement is good enough and we continued with the combined Gaia-RV fit.

The result of the combined fit gives an even higher eccentricity compared to the RV fit. The posterior distribution shows to be concentrated at $e=1$ with a downward tail. This behavior is unexpected as the eccentricity is well constrained by the RV data at $e=0.9464 \pm 0.0041$ and the prior information from Gaia should push it to an even lower value. A comparison of the Bayesian Information Criterion (BIC) between a model with an eccentricity of 0.94 and another with 0.98 resulted in a difference of -4, in favor of the former. This difference, however, is not deemed statistically significant. In really high eccentricities it's difficult to explore the peaks of RV and thus higher eccentricities are always possible, however, this doesn't explain why the eccentricity posterior looks the way it does.

In the combined Gaia-RV fit we obtain an orbital inclination of $43.8 \pm 1.1$, which results in a true mass of $105.7^{+1.2}_{-1.1}$ \Mjup, but due to the discrepancies in the estimation of the eccentricity the solution should be taken with caution.


\subsection{HD3277}

HD3277 is a late G type star at a distance of 29 pc from the Sun with an estimated mass of $0.942^{+0.110}_{-0.096} \Msun$. A potential brown dwarf candidate was found by \cite{2011A&A...525A..95S} on a 46-day period orbit using radial velocity data from the CORALIE spectrograph. In that same work, they conducted a comparative study with Hipparcos astrometry and determined that the companion to HD3277 was in fact a star with an estimated true mass of $344\pm76 \Mjup$.

Since \cite{2011A&A...525A..95S} we have collected 10 additional RV measurements with the CORALIE spectrograph over a timespan of 12 years, which already provides us with an improved precision in the orbital parameters from the RV only fit by close to an order of magnitude (see Table \ref{tab:rvonlyMCMC}).

We performed the combined Gaia-RV fit with success, obtaining a period of $46.151350^{+8.2e-5}_{-6.2e-5}$ days, an inclination of $169.810^{+0.040}_{-0.035}$ degrees which results in a true mass of $489.4^{+2.5}_{-2.1} \Mjup$, considerably higher than the one reported by \cite{2011A&A...525A..95S}. 

However, as reported in \cite{2011A&A...525A..95S}, the presence of a $\sim\!500$ \Mjup\ companion results in its light contaminating the fiber and subsequently altering the cross-correlation function (CCF) and hence the RVs. This can be seen by the large scatter in the Full Width Half Maximum of the CCF, indicating the blended SB2 nature of this target. A similar phenomenon occurs in astrometry, as the binary nature of the system results in the photocenter being influenced by the companion and shifted away from the primary, thus affecting the orbital solution. If these effects are not accounted for, they introduce biases into the orbital solutions derived with both the RVs and Gaia astrometry. Specifically, the RV semi-amplitude and the astrometric amplitude are underestimated, which leads to an underestimation of the mass of the stellar companion. Fully separating each star's contributions to the CCF requires an in depth analysis, which is beyond the scope of this paper.

For this reason, the results for HD3277 from both the RV only (Table \ref{tab:rvonlyMCMC}) and the joint MCMC fit (Table \ref{tab:jointMCMC}) should be taken with caution, as a more thorough analysis is needed to take into account the shift of the photocenter in Gaia and the affected RVs.


\subsubsection{HD17289}

HD17289 is a G type star at a distance of 50 pc and has an estimated mass of $1.11 \pm 0.17$ \Msun. A companion to HD17289 on a 536-day orbit was first discovered by \cite{2007ApJS..173..137G} using only Hipparcos data. Then, \cite{2011A&A...525A..95S} reanalyzed the system using both Hipparcos and RV data from the CORALIE spectrograph. They find better constrained orbital parameters and discover the companion to be a star with a mass of $547 \pm 47$ \Mjup, even though it would be a BD candidate from the RVs alone. More recently, \cite{2022A&A...667A.172D} analyzed this system again using a fully joint astrometric and RV model, obtaining similar results.

We present 6 new RV measurements taken with the CORALIE spectrograph from 2011 until December 2016. We perform an RV-only fit and obtain a 2 to 7-fold improvement in precision of the orbital parameters thanks to the new Doppler data (see Table \ref{tab:rvonlyMCMC}).

There is good agreement for the period, and time of periastron between the Gaia and RV solutions. However, the eccentricities look close at $0.528\pm0.002$ and $0.492\pm0.005$ for the RV and Gaia solutions, respectively, but the small uncertainty on the Gaia estimate puts them at a distance of 7$\sigma$. Same for the $\omega$, the RV and Gaia estimates are 8$\sigma$ apart. We decided to go forward with the combined Gaia-RV fit with these discrepancies in mind.

From the combined Gaia-RV fit, we obtain an orbital period of $561.877^{+0.100}_{-0.085}$ days, an eccentricity of $e = 0.5152 \pm 0.0037$, and an inclination of $I = 172.917 \pm 0.044$. This puts the true mass of the companion at $540.6^{+2.1}_{-2.0}$ \Mjup. There is still a strong tension (at > 4$\sigma$) in the precise estimate of the eccentricity, $\omega$ and $\Omega$. The origin of this tension is likely due to the impact of the companion on the RVs. 

Similarly to HD3277, the large mass of the stellar companion affects both the RVs and the astrometry. The derived orbital solution will be biased if the non-negligible flux ratio is not properly accounted for, and thus the mass of the companion will be underestimated. Such an analysis is beyond the scope of this paper, and therefore, we warn the readers to exercise caution with the orbital solution of HD17289 provided in both Table \ref{tab:rvonlyMCMC} and Table \ref{tab:jointMCMC}.




\section{Discussion}
\label{sec:discussion}

The sample presented in this study is limited in its representativeness, as it consists only of the giant planets and brown dwarf candidates that were verified and validated by Gaia. Hence, the general population-level insights that can be drawn from these results are of modest significance. Nonetheless, we can still present some general conclusions.

In Table \ref{tab:jointMCMC} the results for the combined Gaia-RV fit are shown for all targets. We split up the table in two, for the robust solutions and for the systems where the solution might be unreliable because of discrepancies between the Gaia solution and the RV solution. For this discussion, we will only consider the systems where a reliable combined solution was found.

The use of Gaia astrometry in recent years has significantly enhanced the ability to accurately determine the true masses of giant planets and brown dwarfs. Using the PMa technique, \cite{2023A&A...674A.114B} showed that 11 BD candidate companions from the CORALIE sample turn out to be stars. In this study, we add another eight to that list, as HD17155 was already found to be a binary by \cite{2023A&A...674A.114B}. For HD3277, HD89707, HD151528, and HD164427A, \cite{2023A&A...674A.114B} didn't manage to get additional constraints as the semi-major axis of their orbits are less than 1 au which prevents a reliable analysis using PMa. In another case, \cite{2021AJ....162...12V} show that the low mass BD companion candidate to HD92987 \citep{2019A&A...625A..71R} is actually an M dwarf.

\begin{figure}
    \centering
    \includegraphics[width=\columnwidth]{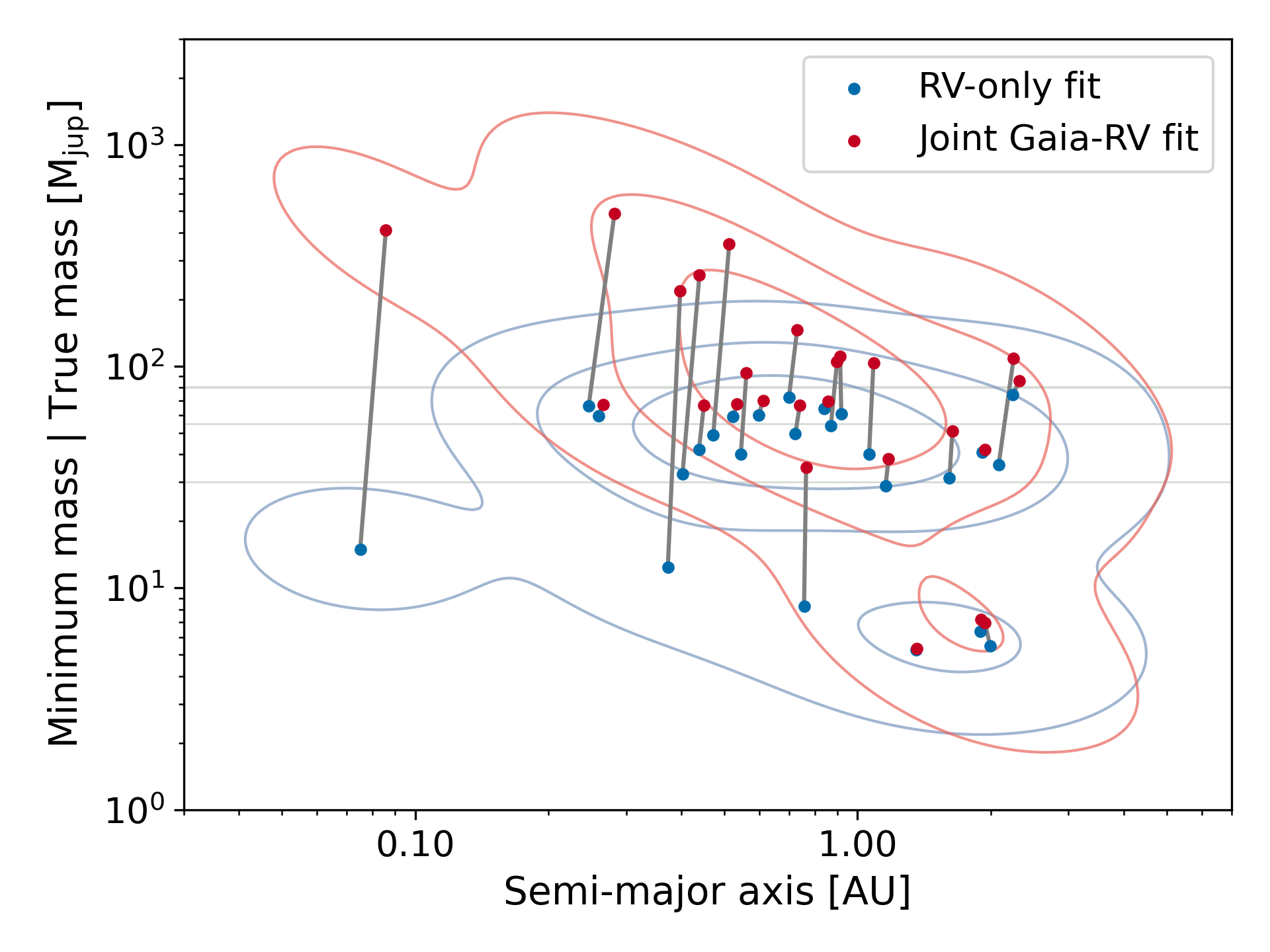}
    \caption{[Minimum/True] mass of the companion and the semi-major axis for the RV-only solutions (blue) and the joint Gaia-RV solution (red). The horizontal solid lines show the 30, 55, and 80 \Mjup\ mass limits, respectively. A kernel density estimate (KDE) plot is shown on top for each set of solutions. Only companions with robust solutions are shown, i.e, targets from the first section of Table \ref{tab:jointMCMC}.}
    \label{fig:before_vs_after}
\end{figure}

\begin{figure}
    \centering
    \includegraphics[width=\columnwidth]{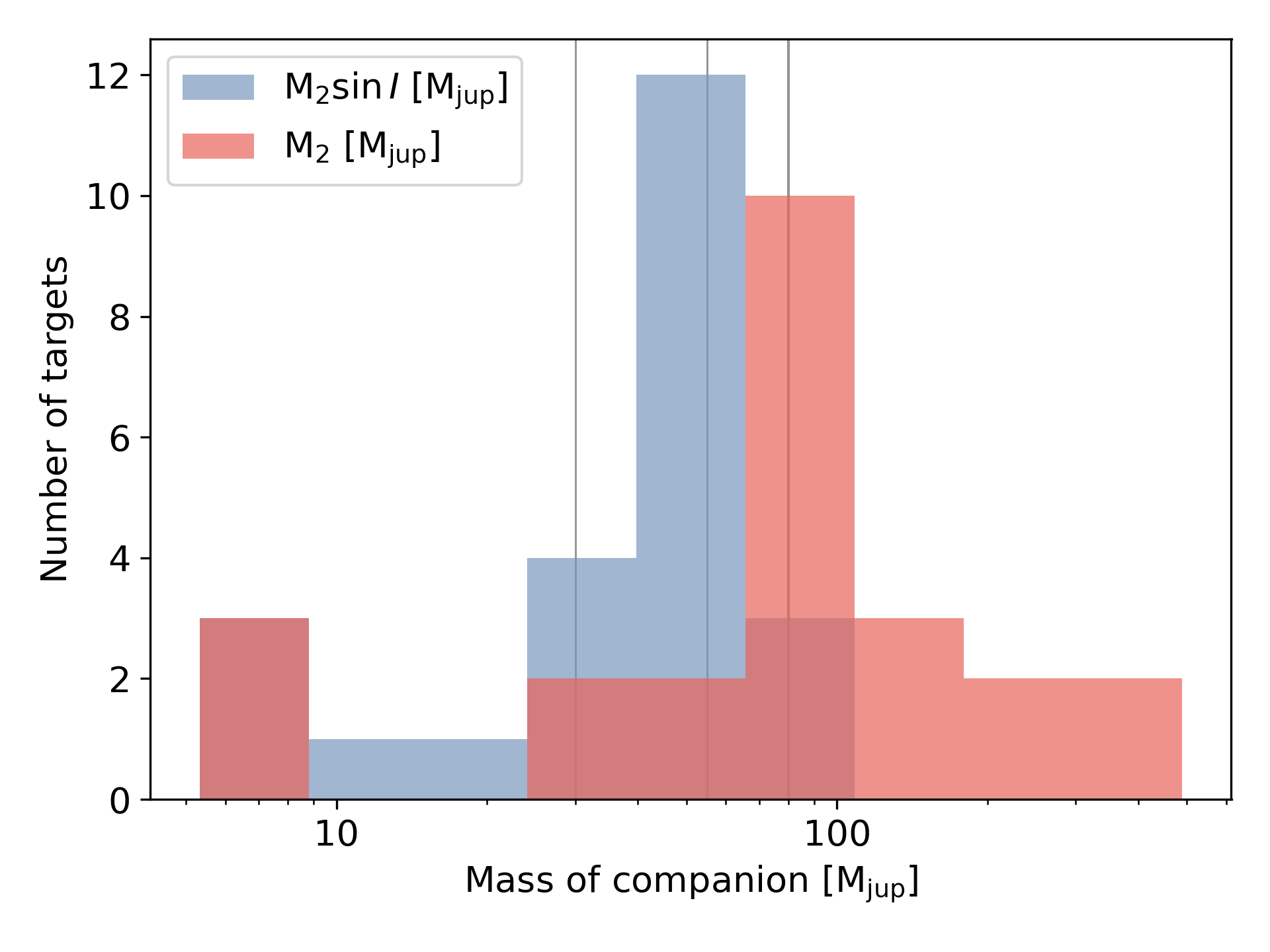}
    \caption{Histogram of companion masses for minimum and true masses. The horizontal solid lines show the 30, 55, and 80 \Mjup\ mass limits, respectively.}
    \label{fig:masses_hist}
\end{figure}

\begin{figure}
    \centering
    \includegraphics[width=\columnwidth]{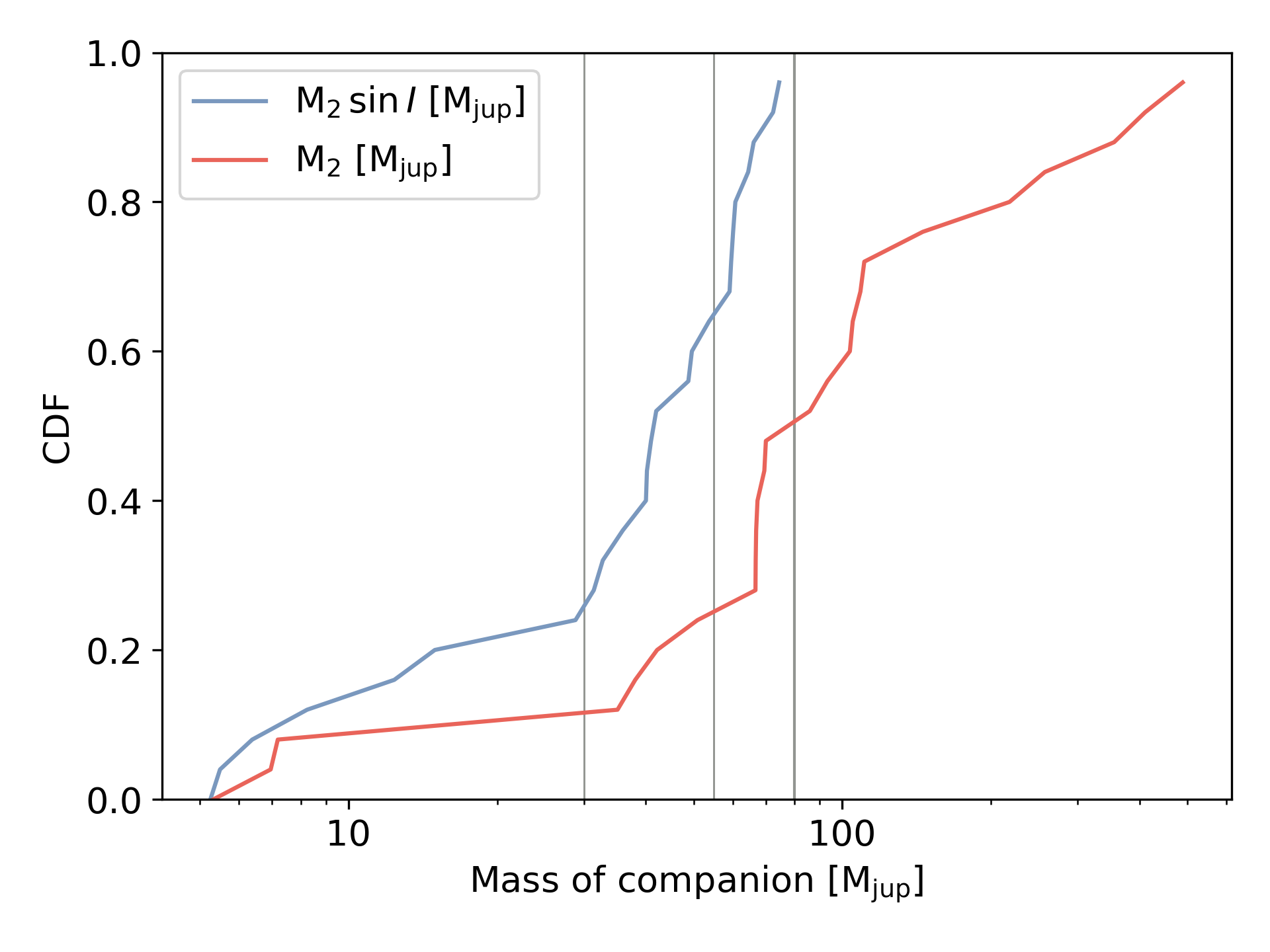}
    \caption{Cumulative Distribution Function for the masses of the companions in minimum (blue) and true (red) masses. The horizontal solid lines show the 30, 55, and 80 \Mjup\ mass limits, respectively.}
    \label{fig:masses_cdf}
\end{figure}

This highlights the prominence of the scarcity of BDs and that the occurrence rate of BDs in the $30-55 \Mjup$\ mass range might be lower than we thought. Identifying these false BD candidates is crucial if we really want to have robust observational constraints for planet formation theories. In Fig. \ref{fig:before_vs_after}, \ref{fig:masses_hist}, and \ref{fig:masses_cdf} we show the change in the distribution of masses from the RV only to the combined Gaia-RV mass, which all show a clear general increase in the masses of the companion. The median mass increase factor is of 1.88, which is higher than the expected factor of 1.15 assuming a median inclination of $I=60^{\circ}$ from randomly oriented orbits. This is a consequence of Gaia being more likely to detect face-on orbits due to the properties of its scanning law \citep{2023A&A...674A..34G}.

\begin{figure}
    \centering
    \includegraphics[width=\columnwidth]{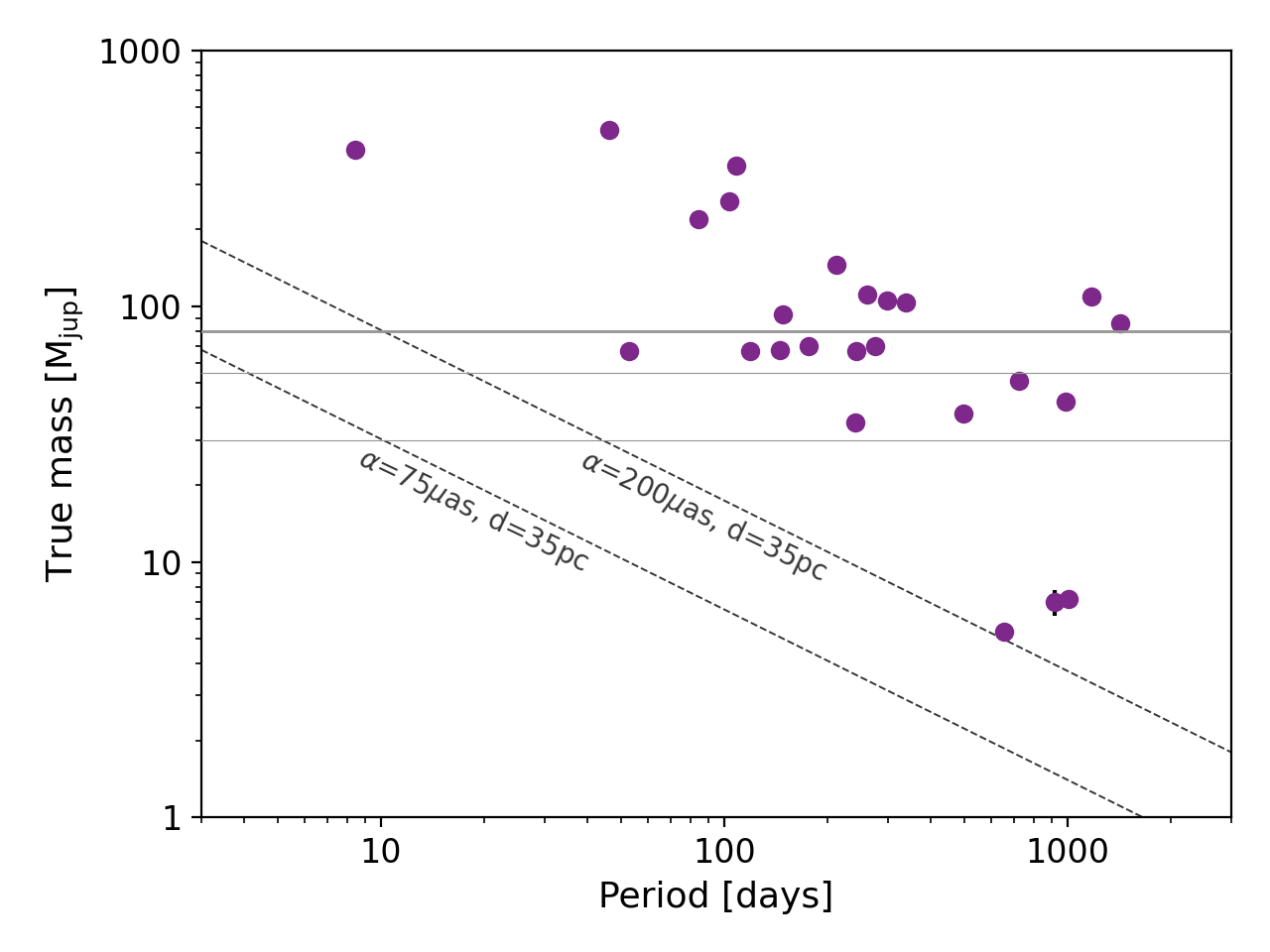}
    \caption{True mass of the companions versus the orbital period. The horizontal solid lines show the 30, 55, and 80 \Mjup\ mass limits, respectively. The dashed lines show the expected detection mass limits for companions around solar like stars (1\Msun) at a distance of 35pc considering astrometric signatures of $\alpha=200\mu\mathrm{as}$ (approximate current capacity) and $\alpha=75\mu\mathrm{as}$ (approximate astrometric signature expected to be detectable in Gaia DR4 considering a base uncertainty of 25$\mu$as and an SNR of 3).}
    \label{fig:mass_vs_period}
\end{figure}

In Fig. \ref{fig:mass_vs_period} we show the distribution of true masses and orbital periods. There is a clear downward trend in the masses for longer orbital periods. This can be easily associated to the detection capabilities of Gaia, where longer orbits produce larger astrometric signatures and thus smaller masses are detectable. For short orbits, the mass of the companion has to be larger for the astrometric signature to be of similar magnitude. The smallest orbit found in this sample is that of HD162020B, whose period is only 8.4 days. However, its detection by Gaia was possible solely due to its stellar nature.

When comparing the results obtained from Gaia only to the results obtained from RV only, it is apparent that Gaia still struggles to accurately determine the eccentricity of the orbits, typically reporting lower values than the ones we obtain from the RV analysis. This is possibly due to the limited time span and sampling of Gaia observations for DR3 and a still not-perfect outlier rejection scheme in the Gaia time series analysis. This issue was previously acknowledged by \citep[][Fig. 3]{2023A&A...674A..10H}, who noted that Gaia tends to fit lower eccentricities than the reference values from RV surveys. Additionally, some of the large Z-scores seen in Fig. \ref{fig:app-phasefolded} could be due to underestimated uncertainties in the Gaia solution. \cite{2023A&A...674A..10H} described that the covariance matrix of the parameters was estimated locally around the best fit solution (from the partial derivatives of the model) which could lead to underestimation.

In contrast, the period is typically recovered with high accuracy by Gaia and in agreement with the period derived from RV only. The high accuracy of the period measurements obtained by Gaia highlights the potential of this data source to contribute to our understanding of companion populations, despite its current limitations in determining the eccentricity of the orbits. 

\subsection{Brown-dwarf companions in the CORALIE survey}

These updated masses of brown dwarf companions allows us to revisit their occurrence rate in the CORALIE survey. The CORALIE survey focuses on a volume-limited sample consisting of 1647 solar-type stars (FGK) located up to a distance of 50 pc \citep{2002A&A...388..632P, 2002A&A...390..267U}. Within this sample, \cite{2011A&A...525A..95S} identified 11 brown dwarf candidates with masses ranging from 13 to 80 \Mjup\ that were found in close orbits up to 10 au. However, it should be noted that the completeness of the survey up to 10 au was not guaranteed at that time. The CORALIE survey had been running for 12 years, providing complete orbit data only for companions up to $a\!\sim\!5$~au. Some companions with larger separations could be inferred from incomplete orbit data, but the detection completeness up to 10 au was not ensured. Now, with the CORALIE survey having run for 25 years, we can confidently assert the detection of companions with complete orbits at a separation of 10 au.

Among the 11 brown dwarfs initially considered by \cite{2011A&A...525A..95S} (HD4747, HD52756, HD74014, HD89707, HD154697, HD162020, HD167665, HD168443, HD189310, HD202206, HD211847), subsequent studies have revealed that five of them are actually of stellar nature. These stars are HD89707 and HD162020 (this work), HD154697 \citep{2023A&A...674A.114B}, HD211847 \citep{2017A&A...602A..87M}, and HD202206 \citep{2017AJ....153..258B}. On the other hand, for the remaining six stars, some of their companions have been confirmed as brown dwarfs: HD74014, HD167665, HD4747 \citep{2023A&A...674A.114B}, and HD52756 (this work).

Since the publication of \cite{2011A&A...525A..95S}, additional close-in brown dwarfs have been discovered in the CORALIE sample. These include HD28454, HD30774, HD112863, HD153284, HD184860A, HD206505, and HD219709 \citep{2023A&A...674A.114B}. Among these, the companions of HD112863 and HD206505 were confirmed as brown dwarfs through an analysis of proper motion anomalies, while for the others either the period of the companion is too short or it's not available in the Hipparcos catalog which prevents this type of analysis.

The revised count of brown dwarfs with masses ranging from 13 to 80 \Mjup\ and orbits below 10 au in the CORALIE survey now stands at 13, namely HD28454, HD30774, HD4747, HD52756, HD74014, HD112863, HD153284, HD167665, HD168443, HD184860A, HD189310, HD206505, and HD219709. Among these companions, six (HD28454, HD30774, HD153284, HD184860A, HD189310, and HD219709) still require determination of their true masses. For three of them (HD28454, HD153284, and HD219709), their minimum mass is above 70 \Mjup, thus it's likely that some of these companions are actually stars. Taking into account this updated list, we can estimate a tentative occurrence rate of close-in (<10AU) brown-dwarf companions around Sun-like stars within 50 pc. Using binomial statistics, the rate is 13/1647 = $0.8^{+0.3}_{-0.2} \%$ which is slightly higher than the estimate in \cite{2011A&A...525A..95S} but compatible. The increase is explained by the longer time range of the observations, covering better out to 10AU. It is important to highlight that there remains a possibility that some candidates from the CORALIE sample initially identified as planetary companions may turn out to be brown dwarfs, or for some of these brown dwarfs to be stars.

Of these 13 brown dwarf companions, only 1 has a mass in the 30 to 55 \Mjup\ range, namely HD30774 with $m\sin(I)=41\Mjup$ \citep{2023A&A...674A.114B}. However, its true mass is still to be determined.


\section{Summary and Conclusions}
\label{sec:conclusion}

We revisited stars known to have substellar companion candidates that were initially detected through radial velocities and later validated using astrometry data from Gaia. By using the orbital solution provided by Gaia DR3, we incorporated it as a prior in the analysis of the available RV data. This joint Gaia-RV analysis allowed us to derive a comprehensive 3D representation of the companion's orbit, enabling us to determine their true mass and unveil their actual nature.

Our analysis focused on 32 targets deemed suitable for this investigation, specifically those with substellar companion candidates and no additional companions that would hinder a comparison with the Gaia orbital solution. Among these, 2 companions were deemed unsuitable for a joint analysis because of large inconsistencies between the RVs and the Gaia solutions, as well as unreliable RVs. An additional 6 companions yielded joint solutions that exhibited some discrepancies between the RVs and Gaia; these solutions should be approached with caution. Thus, 24 companions were found to have robust joint Gaia-RV solutions.

We examined the companions previously classified as planets based on their RV analysis, which had already been reevaluated by \cite{2022AJ....164..196W} using a similar approach. Our results largely aligned with theirs, but thanks to the inclusion of new RV data, we obtained improved orbital constraints for BD-170063. We also determined that the companion HD68638A is a BD, and once again confirmed the stellar nature of HD114762B.

Among the BD candidates, ten previously considered to have minimum masses falling between the 30-55 \Mjup\ mass range were found to be either high-mass BDs (HD77065, CD-4610046) or low-mass stars (HD140913, HD148284, HD89707, HD48679, HD112758, HD164427). This significant reduction in BD candidates within the desert region suggests an even lower occurrence rate of these companions. The improved constraints and identification of the true nature of many of these companions will greatly enhance the refinement and reliability of high-mass planet and BD populations. This in turn will lead to tighter constraints on future planet formation theories.

The main changes in companion masses can be summarized by categorizing them into the three ranges based on their minimum masses:

\begin{itemize}
    \item $[<30 \Mjup]$: 7 companions in the sample $\Rightarrow$ 2 are within 30-55\Mjup\ and 2 are stars (>80\Mjup).
    \item $[30-55 \Mjup]$: 9 companions in the sample $\Rightarrow$ 2 remain in this range, 2 are high mass BD (55-80\Mjup), and 5 are stars (>80\Mjup).
    \item $[55-80 \Mjup]$: 8 companions in the sample $\Rightarrow$ 4 are stars (>80\Mjup).
\end{itemize}

We updated the analysis of close-in brown dwarf companions ($13-80\Mjup$) within 10 au of solar type stars in the volume limited CORALIE sample. \cite{2011A&A...525A..95S} reported eleven brown dwarfs, but later studies showed that five of them were actually stars. \cite{2023A&A...674A.114B} added seven new brown dwarfs to the sample. Out of 1647 stars in the CORALIE sample, we found 13 brown dwarf candidates, giving an upper limit for their occurrence rate of $0.08^{+0.03}_{-0.02}$.

There are still areas in need of improvement, as some joint solutions do not align with the RV-only solutions. We anticipate that Gaia DR4 will address most of these issues by providing not only a doubled observation baseline but also complete epoch astrometry, facilitating comprehensive combined analyses of RV and astrometry datasets \citep{2022A&A...667A.172D}. This upcoming release will significantly enhance our ability to detect objects with lower masses than is currently possible and provide a clearer understanding of the populations existing at the transitional boundary between planets and low-mass brown dwarfs within 5 au. Gaia DR4 will also prove invaluable in studying active and young stars, where the presence of planets poses challenges to RV measurements. These advancements will enable us to study brown dwarf populations and their orbital architectures as a function of the host stellar parameters with an unprecedented level of detail.

As mentioned previously, we had to exclude multi-planet systems from the analysis because Gaia DR3's search was focused on a single companion, rendering a reliable joint analysis for multi-planet systems unfeasible. However, with the forthcoming Gaia DR4, we are confident that the study of multi-planet systems will become possible, enabling the detection of multi-giant systems.


\begin{acknowledgements}
This work has been carried out within the framework of the National Center of Competence in Research PlanetS supported by the Swiss National Science Foundation under grants 51NF40\_182901 and 51NF40\_205606. The authors acknowledge the financial support of the SNSF. \\
This publication makes use of The Data \& Analysis Center for Exoplanets (DACE), which is a facility based at the University of Geneva (CH) dedicated to extrasolar planets data visualisation, exchange and analysis. DACE is a platform of the Swiss National Centre of Competence in Research (NCCR) PlanetS, federating the Swiss expertise in Exoplanet research. The DACE platform is available at \url{https://dace.unige.ch.}\\
This work has made use of data from the European Space Agency (ESA) mission {\it Gaia} (\url{https://www.cosmos.esa.int/gaia}), processed by the {\it Gaia} Data Processing and Analysis Consortium (DPAC, \url{https://www.cosmos.esa.int/web/gaia/dpac/consortium}). Funding for the DPAC has been provided by national institutions, in particular the institutions participating in the {\it Gaia} Multilateral Agreement. \\
This work made use of Astropy:\footnote{http://www.astropy.org} a community-developed core Python package and an ecosystem of tools and resources for astronomy \citep{2013A&A...558A..33A, 2018AJ....156..123A, 2022ApJ...935..167A}.
\end{acknowledgements}


\bibliographystyle{aa} 
\bibliography{ads_bibliography.bib}

\begin{thebibliography}{93}
\expandafter\ifx\csname natexlab\endcsname\relax\def\natexlab#1{#1}\fi

\bibitem[{{Anders} {et~al.}(2019){Anders}, {Khalatyan}, {Chiappini}, {Queiroz},
  {Santiago}, {Jordi}, {Girardi}, {Brown}, {Matijevi{\v{c}}}, {Monari},
  {Cantat-Gaudin}, {Weiler}, {Khan}, {Miglio}, {Carrillo}, {Romero-G{\'o}mez},
  {Minchev}, {de Jong}, {Antoja}, {Ramos}, {Steinmetz}, \&
  {Enke}}]{2019A&A...628A..94A}
{Anders}, F., {Khalatyan}, A., {Chiappini}, C., {et~al.} 2019, \aap, 628, A94

\bibitem[{{Armitage} \& {Bonnell}(2002)}]{2002MNRAS.330L..11A}
{Armitage}, P.~J. \& {Bonnell}, I.~A. 2002, \mnras, 330, L11

\bibitem[{{Arriagada} {et~al.}(2010){Arriagada}, {Butler}, {Minniti},
  {L{\'o}pez-Morales}, {Shectman}, {Adams}, {Boss}, \&
  {Chambers}}]{2010ApJ...711.1229A}
{Arriagada}, P., {Butler}, R.~P., {Minniti}, D., {et~al.} 2010, \apj, 711, 1229

\bibitem[{{Astropy Collaboration} {et~al.}(2022){Astropy Collaboration},
  {Price-Whelan}, {Lim}, {Earl}, {Starkman}, {Bradley}, {Shupe}, {Patil},
  {Corrales}, {Brasseur}, {N{\"o}the}, {Donath}, {Tollerud}, {Morris},
  {Ginsburg}, {Vaher}, {Weaver}, {Tocknell}, {Jamieson}, {van Kerkwijk},
  {Robitaille}, {Merry}, {Bachetti}, {G{\"u}nther}, {Aldcroft},
  {Alvarado-Montes}, {Archibald}, {B{\'o}di}, {Bapat}, {Barentsen},
  {Baz{\'a}n}, {Biswas}, {Boquien}, {Burke}, {Cara}, {Cara}, {Conroy},
  {Conseil}, {Craig}, {Cross}, {Cruz}, {D'Eugenio}, {Dencheva}, {Devillepoix},
  {Dietrich}, {Eigenbrot}, {Erben}, {Ferreira}, {Foreman-Mackey}, {Fox},
  {Freij}, {Garg}, {Geda}, {Glattly}, {Gondhalekar}, {Gordon}, {Grant},
  {Greenfield}, {Groener}, {Guest}, {Gurovich}, {Handberg}, {Hart},
  {Hatfield-Dodds}, {Homeier}, {Hosseinzadeh}, {Jenness}, {Jones}, {Joseph},
  {Kalmbach}, {Karamehmetoglu}, {Ka{\l}uszy{\'n}ski}, {Kelley}, {Kern},
  {Kerzendorf}, {Koch}, {Kulumani}, {Lee}, {Ly}, {Ma}, {MacBride}, {Maljaars},
  {Muna}, {Murphy}, {Norman}, {O'Steen}, {Oman}, {Pacifici}, {Pascual},
  {Pascual-Granado}, {Patil}, {Perren}, {Pickering}, {Rastogi}, {Roulston},
  {Ryan}, {Rykoff}, {Sabater}, {Sakurikar}, {Salgado}, {Sanghi}, {Saunders},
  {Savchenko}, {Schwardt}, {Seifert-Eckert}, {Shih}, {Jain}, {Shukla}, {Sick},
  {Simpson}, {Singanamalla}, {Singer}, {Singhal}, {Sinha}, {Sip{\H{o}}cz},
  {Spitler}, {Stansby}, {Streicher}, {{\v{S}}umak}, {Swinbank}, {Taranu},
  {Tewary}, {Tremblay}, {de Val-Borro}, {Van Kooten}, {Vasovi{\'c}}, {Verma},
  {de Miranda Cardoso}, {Williams}, {Wilson}, {Winkel}, {Wood-Vasey}, {Xue},
  {Yoachim}, {Zhang}, {Zonca}, \& {Astropy Project
  Contributors}}]{2022ApJ...935..167A}
{Astropy Collaboration}, {Price-Whelan}, A.~M., {Lim}, P.~L., {et~al.} 2022,
  \apj, 935, 167

\bibitem[{{Astropy Collaboration} {et~al.}(2018){Astropy Collaboration},
  {Price-Whelan}, {Sip{\H{o}}cz}, {G{\"u}nther}, {Lim}, {Crawford}, {Conseil},
  {Shupe}, {Craig}, {Dencheva}, {Ginsburg}, {VanderPlas}, {Bradley},
  {P{\'e}rez-Su{\'a}rez}, {de Val-Borro}, {Aldcroft}, {Cruz}, {Robitaille},
  {Tollerud}, {Ardelean}, {Babej}, {Bach}, {Bachetti}, {Bakanov}, {Bamford},
  {Barentsen}, {Barmby}, {Baumbach}, {Berry}, {Biscani}, {Boquien}, {Bostroem},
  {Bouma}, {Brammer}, {Bray}, {Breytenbach}, {Buddelmeijer}, {Burke},
  {Calderone}, {Cano Rodr{\'\i}guez}, {Cara}, {Cardoso}, {Cheedella}, {Copin},
  {Corrales}, {Crichton}, {D'Avella}, {Deil}, {Depagne}, {Dietrich}, {Donath},
  {Droettboom}, {Earl}, {Erben}, {Fabbro}, {Ferreira}, {Finethy}, {Fox},
  {Garrison}, {Gibbons}, {Goldstein}, {Gommers}, {Greco}, {Greenfield},
  {Groener}, {Grollier}, {Hagen}, {Hirst}, {Homeier}, {Horton}, {Hosseinzadeh},
  {Hu}, {Hunkeler}, {Ivezi{\'c}}, {Jain}, {Jenness}, {Kanarek}, {Kendrew},
  {Kern}, {Kerzendorf}, {Khvalko}, {King}, {Kirkby}, {Kulkarni}, {Kumar},
  {Lee}, {Lenz}, {Littlefair}, {Ma}, {Macleod}, {Mastropietro}, {McCully},
  {Montagnac}, {Morris}, {Mueller}, {Mumford}, {Muna}, {Murphy}, {Nelson},
  {Nguyen}, {Ninan}, {N{\"o}the}, {Ogaz}, {Oh}, {Parejko}, {Parley}, {Pascual},
  {Patil}, {Patil}, {Plunkett}, {Prochaska}, {Rastogi}, {Reddy Janga},
  {Sabater}, {Sakurikar}, {Seifert}, {Sherbert}, {Sherwood-Taylor}, {Shih},
  {Sick}, {Silbiger}, {Singanamalla}, {Singer}, {Sladen}, {Sooley},
  {Sornarajah}, {Streicher}, {Teuben}, {Thomas}, {Tremblay}, {Turner},
  {Terr{\'o}n}, {van Kerkwijk}, {de la Vega}, {Watkins}, {Weaver}, {Whitmore},
  {Woillez}, {Zabalza}, \& {Astropy Contributors}}]{2018AJ....156..123A}
{Astropy Collaboration}, {Price-Whelan}, A.~M., {Sip{\H{o}}cz}, B.~M., {et~al.}
  2018, \aj, 156, 123

\bibitem[{{Astropy Collaboration} {et~al.}(2013){Astropy Collaboration},
  {Robitaille}, {Tollerud}, {Greenfield}, {Droettboom}, {Bray}, {Aldcroft},
  {Davis}, {Ginsburg}, {Price-Whelan}, {Kerzendorf}, {Conley}, {Crighton},
  {Barbary}, {Muna}, {Ferguson}, {Grollier}, {Parikh}, {Nair}, {Unther},
  {Deil}, {Woillez}, {Conseil}, {Kramer}, {Turner}, {Singer}, {Fox}, {Weaver},
  {Zabalza}, {Edwards}, {Azalee Bostroem}, {Burke}, {Casey}, {Crawford},
  {Dencheva}, {Ely}, {Jenness}, {Labrie}, {Lim}, {Pierfederici}, {Pontzen},
  {Ptak}, {Refsdal}, {Servillat}, \& {Streicher}}]{2013A&A...558A..33A}
{Astropy Collaboration}, {Robitaille}, T.~P., {Tollerud}, E.~J., {et~al.} 2013,
  \aap, 558, A33

\bibitem[{{Baranne} {et~al.}(1979){Baranne}, {Mayor}, \&
  {Poncet}}]{1979VA.....23..279B}
{Baranne}, A., {Mayor}, M., \& {Poncet}, J.~L. 1979, Vistas in Astronomy, 23,
  279

\bibitem[{{Baranne} {et~al.}(1996){Baranne}, {Queloz}, {Mayor}, {Adrianzyk},
  {Knispel}, {Kohler}, {Lacroix}, {Meunier}, {Rimbaud}, \&
  {Vin}}]{1996A&AS..119..373B}
{Baranne}, A., {Queloz}, D., {Mayor}, M., {et~al.} 1996, \aaps, 119, 373

\bibitem[{{Barbato} {et~al.}(2023){Barbato}, {S{\'e}gransan}, {Udry}, {Unger},
  {Bouchy}, {Lovis}, {Mayor}, {Pepe}, {Queloz}, {Santos}, {Delisle},
  {Figueira}, {Marmier}, {Matthews}, {Lo Curto}, {Venturini}, {Chaverot},
  {Cretignier}, {Otegi}, \& {Stalport}}]{2023A&A...674A.114B}
{Barbato}, D., {S{\'e}gransan}, D., {Udry}, S., {et~al.} 2023, \aap, 674, A114

\bibitem[{{Benedict} \& {Harrison}(2017)}]{2017AJ....153..258B}
{Benedict}, G.~F. \& {Harrison}, T.~E. 2017, \aj, 153, 258

\bibitem[{{Bernstein} {et~al.}(2003){Bernstein}, {Shectman}, {Gunnels},
  {Mochnacki}, \& {Athey}}]{2003SPIE.4841.1694B}
{Bernstein}, R., {Shectman}, S.~A., {Gunnels}, S.~M., {Mochnacki}, S., \&
  {Athey}, A.~E. 2003, in Society of Photo-Optical Instrumentation Engineers
  (SPIE) Conference Series, Vol. 4841, Instrument Design and Performance for
  Optical/Infrared Ground-based Telescopes, ed. M.~{Iye} \& A.~F.~M.
  {Moorwood}, 1694--1704

\bibitem[{{Boss}(1997)}]{1997Sci...276.1836B}
{Boss}, A.~P. 1997, Science, 276, 1836

\bibitem[{{Brandt}(2021)}]{2021ApJS..254...42B}
{Brandt}, T.~D. 2021, \apjs, 254, 42

\bibitem[{{Brandt} {et~al.}(2021){Brandt}, {Dupuy}, {Li}, {Brandt}, {Zeng},
  {Michalik}, {Bardalez Gagliuffi}, \& {Raposo-Pulido}}]{2021AJ....162..186B}
{Brandt}, T.~D., {Dupuy}, T.~J., {Li}, Y., {et~al.} 2021, \aj, 162, 186

\bibitem[{{Brown}(2021)}]{2021ARA&A..59...59B}
{Brown}, A. G.~A. 2021, \araa, 59, 59

\bibitem[{{Butler} {et~al.}(2017){Butler}, {Vogt}, {Laughlin}, {Burt},
  {Rivera}, {Tuomi}, {Teske}, {Arriagada}, {Diaz}, {Holden}, \&
  {Keiser}}]{2017AJ....153..208B}
{Butler}, R.~P., {Vogt}, S.~S., {Laughlin}, G., {et~al.} 2017, \aj, 153, 208

\bibitem[{{Butler} {et~al.}(2006){Butler}, {Wright}, {Marcy}, {Fischer},
  {Vogt}, {Tinney}, {Jones}, {Carter}, {Johnson}, {McCarthy}, \&
  {Penny}}]{2006ApJ...646..505B}
{Butler}, R.~P., {Wright}, J.~T., {Marcy}, G.~W., {et~al.} 2006, \apj, 646, 505

\bibitem[{{Cameron}(1978)}]{1978M&P....18....5C}
{Cameron}, A.~G.~W. 1978, Moon and Planets, 18, 5

\bibitem[{{Casertano} {et~al.}(2008){Casertano}, {Lattanzi}, {Sozzetti},
  {Spagna}, {Jancart}, {Morbidelli}, {Pannunzio}, {Pourbaix}, \&
  {Queloz}}]{2008A&A...482..699C}
{Casertano}, S., {Lattanzi}, M.~G., {Sozzetti}, A., {et~al.} 2008, \aap, 482,
  699

\bibitem[{{Chabrier} {et~al.}(2014){Chabrier}, {Johansen}, {Janson}, \&
  {Rafikov}}]{2014prpl.conf..619C}
{Chabrier}, G., {Johansen}, A., {Janson}, M., \& {Rafikov}, R. 2014, in
  Protostars and Planets VI, ed. H.~{Beuther}, R.~S. {Klessen}, C.~P.
  {Dullemond}, \& T.~{Henning}, 619--642

\bibitem[{{Choi} {et~al.}(2016){Choi}, {Dotter}, {Conroy}, {Cantiello},
  {Paxton}, \& {Johnson}}]{2016ApJ...823..102C}
{Choi}, J., {Dotter}, A., {Conroy}, C., {et~al.} 2016, \apj, 823, 102

\bibitem[{{Cochran} \& {Hatzes}(1993)}]{1993ASPC...36..267C}
{Cochran}, W.~D. \& {Hatzes}, A.~P. 1993, in Astronomical Society of the
  Pacific Conference Series, Vol.~36, Planets Around Pulsars, ed. J.~A.
  {Phillips}, S.~E. {Thorsett}, \& S.~R. {Kulkarni}, 267--273

\bibitem[{{Cochran} {et~al.}(1991){Cochran}, {Hatzes}, \&
  {Hancock}}]{1991ApJ...380L..35C}
{Cochran}, W.~D., {Hatzes}, A.~P., \& {Hancock}, T.~J. 1991, \apjl, 380, L35

\bibitem[{{Dalal} {et~al.}(2021){Dalal}, {Kiefer}, {H{\'e}brard}, {Sahlmann},
  {Sousa}, {Forveille}, {Delfosse}, {Arnold}, {Astudillo-Defru}, {Bonfils},
  {Boisse}, {Bouchy}, {Bourrier}, {Brugger}, {Cort{\'e}s-Zuleta}, {Deleuil},
  {Demangeon}, {D{\'\i}az}, {Hara}, {Heidari}, {Hobson}, {Lopez}, {Lovis},
  {Martioli}, {Mignon}, {Mousis}, {Moutou}, {Rey}, {Santerne}, {Santos},
  {S{\'e}gransan}, {Str{\o}m}, \& {Udry}}]{2021A&A...651A..11D}
{Dalal}, S., {Kiefer}, F., {H{\'e}brard}, G., {et~al.} 2021, \aap, 651, A11

\bibitem[{{Delisle}(2022)}]{2022ascl.soft07011D}
{Delisle}, J.-B. 2022, {samsam: Scaled Adaptive Metropolis SAMpler},
  Astrophysics Source Code Library, record ascl:2207.011

\bibitem[{{Delisle} \& {S{\'e}gransan}(2022)}]{2022A&A...667A.172D}
{Delisle}, J.~B. \& {S{\'e}gransan}, D. 2022, \aap, 667, A172

\bibitem[{{D{\'\i}az} {et~al.}(2012){D{\'\i}az}, {Santerne}, {Sahlmann},
  {H{\'e}brard}, {Eggenberger}, {Santos}, {Moutou}, {Arnold}, {Boisse},
  {Bonfils}, {Bouchy}, {Delfosse}, {Desort}, {Ehrenreich}, {Forveille},
  {Lagrange}, {Lovis}, {Pepe}, {Perrier}, {Queloz}, {S{\'e}gransan}, {Udry}, \&
  {Vidal-Madjar}}]{2012A&A...538A.113D}
{D{\'\i}az}, R.~F., {Santerne}, A., {Sahlmann}, J., {et~al.} 2012, \aap, 538,
  A113

\bibitem[{{Diego} {et~al.}(1990){Diego}, {Charalambous}, {Fish}, \&
  {Walker}}]{1990SPIE.1235..562D}
{Diego}, F., {Charalambous}, A., {Fish}, A.~C., \& {Walker}, D.~D. 1990, in
  Society of Photo-Optical Instrumentation Engineers (SPIE) Conference Series,
  Vol. 1235, Instrumentation in Astronomy VII, ed. D.~L. {Crawford}, 562--576

\bibitem[{{Dotter}(2016)}]{2016ApJS..222....8D}
{Dotter}, A. 2016, \apjs, 222, 8

\bibitem[{{Duquennoy} \& {Mayor}(1991)}]{1991A&A...248..485D}
{Duquennoy}, A. \& {Mayor}, M. 1991, \aap, 248, 485

\bibitem[{{Eastman} {et~al.}(2019){Eastman}, {Rodriguez}, {Agol}, {Stassun},
  {Beatty}, {Vanderburg}, {Gaudi}, {Collins}, \& {Luger}}]{2019arXiv190709480E}
{Eastman}, J.~D., {Rodriguez}, J.~E., {Agol}, E., {et~al.} 2019, arXiv
  e-prints, arXiv:1907.09480

\bibitem[{{Feng} {et~al.}(2022){Feng}, {Butler}, {Vogt}, {Clement}, {Tinney},
  {Cui}, {Aizawa}, {Jones}, {Bailey}, {Burt}, {Carter}, {Crane}, {Flammini
  Dotti}, {Holden}, {Ma}, {Ogihara}, {Oppenheimer}, {O'Toole}, {Shectman},
  {Wittenmyer}, {Wang}, {Wright}, \& {Xuan}}]{2022ApJS..262...21F}
{Feng}, F., {Butler}, R.~P., {Vogt}, S.~S., {et~al.} 2022, \apjs, 262, 21

\bibitem[{{Fischer} {et~al.}(2014){Fischer}, {Marcy}, \&
  {Spronck}}]{2014ApJS..210....5F}
{Fischer}, D.~A., {Marcy}, G.~W., \& {Spronck}, J. F.~P. 2014, \apjs, 210, 5

\bibitem[{{Gaia Collaboration} {et~al.}(2023{\natexlab{a}}){Gaia
  Collaboration}, {Arenou}, {Babusiaux}, {Barstow}, {Faigler}, {Jorissen},
  {Kervella}, {Mazeh}, {Mowlavi}, {Panuzzo}, {Sahlmann}, {Shahaf}, {Sozzetti},
  {Bauchet}, {Damerdji}, {Gavras}, {Giacobbe}, {Gosset}, {Halbwachs}, {Holl},
  {Lattanzi}, {Leclerc}, {Morel}, {Pourbaix}, {Re Fiorentin}, {Sadowski},
  {S{\'e}gransan}, {Siopis}, {Teyssier}, {Zwitter}, {Planquart}, {Brown},
  {Vallenari}, {Prusti}, {de Bruijne}, {Biermann}, {Creevey}, {Ducourant},
  {Evans}, {Eyer}, {Guerra}, {Hutton}, {Jordi}, {Klioner}, {Lammers},
  {Lindegren}, {Luri}, {Mignard}, {Panem}, {Randich}, {Sartoretti}, {Soubiran},
  {Tanga}, {Walton}, {Bailer-Jones}, {Bastian}, {Drimmel}, {Jansen}, {Katz},
  {van Leeuwen}, {Bakker}, {Cacciari}, {Casta{\~n}eda}, {De Angeli},
  {Fabricius}, {Fouesneau}, {Fr{\'e}mat}, {Galluccio}, {Guerrier}, {Heiter},
  {Masana}, {Messineo}, {Nicolas}, {Nienartowicz}, {Pailler}, {Riclet}, {Roux},
  {Seabroke}, {Sordo}, {Th{\'e}venin}, {Gracia-Abril}, {Portell}, {Altmann},
  {Andrae}, {Audard}, {Bellas-Velidis}, {Benson}, {Berthier}, {Blomme},
  {Burgess}, {Busonero}, {Busso}, {C{\'a}novas}, {Carry}, {Cellino}, {Cheek},
  {Clementini}, {Davidson}, {de Teodoro}, {Nu{\~n}ez Campos}, {Delchambre},
  {Dell'Oro}, {Esquej}, {Fern{\'a}ndez-Hern{\'a}ndez}, {Fraile}, {Garabato},
  {Garc{\'\i}a-Lario}, {Haigron}, {Hambly}, {Harrison}, {Hern{\'a}ndez},
  {Hestroffer}, {Hodgkin}, {Jan{\ss}en}, {Jevardat de Fombelle}, {Jordan},
  {Krone-Martins}, {Lanzafame}, {L{\"o}ffler}, {Marchal}, {Marrese},
  {Moitinho}, {Muinonen}, {Osborne}, {Pancino}, {Pauwels}, {Recio-Blanco},
  {Reyl{\'e}}, {Riello}, {Rimoldini}, {Roegiers}, {Rybizki}, {Sarro}, {Smith},
  {Utrilla}, {van Leeuwen}, {Abbas}, {{\'A}brah{\'a}m}, {Abreu Aramburu},
  {Aerts}, {Aguado}, {Ajaj}, {Aldea-Montero}, {Altavilla}, {{\'A}lvarez},
  {Alves}, {Anders}, {Anderson}, {Anglada Varela}, {Antoja}, {Baines}, {Baker},
  {Balaguer-N{\'u}{\~n}ez}, {Balbinot}, {Balog}, {Barache}, {Barbato},
  {Barros}, {Bartolom{\'e}}, {Bassilana}, {Becciani}, {Bellazzini},
  {Berihuete}, {Bernet}, {Bertone}, {Bianchi}, {Binnenfeld}, {Blanco-Cuaresma},
  {Blazere}, {Boch}, {Bombrun}, {Bossini}, {Bouquillon}, {Bragaglia},
  {Bramante}, {Breedt}, {Bressan}, {Brouillet}, {Brugaletta}, {Bucciarelli},
  {Burlacu}, {Butkevich}, {Buzzi}, {Caffau}, {Cancelliere}, {Cantat-Gaudin},
  {Carballo}, {Carlucci}, {Carnerero}, {Carrasco}, {Casamiquela}, {Castellani},
  {Castro-Ginard}, {Chaoul}, {Charlot}, {Chemin}, {Chiaramida}, {Chiavassa},
  {Chornay}, {Comoretto}, {Contursi}, {Cooper}, {Cornez}, {Cowell}, {Crifo},
  {Cropper}, {Crosta}, {Crowley}, {Dafonte}, {Dapergolas}, {David}, {de
  Laverny}, {De Luise}, {De March}, {De Ridder}, {de Souza}, {de Torres}, {del
  Peloso}, {del Pozo}, {Delbo}, {Delgado}, {Delisle}, {Demouchy},
  {Dharmawardena}, {Diakite}, {Diener}, {Distefano}, {Dolding}, {Enke},
  {Fabre}, {Fabrizio}, {Fedorets}, {Fernique}, {Figueras}, {Fournier},
  {Fouron}, {Fragkoudi}, {Gai}, {Garcia-Gutierrez}, {Garcia-Reinaldos},
  {Garc{\'\i}a-Torres}, {Garofalo}, {Gavel}, {Gerlach}, {Geyer}, {Gilmore},
  {Girona}, {Giuffrida}, {Gomel}, {Gomez}, {Gonz{\'a}lez-N{\'u}{\~n}ez},
  {Gonz{\'a}lez-Santamar{\'\i}a}, {Gonz{\'a}lez-Vidal}, {Granvik}, {Guillout},
  {Guiraud}, {Guti{\'e}rrez-S{\'a}nchez}, {Guy}, {Hatzidimitriou}, {Hauser},
  {Haywood}, {Helmer}, {Helmi}, {Sarmiento}, {Hidalgo}, {Hilger},
  {H{\l}adczuk}, {Hobbs}, {Holland}, {Huckle}, {Jardine}, {Jasniewicz},
  {Jean-Antoine Piccolo}, {Jim{\'e}nez-Arranz}, {Juaristi Campillo}, {Julbe},
  {Karbevska}, {Khanna}, {Kordopatis}, {Korn}, {K{\'o}sp{\'a}l},
  {Kostrzewa-Rutkowska}, {Kruszy{\'n}ska}, {Kun}, {Laizeau}, {Lambert},
  {Lanza}, {Lasne}, {Le Campion}, {Lebreton}, {Lebzelter}, {Leccia},
  {Lecoeur-Taibi}, {Liao}, {Licata}, {Lindstr{\o}m}, {Lister}, {Livanou},
  {Lobel}, {Lorca}, {Loup}, {Madrero Pardo}, {Magdaleno Romeo}, {Managau},
  {Mann}, {Manteiga}, {Marchant}, {Marconi}, {Marcos}, {Marcos Santos},
  {Mar{\'\i}n Pina}, {Marinoni}, {Marocco}, {Marshall}, {Martin Polo},
  {Mart{\'\i}n-Fleitas}, {Marton}, {Mary}, {Masip}, {Massari},
  {Mastrobuono-Battisti}, {McMillan}, {Messina}, {Michalik}, {Millar}, {Mints},
  {Molina}, {Molinaro}, {Moln{\'a}r}, {Monari}, {Mongui{\'o}}, {Montegriffo},
  {Montero}, {Mor}, {Mora}, {Morbidelli}, {Morris}, {Muraveva}, {Murphy},
  {Musella}, {Nagy}, {Noval}, {Oca{\~n}a}, {Ogden}, {Ordenovic}, {Osinde},
  {Pagani}, {Pagano}, {Palaversa}, {Palicio}, {Pallas-Quintela}, {Panahi},
  {Payne-Wardenaar}, {Pe{\~n}alosa Esteller}, {Penttil{\"a}}, {Pichon},
  {Piersimoni}, {Pineau}, {Plachy}, {Plum}, {Poggio}, {Pr{\v{s}}a}, {Pulone},
  {Racero}, {Ragaini}, {Rainer}, {Raiteri}, {Ramos}, {Ramos-Lerate}, {Regibo},
  {Richards}, {Rios Diaz}, {Ripepi}, {Riva}, {Rix}, {Rixon}, {Robichon},
  {Robin}, {Robin}, {Roelens}, {Rogues}, {Rohrbasser}, {Romero-G{\'o}mez},
  {Rowell}, {Royer}, {Ruz Mieres}, {Rybicki}, {S{\'a}ez N{\'u}{\~n}ez},
  {Sagrist{\`a} Sell{\'e}s}, {Salguero}, {Samaras}, {Sanchez Gimenez}, {Sanna},
  {Santove{\~n}a}, {Sarasso}, {Schultheis}, {Sciacca}, {Segol}, {Segovia},
  {Semeux}, {Siddiqui}, {Siebert}, {Siltala}, {Silvelo}, {Slezak}, {Slezak},
  {Smart}, {Snaith}, {Solano}, {Solitro}, {Souami}, {Souchay}, {Spagna},
  {Spina}, {Spoto}, {Steele}, {Steidelm{\"u}ller}, {Stephenson}, {S{\"u}veges},
  {Surdej}, {Szabados}, {Szegedi-Elek}, {Taris}, {Taylor}, {Teixeira},
  {Tolomei}, {Tonello}, {Torra}, {Torra}, {Torralba Elipe}, {Trabucchi},
  {Tsounis}, {Turon}, {Ulla}, {Unger}, {Vaillant}, {van Dillen}, {van Reeven},
  {Vanel}, {Vecchiato}, {Viala}, {Vicente}, {Voutsinas}, {Weiler}, {Wevers},
  {Wyrzykowski}, {Yoldas}, {Yvard}, {Zhao}, {Zorec}, \&
  {Zucker}}]{2023A&A...674A..34G}
{Gaia Collaboration}, {Arenou}, F., {Babusiaux}, C., {et~al.}
  2023{\natexlab{a}}, \aap, 674, A34

\bibitem[{{Gaia Collaboration} {et~al.}(2021){Gaia Collaboration}, {Brown},
  {Vallenari}, {Prusti}, {de Bruijne}, {Babusiaux}, {Biermann}, {Creevey},
  {Evans}, {Eyer}, {Hutton}, {Jansen}, {Jordi}, {Klioner}, {Lammers},
  {Lindegren}, {Luri}, {Mignard}, {Panem}, {Pourbaix}, {Randich}, {Sartoretti},
  {Soubiran}, {Walton}, {Arenou}, {Bailer-Jones}, {Bastian}, {Cropper},
  {Drimmel}, {Katz}, {Lattanzi}, {van Leeuwen}, {Bakker}, {Cacciari},
  {Casta{\~n}eda}, {De Angeli}, {Ducourant}, {Fabricius}, {Fouesneau},
  {Fr{\'e}mat}, {Guerra}, {Guerrier}, {Guiraud}, {Jean-Antoine Piccolo},
  {Masana}, {Messineo}, {Mowlavi}, {Nicolas}, {Nienartowicz}, {Pailler},
  {Panuzzo}, {Riclet}, {Roux}, {Seabroke}, {Sordo}, {Tanga}, {Th{\'e}venin},
  {Gracia-Abril}, {Portell}, {Teyssier}, {Altmann}, {Andrae}, {Bellas-Velidis},
  {Benson}, {Berthier}, {Blomme}, {Brugaletta}, {Burgess}, {Busso}, {Carry},
  {Cellino}, {Cheek}, {Clementini}, {Damerdji}, {Davidson}, {Delchambre},
  {Dell'Oro}, {Fern{\'a}ndez-Hern{\'a}ndez}, {Galluccio}, {Garc{\'\i}a-Lario},
  {Garcia-Reinaldos}, {Gonz{\'a}lez-N{\'u}{\~n}ez}, {Gosset}, {Haigron},
  {Halbwachs}, {Hambly}, {Harrison}, {Hatzidimitriou}, {Heiter},
  {Hern{\'a}ndez}, {Hestroffer}, {Hodgkin}, {Holl}, {Jan{\ss}en}, {Jevardat de
  Fombelle}, {Jordan}, {Krone-Martins}, {Lanzafame}, {L{\"o}ffler}, {Lorca},
  {Manteiga}, {Marchal}, {Marrese}, {Moitinho}, {Mora}, {Muinonen}, {Osborne},
  {Pancino}, {Pauwels}, {Petit}, {Recio-Blanco}, {Richards}, {Riello},
  {Rimoldini}, {Robin}, {Roegiers}, {Rybizki}, {Sarro}, {Siopis}, {Smith},
  {Sozzetti}, {Ulla}, {Utrilla}, {van Leeuwen}, {van Reeven}, {Abbas}, {Abreu
  Aramburu}, {Accart}, {Aerts}, {Aguado}, {Ajaj}, {Altavilla}, {{\'A}lvarez},
  {{\'A}lvarez Cid-Fuentes}, {Alves}, {Anderson}, {Anglada Varela}, {Antoja},
  {Audard}, {Baines}, {Baker}, {Balaguer-N{\'u}{\~n}ez}, {Balbinot}, {Balog},
  {Barache}, {Barbato}, {Barros}, {Barstow}, {Bartolom{\'e}}, {Bassilana},
  {Bauchet}, {Baudesson-Stella}, {Becciani}, {Bellazzini}, {Bernet}, {Bertone},
  {Bianchi}, {Blanco-Cuaresma}, {Boch}, {Bombrun}, {Bossini}, {Bouquillon},
  {Bragaglia}, {Bramante}, {Breedt}, {Bressan}, {Brouillet}, {Bucciarelli},
  {Burlacu}, {Busonero}, {Butkevich}, {Buzzi}, {Caffau}, {Cancelliere},
  {C{\'a}novas}, {Cantat-Gaudin}, {Carballo}, {Carlucci}, {Carnerero},
  {Carrasco}, {Casamiquela}, {Castellani}, {Castro-Ginard}, {Castro Sampol},
  {Chaoul}, {Charlot}, {Chemin}, {Chiavassa}, {Cioni}, {Comoretto}, {Cooper},
  {Cornez}, {Cowell}, {Crifo}, {Crosta}, {Crowley}, {Dafonte}, {Dapergolas},
  {David}, {David}, {de Laverny}, {De Luise}, {De March}, {De Ridder}, {de
  Souza}, {de Teodoro}, {de Torres}, {del Peloso}, {del Pozo}, {Delbo},
  {Delgado}, {Delgado}, {Delisle}, {Di Matteo}, {Diakite}, {Diener},
  {Distefano}, {Dolding}, {Eappachen}, {Edvardsson}, {Enke}, {Esquej}, {Fabre},
  {Fabrizio}, {Faigler}, {Fedorets}, {Fernique}, {Fienga}, {Figueras},
  {Fouron}, {Fragkoudi}, {Fraile}, {Franke}, {Gai}, {Garabato},
  {Garcia-Gutierrez}, {Garc{\'\i}a-Torres}, {Garofalo}, {Gavras}, {Gerlach},
  {Geyer}, {Giacobbe}, {Gilmore}, {Girona}, {Giuffrida}, {Gomel}, {Gomez},
  {Gonzalez-Santamaria}, {Gonz{\'a}lez-Vidal}, {Granvik},
  {Guti{\'e}rrez-S{\'a}nchez}, {Guy}, {Hauser}, {Haywood}, {Helmi}, {Hidalgo},
  {Hilger}, {H{\l}adczuk}, {Hobbs}, {Holland}, {Huckle}, {Jasniewicz},
  {Jonker}, {Juaristi Campillo}, {Julbe}, {Karbevska}, {Kervella}, {Khanna},
  {Kochoska}, {Kontizas}, {Kordopatis}, {Korn}, {Kostrzewa-Rutkowska},
  {Kruszy{\'n}ska}, {Lambert}, {Lanza}, {Lasne}, {Le Campion}, {Le Fustec},
  {Lebreton}, {Lebzelter}, {Leccia}, {Leclerc}, {Lecoeur-Taibi}, {Liao},
  {Licata}, {Lindstr{\o}m}, {Lister}, {Livanou}, {Lobel}, {Madrero Pardo},
  {Managau}, {Mann}, {Marchant}, {Marconi}, {Marcos Santos}, {Marinoni},
  {Marocco}, {Marshall}, {Martin Polo}, {Mart{\'\i}n-Fleitas}, {Masip},
  {Massari}, {Mastrobuono-Battisti}, {Mazeh}, {McMillan}, {Messina},
  {Michalik}, {Millar}, {Mints}, {Molina}, {Molinaro}, {Moln{\'a}r},
  {Montegriffo}, {Mor}, {Morbidelli}, {Morel}, {Morris}, {Mulone}, {Munoz},
  {Muraveva}, {Murphy}, {Musella}, {Noval}, {Ord{\'e}novic}, {Orr{\`u}},
  {Osinde}, {Pagani}, {Pagano}, {Palaversa}, {Palicio}, {Panahi}, {Pawlak},
  {Pe{\~n}alosa Esteller}, {Penttil{\"a}}, {Piersimoni}, {Pineau}, {Plachy},
  {Plum}, {Poggio}, {Poretti}, {Poujoulet}, {Pr{\v{s}}a}, {Pulone}, {Racero},
  {Ragaini}, {Rainer}, {Raiteri}, {Rambaux}, {Ramos}, {Ramos-Lerate}, {Re
  Fiorentin}, {Regibo}, {Reyl{\'e}}, {Ripepi}, {Riva}, {Rixon}, {Robichon},
  {Robin}, {Roelens}, {Rohrbasser}, {Romero-G{\'o}mez}, {Rowell}, {Royer},
  {Rybicki}, {Sadowski}, {Sagrist{\`a} Sell{\'e}s}, {Sahlmann}, {Salgado},
  {Salguero}, {Samaras}, {Sanchez Gimenez}, {Sanna}, {Santove{\~n}a},
  {Sarasso}, {Schultheis}, {Sciacca}, {Segol}, {Segovia}, {S{\'e}gransan},
  {Semeux}, {Shahaf}, {Siddiqui}, {Siebert}, {Siltala}, {Slezak}, {Smart},
  {Solano}, {Solitro}, {Souami}, {Souchay}, {Spagna}, {Spoto}, {Steele},
  {Steidelm{\"u}ller}, {Stephenson}, {S{\"u}veges}, {Szabados}, {Szegedi-Elek},
  {Taris}, {Tauran}, {Taylor}, {Teixeira}, {Thuillot}, {Tonello}, {Torra},
  {Torra}, {Turon}, {Unger}, {Vaillant}, {van Dillen}, {Vanel}, {Vecchiato},
  {Viala}, {Vicente}, {Voutsinas}, {Weiler}, {Wevers}, {Wyrzykowski}, {Yoldas},
  {Yvard}, {Zhao}, {Zorec}, {Zucker}, {Zurbach}, \&
  {Zwitter}}]{2021A&A...649A...1G}
{Gaia Collaboration}, {Brown}, A.~G.~A., {Vallenari}, A., {et~al.} 2021, \aap,
  649, A1

\bibitem[{{Gaia Collaboration} {et~al.}(2016){Gaia Collaboration}, {Prusti},
  {de Bruijne}, {Brown}, {Vallenari}, {Babusiaux}, {Bailer-Jones}, {Bastian},
  {Biermann}, {Evans}, {Eyer}, {Jansen}, {Jordi}, {Klioner}, {Lammers},
  {Lindegren}, {Luri}, {Mignard}, {Milligan}, {Panem}, {Poinsignon},
  {Pourbaix}, {Randich}, {Sarri}, {Sartoretti}, {Siddiqui}, {Soubiran},
  {Valette}, {van Leeuwen}, {Walton}, {Aerts}, {Arenou}, {Cropper}, {Drimmel},
  {H{\o}g}, {Katz}, {Lattanzi}, {O'Mullane}, {Grebel}, {Holland}, {Huc},
  {Passot}, {Bramante}, {Cacciari}, {Casta{\~n}eda}, {Chaoul}, {Cheek}, {De
  Angeli}, {Fabricius}, {Guerra}, {Hern{\'a}ndez}, {Jean-Antoine-Piccolo},
  {Masana}, {Messineo}, {Mowlavi}, {Nienartowicz}, {Ord{\'o}{\~n}ez-Blanco},
  {Panuzzo}, {Portell}, {Richards}, {Riello}, {Seabroke}, {Tanga},
  {Th{\'e}venin}, {Torra}, {Els}, {Gracia-Abril}, {Comoretto},
  {Garcia-Reinaldos}, {Lock}, {Mercier}, {Altmann}, {Andrae}, {Astraatmadja},
  {Bellas-Velidis}, {Benson}, {Berthier}, {Blomme}, {Busso}, {Carry},
  {Cellino}, {Clementini}, {Cowell}, {Creevey}, {Cuypers}, {Davidson}, {De
  Ridder}, {de Torres}, {Delchambre}, {Dell'Oro}, {Ducourant}, {Fr{\'e}mat},
  {Garc{\'\i}a-Torres}, {Gosset}, {Halbwachs}, {Hambly}, {Harrison}, {Hauser},
  {Hestroffer}, {Hodgkin}, {Huckle}, {Hutton}, {Jasniewicz}, {Jordan},
  {Kontizas}, {Korn}, {Lanzafame}, {Manteiga}, {Moitinho}, {Muinonen},
  {Osinde}, {Pancino}, {Pauwels}, {Petit}, {Recio-Blanco}, {Robin}, {Sarro},
  {Siopis}, {Smith}, {Smith}, {Sozzetti}, {Thuillot}, {van Reeven}, {Viala},
  {Abbas}, {Abreu Aramburu}, {Accart}, {Aguado}, {Allan}, {Allasia},
  {Altavilla}, {{\'A}lvarez}, {Alves}, {Anderson}, {Andrei}, {Anglada Varela},
  {Antiche}, {Antoja}, {Ant{\'o}n}, {Arcay}, {Atzei}, {Ayache}, {Bach},
  {Baker}, {Balaguer-N{\'u}{\~n}ez}, {Barache}, {Barata}, {Barbier}, {Barblan},
  {Baroni}, {Barrado y Navascu{\'e}s}, {Barros}, {Barstow}, {Becciani},
  {Bellazzini}, {Bellei}, {Bello Garc{\'\i}a}, {Belokurov}, {Bendjoya},
  {Berihuete}, {Bianchi}, {Bienaym{\'e}}, {Billebaud}, {Blagorodnova},
  {Blanco-Cuaresma}, {Boch}, {Bombrun}, {Borrachero}, {Bouquillon}, {Bourda},
  {Bouy}, {Bragaglia}, {Breddels}, {Brouillet}, {Br{\"u}semeister},
  {Bucciarelli}, {Budnik}, {Burgess}, {Burgon}, {Burlacu}, {Busonero}, {Buzzi},
  {Caffau}, {Cambras}, {Campbell}, {Cancelliere}, {Cantat-Gaudin}, {Carlucci},
  {Carrasco}, {Castellani}, {Charlot}, {Charnas}, {Charvet}, {Chassat},
  {Chiavassa}, {Clotet}, {Cocozza}, {Collins}, {Collins}, {Costigan}, {Crifo},
  {Cross}, {Crosta}, {Crowley}, {Dafonte}, {Damerdji}, {Dapergolas}, {David},
  {David}, {De Cat}, {de Felice}, {de Laverny}, {De Luise}, {De March}, {de
  Martino}, {de Souza}, {Debosscher}, {del Pozo}, {Delbo}, {Delgado},
  {Delgado}, {di Marco}, {Di Matteo}, {Diakite}, {Distefano}, {Dolding}, {Dos
  Anjos}, {Drazinos}, {Dur{\'a}n}, {Dzigan}, {Ecale}, {Edvardsson}, {Enke},
  {Erdmann}, {Escolar}, {Espina}, {Evans}, {Eynard Bontemps}, {Fabre},
  {Fabrizio}, {Faigler}, {Falc{\~a}o}, {Farr{\`a}s Casas}, {Faye}, {Federici},
  {Fedorets}, {Fern{\'a}ndez-Hern{\'a}ndez}, {Fernique}, {Fienga}, {Figueras},
  {Filippi}, {Findeisen}, {Fonti}, {Fouesneau}, {Fraile}, {Fraser}, {Fuchs},
  {Furnell}, {Gai}, {Galleti}, {Galluccio}, {Garabato}, {Garc{\'\i}a-Sedano},
  {Gar{\'e}}, {Garofalo}, {Garralda}, {Gavras}, {Gerssen}, {Geyer}, {Gilmore},
  {Girona}, {Giuffrida}, {Gomes}, {Gonz{\'a}lez-Marcos},
  {Gonz{\'a}lez-N{\'u}{\~n}ez}, {Gonz{\'a}lez-Vidal}, {Granvik}, {Guerrier},
  {Guillout}, {Guiraud}, {G{\'u}rpide}, {Guti{\'e}rrez-S{\'a}nchez}, {Guy},
  {Haigron}, {Hatzidimitriou}, {Haywood}, {Heiter}, {Helmi}, {Hobbs},
  {Hofmann}, {Holl}, {Holland}, {Hunt}, {Hypki}, {Icardi}, {Irwin}, {Jevardat
  de Fombelle}, {Jofr{\'e}}, {Jonker}, {Jorissen}, {Julbe}, {Karampelas},
  {Kochoska}, {Kohley}, {Kolenberg}, {Kontizas}, {Koposov}, {Kordopatis},
  {Koubsky}, {Kowalczyk}, {Krone-Martins}, {Kudryashova}, {Kull}, {Bachchan},
  {Lacoste-Seris}, {Lanza}, {Lavigne}, {Le Poncin-Lafitte}, {Lebreton},
  {Lebzelter}, {Leccia}, {Leclerc}, {Lecoeur-Taibi}, {Lemaitre}, {Lenhardt},
  {Leroux}, {Liao}, {Licata}, {Lindstr{\o}m}, {Lister}, {Livanou}, {Lobel},
  {L{\"o}ffler}, {L{\'o}pez}, {Lopez-Lozano}, {Lorenz}, {Loureiro},
  {MacDonald}, {Magalh{\~a}es Fernandes}, {Managau}, {Mann}, {Mantelet},
  {Marchal}, {Marchant}, {Marconi}, {Marie}, {Marinoni}, {Marrese},
  {Marschalk{\'o}}, {Marshall}, {Mart{\'\i}n-Fleitas}, {Martino}, {Mary},
  {Matijevi{\v{c}}}, {Mazeh}, {McMillan}, {Messina}, {Mestre}, {Michalik},
  {Millar}, {Miranda}, {Molina}, {Molinaro}, {Molinaro}, {Moln{\'a}r},
  {Moniez}, {Montegriffo}, {Monteiro}, {Mor}, {Mora}, {Morbidelli}, {Morel},
  {Morgenthaler}, {Morley}, {Morris}, {Mulone}, {Muraveva}, {Musella},
  {Narbonne}, {Nelemans}, {Nicastro}, {Noval}, {Ord{\'e}novic},
  {Ordieres-Mer{\'e}}, {Osborne}, {Pagani}, {Pagano}, {Pailler}, {Palacin},
  {Palaversa}, {Parsons}, {Paulsen}, {Pecoraro}, {Pedrosa}, {Pentik{\"a}inen},
  {Pereira}, {Pichon}, {Piersimoni}, {Pineau}, {Plachy}, {Plum}, {Poujoulet},
  {Pr{\v{s}}a}, {Pulone}, {Ragaini}, {Rago}, {Rambaux}, {Ramos-Lerate},
  {Ranalli}, {Rauw}, {Read}, {Regibo}, {Renk}, {Reyl{\'e}}, {Ribeiro},
  {Rimoldini}, {Ripepi}, {Riva}, {Rixon}, {Roelens}, {Romero-G{\'o}mez},
  {Rowell}, {Royer}, {Rudolph}, {Ruiz-Dern}, {Sadowski}, {Sagrist{\`a}
  Sell{\'e}s}, {Sahlmann}, {Salgado}, {Salguero}, {Sarasso}, {Savietto},
  {Schnorhk}, {Schultheis}, {Sciacca}, {Segol}, {Segovia}, {Segransan},
  {Serpell}, {Shih}, {Smareglia}, {Smart}, {Smith}, {Solano}, {Solitro},
  {Sordo}, {Soria Nieto}, {Souchay}, {Spagna}, {Spoto}, {Stampa}, {Steele},
  {Steidelm{\"u}ller}, {Stephenson}, {Stoev}, {Suess}, {S{\"u}veges}, {Surdej},
  {Szabados}, {Szegedi-Elek}, {Tapiador}, {Taris}, {Tauran}, {Taylor},
  {Teixeira}, {Terrett}, {Tingley}, {Trager}, {Turon}, {Ulla}, {Utrilla},
  {Valentini}, {van Elteren}, {Van Hemelryck}, {van Leeuwen}, {Varadi},
  {Vecchiato}, {Veljanoski}, {Via}, {Vicente}, {Vogt}, {Voss}, {Votruba},
  {Voutsinas}, {Walmsley}, {Weiler}, {Weingrill}, {Werner}, {Wevers},
  {Whitehead}, {Wyrzykowski}, {Yoldas}, {{\v{Z}}erjal}, {Zucker}, {Zurbach},
  {Zwitter}, {Alecu}, {Allen}, {Allende Prieto}, {Amorim},
  {Anglada-Escud{\'e}}, {Arsenijevic}, {Azaz}, {Balm}, {Beck}, {Bernstein},
  {Bigot}, {Bijaoui}, {Blasco}, {Bonfigli}, {Bono}, {Boudreault}, {Bressan},
  {Brown}, {Brunet}, {Bunclark}, {Buonanno}, {Butkevich}, {Carret}, {Carrion},
  {Chemin}, {Ch{\'e}reau}, {Corcione}, {Darmigny}, {de Boer}, {de Teodoro}, {de
  Zeeuw}, {Delle Luche}, {Domingues}, {Dubath}, {Fodor}, {Fr{\'e}zouls},
  {Fries}, {Fustes}, {Fyfe}, {Gallardo}, {Gallegos}, {Gardiol}, {Gebran},
  {Gomboc}, {G{\'o}mez}, {Grux}, {Gueguen}, {Heyrovsky}, {Hoar}, {Iannicola},
  {Isasi Parache}, {Janotto}, {Joliet}, {Jonckheere}, {Keil}, {Kim},
  {Klagyivik}, {Klar}, {Knude}, {Kochukhov}, {Kolka}, {Kos}, {Kutka}, {Lainey},
  {LeBouquin}, {Liu}, {Loreggia}, {Makarov}, {Marseille}, {Martayan},
  {Martinez-Rubi}, {Massart}, {Meynadier}, {Mignot}, {Munari}, {Nguyen},
  {Nordlander}, {Ocvirk}, {O'Flaherty}, {Olias Sanz}, {Ortiz}, {Osorio},
  {Oszkiewicz}, {Ouzounis}, {Palmer}, {Park}, {Pasquato}, {Peltzer}, {Peralta},
  {P{\'e}turaud}, {Pieniluoma}, {Pigozzi}, {Poels}, {Prat}, {Prod'homme},
  {Raison}, {Rebordao}, {Risquez}, {Rocca-Volmerange}, {Rosen}, {Ruiz-Fuertes},
  {Russo}, {Sembay}, {Serraller Vizcaino}, {Short}, {Siebert}, {Silva},
  {Sinachopoulos}, {Slezak}, {Soffel}, {Sosnowska}, {Strai{\v{z}}ys}, {ter
  Linden}, {Terrell}, {Theil}, {Tiede}, {Troisi}, {Tsalmantza}, {Tur},
  {Vaccari}, {Vachier}, {Valles}, {Van Hamme}, {Veltz}, {Virtanen}, {Wallut},
  {Wichmann}, {Wilkinson}, {Ziaeepour}, \& {Zschocke}}]{2016A&A...595A...1G}
{Gaia Collaboration}, {Prusti}, T., {de Bruijne}, J.~H.~J., {et~al.} 2016,
  \aap, 595, A1

\bibitem[{{Gaia Collaboration} {et~al.}(2023{\natexlab{b}}){Gaia
  Collaboration}, {Vallenari}, {Brown}, {Prusti}, {de Bruijne}, {Arenou},
  {Babusiaux}, {Biermann}, {Creevey}, {Ducourant}, {Evans}, {Eyer}, {Guerra},
  {Hutton}, {Jordi}, {Klioner}, {Lammers}, {Lindegren}, {Luri}, {Mignard},
  {Panem}, {Pourbaix}, {Randich}, {Sartoretti}, {Soubiran}, {Tanga}, {Walton},
  {Bailer-Jones}, {Bastian}, {Drimmel}, {Jansen}, {Katz}, {Lattanzi}, {van
  Leeuwen}, {Bakker}, {Cacciari}, {Casta{\~n}eda}, {De Angeli}, {Fabricius},
  {Fouesneau}, {Fr{\'e}mat}, {Galluccio}, {Guerrier}, {Heiter}, {Masana},
  {Messineo}, {Mowlavi}, {Nicolas}, {Nienartowicz}, {Pailler}, {Panuzzo},
  {Riclet}, {Roux}, {Seabroke}, {Sordo}, {Th{\'e}venin}, {Gracia-Abril},
  {Portell}, {Teyssier}, {Altmann}, {Andrae}, {Audard}, {Bellas-Velidis},
  {Benson}, {Berthier}, {Blomme}, {Burgess}, {Busonero}, {Busso},
  {C{\'a}novas}, {Carry}, {Cellino}, {Cheek}, {Clementini}, {Damerdji},
  {Davidson}, {de Teodoro}, {Nu{\~n}ez Campos}, {Delchambre}, {Dell'Oro},
  {Esquej}, {Fern{\'a}ndez-Hern{\'a}ndez}, {Fraile}, {Garabato},
  {Garc{\'\i}a-Lario}, {Gosset}, {Haigron}, {Halbwachs}, {Hambly}, {Harrison},
  {Hern{\'a}ndez}, {Hestroffer}, {Hodgkin}, {Holl}, {Jan{\ss}en}, {Jevardat de
  Fombelle}, {Jordan}, {Krone-Martins}, {Lanzafame}, {L{\"o}ffler}, {Marchal},
  {Marrese}, {Moitinho}, {Muinonen}, {Osborne}, {Pancino}, {Pauwels},
  {Recio-Blanco}, {Reyl{\'e}}, {Riello}, {Rimoldini}, {Roegiers}, {Rybizki},
  {Sarro}, {Siopis}, {Smith}, {Sozzetti}, {Utrilla}, {van Leeuwen}, {Abbas},
  {{\'A}brah{\'a}m}, {Abreu Aramburu}, {Aerts}, {Aguado}, {Ajaj},
  {Aldea-Montero}, {Altavilla}, {{\'A}lvarez}, {Alves}, {Anders}, {Anderson},
  {Anglada Varela}, {Antoja}, {Baines}, {Baker}, {Balaguer-N{\'u}{\~n}ez},
  {Balbinot}, {Balog}, {Barache}, {Barbato}, {Barros}, {Barstow},
  {Bartolom{\'e}}, {Bassilana}, {Bauchet}, {Becciani}, {Bellazzini},
  {Berihuete}, {Bernet}, {Bertone}, {Bianchi}, {Binnenfeld}, {Blanco-Cuaresma},
  {Blazere}, {Boch}, {Bombrun}, {Bossini}, {Bouquillon}, {Bragaglia},
  {Bramante}, {Breedt}, {Bressan}, {Brouillet}, {Brugaletta}, {Bucciarelli},
  {Burlacu}, {Butkevich}, {Buzzi}, {Caffau}, {Cancelliere}, {Cantat-Gaudin},
  {Carballo}, {Carlucci}, {Carnerero}, {Carrasco}, {Casamiquela}, {Castellani},
  {Castro-Ginard}, {Chaoul}, {Charlot}, {Chemin}, {Chiaramida}, {Chiavassa},
  {Chornay}, {Comoretto}, {Contursi}, {Cooper}, {Cornez}, {Cowell}, {Crifo},
  {Cropper}, {Crosta}, {Crowley}, {Dafonte}, {Dapergolas}, {David}, {David},
  {de Laverny}, {De Luise}, {De March}, {De Ridder}, {de Souza}, {de Torres},
  {del Peloso}, {del Pozo}, {Delbo}, {Delgado}, {Delisle}, {Demouchy},
  {Dharmawardena}, {Di Matteo}, {Diakite}, {Diener}, {Distefano}, {Dolding},
  {Edvardsson}, {Enke}, {Fabre}, {Fabrizio}, {Faigler}, {Fedorets}, {Fernique},
  {Fienga}, {Figueras}, {Fournier}, {Fouron}, {Fragkoudi}, {Gai},
  {Garcia-Gutierrez}, {Garcia-Reinaldos}, {Garc{\'\i}a-Torres}, {Garofalo},
  {Gavel}, {Gavras}, {Gerlach}, {Geyer}, {Giacobbe}, {Gilmore}, {Girona},
  {Giuffrida}, {Gomel}, {Gomez}, {Gonz{\'a}lez-N{\'u}{\~n}ez},
  {Gonz{\'a}lez-Santamar{\'\i}a}, {Gonz{\'a}lez-Vidal}, {Granvik}, {Guillout},
  {Guiraud}, {Guti{\'e}rrez-S{\'a}nchez}, {Guy}, {Hatzidimitriou}, {Hauser},
  {Haywood}, {Helmer}, {Helmi}, {Sarmiento}, {Hidalgo}, {Hilger},
  {H{\l}adczuk}, {Hobbs}, {Holland}, {Huckle}, {Jardine}, {Jasniewicz},
  {Jean-Antoine Piccolo}, {Jim{\'e}nez-Arranz}, {Jorissen}, {Juaristi
  Campillo}, {Julbe}, {Karbevska}, {Kervella}, {Khanna}, {Kontizas},
  {Kordopatis}, {Korn}, {K{\'o}sp{\'a}l}, {Kostrzewa-Rutkowska},
  {Kruszy{\'n}ska}, {Kun}, {Laizeau}, {Lambert}, {Lanza}, {Lasne}, {Le
  Campion}, {Lebreton}, {Lebzelter}, {Leccia}, {Leclerc}, {Lecoeur-Taibi},
  {Liao}, {Licata}, {Lindstr{\o}m}, {Lister}, {Livanou}, {Lobel}, {Lorca},
  {Loup}, {Madrero Pardo}, {Magdaleno Romeo}, {Managau}, {Mann}, {Manteiga},
  {Marchant}, {Marconi}, {Marcos}, {Marcos Santos}, {Mar{\'\i}n Pina},
  {Marinoni}, {Marocco}, {Marshall}, {Martin Polo}, {Mart{\'\i}n-Fleitas},
  {Marton}, {Mary}, {Masip}, {Massari}, {Mastrobuono-Battisti}, {Mazeh},
  {McMillan}, {Messina}, {Michalik}, {Millar}, {Mints}, {Molina}, {Molinaro},
  {Moln{\'a}r}, {Monari}, {Mongui{\'o}}, {Montegriffo}, {Montero}, {Mor},
  {Mora}, {Morbidelli}, {Morel}, {Morris}, {Muraveva}, {Murphy}, {Musella},
  {Nagy}, {Noval}, {Oca{\~n}a}, {Ogden}, {Ordenovic}, {Osinde}, {Pagani},
  {Pagano}, {Palaversa}, {Palicio}, {Pallas-Quintela}, {Panahi},
  {Payne-Wardenaar}, {Pe{\~n}alosa Esteller}, {Penttil{\"a}}, {Pichon},
  {Piersimoni}, {Pineau}, {Plachy}, {Plum}, {Poggio}, {Pr{\v{s}}a}, {Pulone},
  {Racero}, {Ragaini}, {Rainer}, {Raiteri}, {Rambaux}, {Ramos}, {Ramos-Lerate},
  {Re Fiorentin}, {Regibo}, {Richards}, {Rios Diaz}, {Ripepi}, {Riva}, {Rix},
  {Rixon}, {Robichon}, {Robin}, {Robin}, {Roelens}, {Rogues}, {Rohrbasser},
  {Romero-G{\'o}mez}, {Rowell}, {Royer}, {Ruz Mieres}, {Rybicki}, {Sadowski},
  {S{\'a}ez N{\'u}{\~n}ez}, {Sagrist{\`a} Sell{\'e}s}, {Sahlmann}, {Salguero},
  {Samaras}, {Sanchez Gimenez}, {Sanna}, {Santove{\~n}a}, {Sarasso},
  {Schultheis}, {Sciacca}, {Segol}, {Segovia}, {S{\'e}gransan}, {Semeux},
  {Shahaf}, {Siddiqui}, {Siebert}, {Siltala}, {Silvelo}, {Slezak}, {Slezak},
  {Smart}, {Snaith}, {Solano}, {Solitro}, {Souami}, {Souchay}, {Spagna},
  {Spina}, {Spoto}, {Steele}, {Steidelm{\"u}ller}, {Stephenson}, {S{\"u}veges},
  {Surdej}, {Szabados}, {Szegedi-Elek}, {Taris}, {Taylor}, {Teixeira},
  {Tolomei}, {Tonello}, {Torra}, {Torra}, {Torralba Elipe}, {Trabucchi},
  {Tsounis}, {Turon}, {Ulla}, {Unger}, {Vaillant}, {van Dillen}, {van Reeven},
  {Vanel}, {Vecchiato}, {Viala}, {Vicente}, {Voutsinas}, {Weiler}, {Wevers},
  {Wyrzykowski}, {Yoldas}, {Yvard}, {Zhao}, {Zorec}, {Zucker}, \&
  {Zwitter}}]{2023A&A...674A...1G}
{Gaia Collaboration}, {Vallenari}, A., {Brown}, A.~G.~A., {et~al.}
  2023{\natexlab{b}}, \aap, 674, A1

\bibitem[{{Goldin} \& {Makarov}(2007)}]{2007ApJS..173..137G}
{Goldin}, A. \& {Makarov}, V.~V. 2007, \apjs, 173, 137

\bibitem[{{Grether} \& {Lineweaver}(2006)}]{2006ApJ...640.1051G}
{Grether}, D. \& {Lineweaver}, C.~H. 2006, \apj, 640, 1051

\bibitem[{{Grieves} {et~al.}(2017){Grieves}, {Ge}, {Thomas}, {Ma}, {Sithajan},
  {Ghezzi}, {Kimock}, {Willis}, {De Lee}, {Lee}, {Fleming}, {Agol}, {Troup},
  {Paegert}, {Schneider}, {Stassun}, {Varosi}, {Zhao}, {Jian}, {Li}, {Porto de
  Mello}, {Bizyaev}, {Pan}, {Dutra-Ferreira}, {Lorenzo-Oliveira}, {Santiago},
  {da Costa}, {Maia}, {Ogando}, \& {del Peloso}}]{2017MNRAS.467.4264G}
{Grieves}, N., {Ge}, J., {Thomas}, N., {et~al.} 2017, \mnras, 467, 4264

\bibitem[{{Halbwachs} {et~al.}(2000){Halbwachs}, {Arenou}, {Mayor}, {Udry}, \&
  {Queloz}}]{2000A&A...355..581H}
{Halbwachs}, J.~L., {Arenou}, F., {Mayor}, M., {Udry}, S., \& {Queloz}, D.
  2000, \aap, 355, 581

\bibitem[{{Halbwachs} {et~al.}(2018){Halbwachs}, {Mayor}, \&
  {Udry}}]{2018A&A...619A..81H}
{Halbwachs}, J.~L., {Mayor}, M., \& {Udry}, S. 2018, \aap, 619, A81

\bibitem[{{Hilditch}(2001)}]{2001icbs.book.....H}
{Hilditch}, R.~W. 2001, {An Introduction to Close Binary Stars}

\bibitem[{{Holl} {et~al.}(2022){Holl}, {Perryman}, {Lindegren}, {Segransan}, \&
  {Raimbault}}]{2022A&A...661A.151H}
{Holl}, B., {Perryman}, M., {Lindegren}, L., {Segransan}, D., \& {Raimbault},
  M. 2022, \aap, 661, A151

\bibitem[{{Holl} {et~al.}(2023){Holl}, {Sozzetti}, {Sahlmann}, {Giacobbe},
  {S{\'e}gransan}, {Unger}, {Delisle}, {Barbato}, {Lattanzi}, {Morbidelli}, \&
  {Sosnowska}}]{2023A&A...674A..10H}
{Holl}, B., {Sozzetti}, A., {Sahlmann}, J., {et~al.} 2023, \aap, 674, A10

\bibitem[{{Jumper} \& {Fisher}(2013)}]{2013ApJ...769....9J}
{Jumper}, P.~H. \& {Fisher}, R.~T. 2013, \apj, 769, 9

\bibitem[{{Kervella} {et~al.}(2022){Kervella}, {Arenou}, \&
  {Th{\'e}venin}}]{2022A&A...657A...7K}
{Kervella}, P., {Arenou}, F., \& {Th{\'e}venin}, F. 2022, \aap, 657, A7

\bibitem[{{Kiefer} {et~al.}(2021){Kiefer}, {H{\'e}brard}, {Lecavelier des
  Etangs}, {Martioli}, {Dalal}, \& {Vidal-Madjar}}]{2021A&A...645A...7K}
{Kiefer}, F., {H{\'e}brard}, G., {Lecavelier des Etangs}, A., {et~al.} 2021,
  \aap, 645, A7

\bibitem[{{Kiefer} {et~al.}(2019){Kiefer}, {H{\'e}brard}, {Sahlmann}, {Sousa},
  {Forveille}, {Santos}, {Mayor}, {Deleuil}, {Wilson}, {Dalal}, {D{\'\i}az},
  {Henry}, {Hagelberg}, {Hobson}, {Demangeon}, {Bourrier}, {Delfosse},
  {Arnold}, {Astudillo-Defru}, {Beuzit}, {Boisse}, {Bonfils}, {Borgniet},
  {Bouchy}, {Courcol}, {Ehrenreich}, {Hara}, {Lagrange}, {Lovis}, {Montagnier},
  {Moutou}, {Pepe}, {Perrier}, {Rey}, {Santerne}, {S{\'e}gransan}, {Udry}, \&
  {Vidal-Madjar}}]{2019A&A...631A.125K}
{Kiefer}, F., {H{\'e}brard}, G., {Sahlmann}, J., {et~al.} 2019, \aap, 631, A125

\bibitem[{{Kratter} \& {Lodato}(2016)}]{2016ARA&A..54..271K}
{Kratter}, K. \& {Lodato}, G. 2016, \araa, 54, 271

\bibitem[{{K{\"u}rster} {et~al.}(2000){K{\"u}rster}, {Endl}, {Els}, {Hatzes},
  {Cochran}, {D{\"o}bereiner}, \& {Dennerl}}]{2000A&A...353L..33K}
{K{\"u}rster}, M., {Endl}, M., {Els}, S., {et~al.} 2000, \aap, 353, L33

\bibitem[{{Latham} {et~al.}(2002){Latham}, {Stefanik}, {Torres}, {Davis},
  {Mazeh}, {Carney}, {Laird}, \& {Morse}}]{2002AJ....124.1144L}
{Latham}, D.~W., {Stefanik}, R.~P., {Torres}, G., {et~al.} 2002, \aj, 124, 1144

\bibitem[{{Li} {et~al.}(2021){Li}, {Brandt}, {Brandt}, {Dupuy}, {Michalik},
  {Jensen-Clem}, {Zeng}, {Faherty}, \& {Mitra}}]{2021AJ....162..266L}
{Li}, Y., {Brandt}, T.~D., {Brandt}, G.~M., {et~al.} 2021, \aj, 162, 266

\bibitem[{{Ma} \& {Ge}(2014)}]{2014MNRAS.439.2781M}
{Ma}, B. \& {Ge}, J. 2014, \mnras, 439, 2781

\bibitem[{{Marcussen} \& {Albrecht}(2023)}]{2023AJ....165..266M}
{Marcussen}, M.~L. \& {Albrecht}, S.~H. 2023, \aj, 165, 266

\bibitem[{{Marcy} {et~al.}(2001){Marcy}, {Butler}, {Fischer}, {Vogt},
  {Lissauer}, \& {Rivera}}]{2001ApJ...556..296M}
{Marcy}, G.~W., {Butler}, R.~P., {Fischer}, D., {et~al.} 2001, \apj, 556, 296

\bibitem[{{Marois} {et~al.}(2010){Marois}, {Zuckerman}, {Konopacky},
  {Macintosh}, \& {Barman}}]{2010Natur.468.1080M}
{Marois}, C., {Zuckerman}, B., {Konopacky}, Q.~M., {Macintosh}, B., \&
  {Barman}, T. 2010, \nat, 468, 1080

\bibitem[{{Mayor} {et~al.}(2003){Mayor}, {Pepe}, {Queloz}, {Bouchy},
  {Rupprecht}, {Lo Curto}, {Avila}, {Benz}, {Bertaux}, {Bonfils}, {Dall},
  {Dekker}, {Delabre}, {Eckert}, {Fleury}, {Gilliotte}, {Gojak}, {Guzman},
  {Kohler}, {Lizon}, {Longinotti}, {Lovis}, {Megevand}, {Pasquini}, {Reyes},
  {Sivan}, {Sosnowska}, {Soto}, {Udry}, {van Kesteren}, {Weber}, \&
  {Weilenmann}}]{2003Msngr.114...20M}
{Mayor}, M., {Pepe}, F., {Queloz}, D., {et~al.} 2003, The Messenger, 114, 20

\bibitem[{{Mayor} {et~al.}(1997){Mayor}, {Queloz}, {Udry}, \&
  {Halbwachs}}]{1997abos.conf..313M}
{Mayor}, M., {Queloz}, D., {Udry}, S., \& {Halbwachs}, J.-L. 1997, in IAU
  Colloq. 161: Astronomical and Biochemical Origins and the Search for Life in
  the Universe, ed. C.~{Batalli Cosmovici}, S.~{Bowyer}, \& D.~{Werthimer}, 313

\bibitem[{{Mayor} {et~al.}(2004){Mayor}, {Udry}, {Naef}, {Pepe}, {Queloz},
  {Santos}, \& {Burnet}}]{2004A&A...415..391M}
{Mayor}, M., {Udry}, S., {Naef}, D., {et~al.} 2004, \aap, 415, 391

\bibitem[{{Mazeh} {et~al.}(1998){Mazeh}, {Latham}, \&
  {Stefanik}}]{1998AcAau..42..593M}
{Mazeh}, T., {Latham}, D.~W., \& {Stefanik}, P. 1998, Acta Astronautica, 42,
  593

\bibitem[{{Ment} {et~al.}(2018){Ment}, {Fischer}, {Bakos}, {Howard}, \&
  {Isaacson}}]{2018AJ....156..213M}
{Ment}, K., {Fischer}, D.~A., {Bakos}, G., {Howard}, A.~W., \& {Isaacson}, H.
  2018, \aj, 156, 213

\bibitem[{{Mordasini} {et~al.}(2012){Mordasini}, {Alibert}, {Benz}, {Klahr}, \&
  {Henning}}]{2012A&A...541A..97M}
{Mordasini}, C., {Alibert}, Y., {Benz}, W., {Klahr}, H., \& {Henning}, T. 2012,
  \aap, 541, A97

\bibitem[{{Moutou} {et~al.}(2009){Moutou}, {Mayor}, {Lo Curto}, {Udry},
  {Bouchy}, {Benz}, {Lovis}, {Naef}, {Pepe}, {Queloz}, \&
  {Santos}}]{2009A&A...496..513M}
{Moutou}, C., {Mayor}, M., {Lo Curto}, G., {et~al.} 2009, \aap, 496, 513

\bibitem[{{Moutou} {et~al.}(2017){Moutou}, {Vigan}, {Mesa}, {Desidera},
  {Th{\'e}bault}, {Zurlo}, \& {Salter}}]{2017A&A...602A..87M}
{Moutou}, C., {Vigan}, A., {Mesa}, D., {et~al.} 2017, \aap, 602, A87

\bibitem[{{Murphy} {et~al.}(2016){Murphy}, {Bedding}, \&
  {Shibahashi}}]{2016ApJ...827L..17M}
{Murphy}, S.~J., {Bedding}, T.~R., \& {Shibahashi}, H. 2016, \apjl, 827, L17

\bibitem[{{Pepe} {et~al.}(2002){Pepe}, {Mayor}, {Galland}, {Naef}, {Queloz},
  {Santos}, {Udry}, \& {Burnet}}]{2002A&A...388..632P}
{Pepe}, F., {Mayor}, M., {Galland}, F., {et~al.} 2002, \aap, 388, 632

\bibitem[{{Perruchot} {et~al.}(2008){Perruchot}, {Kohler}, {Bouchy}, {Richaud},
  {Richaud}, {Moreaux}, {Merzougui}, {Sottile}, {Hill}, {Knispel}, {Regal},
  {Meunier}, {Ilovaisky}, {Le Coroller}, {Gillet}, {Schmitt}, {Pepe}, {Fleury},
  {Sosnowska}, {Vors}, {M{\'e}gevand}, {Blanc}, {Carol}, {Point}, {Laloge}, \&
  {Brunel}}]{2008SPIE.7014E..0JP}
{Perruchot}, S., {Kohler}, D., {Bouchy}, F., {et~al.} 2008, in Society of
  Photo-Optical Instrumentation Engineers (SPIE) Conference Series, Vol. 7014,
  Ground-based and Airborne Instrumentation for Astronomy II, ed. I.~S.
  {McLean} \& M.~M. {Casali}, 70140J

\bibitem[{{Perryman} {et~al.}(2014){Perryman}, {Hartman}, {Bakos}, \&
  {Lindegren}}]{2014ApJ...797...14P}
{Perryman}, M., {Hartman}, J., {Bakos}, G.~{\'A}., \& {Lindegren}, L. 2014,
  \apj, 797, 14

\bibitem[{{Perryman} {et~al.}(1997){Perryman}, {Lindegren}, {Kovalevsky},
  {Hoeg}, {Bastian}, {Bernacca}, {Cr{\'e}z{\'e}}, {Donati}, {Grenon},
  {Grewing}, {van Leeuwen}, {van der Marel}, {Mignard}, {Murray}, {Le Poole},
  {Schrijver}, {Turon}, {Arenou}, {Froeschl{\'e}}, \&
  {Petersen}}]{1997A&A...323L..49P}
{Perryman}, M.~A.~C., {Lindegren}, L., {Kovalevsky}, J., {et~al.} 1997, \aap,
  323, L49

\bibitem[{{Pollack} {et~al.}(1996){Pollack}, {Hubickyj}, {Bodenheimer},
  {Lissauer}, {Podolak}, \& {Greenzweig}}]{1996Icar..124...62P}
{Pollack}, J.~B., {Hubickyj}, O., {Bodenheimer}, P., {et~al.} 1996, \icarus,
  124, 62

\bibitem[{{Reffert} \& {Quirrenbach}(2011)}]{2011A&A...527A.140R}
{Reffert}, S. \& {Quirrenbach}, A. 2011, \aap, 527, A140

\bibitem[{{Rickman} {et~al.}(2019){Rickman}, {S{\'e}gransan}, {Marmier},
  {Udry}, {Bouchy}, {Lovis}, {Mayor}, {Pepe}, {Queloz}, {Santos}, {Allart},
  {Bonvin}, {Bratschi}, {Cersullo}, {Chazelas}, {Choplin}, {Conod}, {Deline},
  {Delisle}, {Dos Santos}, {Figueira}, {Giles}, {Girard}, {Lavie}, {Martin},
  {Motalebi}, {Nielsen}, {Osborn}, {Ottoni}, {Raimbault}, {Rey}, {Roger},
  {Seidel}, {Stalport}, {Su{\'a}rez Mascare{\~n}o}, {Triaud}, {Turner},
  {Weber}, \& {Wyttenbach}}]{2019A&A...625A..71R}
{Rickman}, E.~L., {S{\'e}gransan}, D., {Marmier}, M., {et~al.} 2019, \aap, 625,
  A71

\bibitem[{{Rosenthal} {et~al.}(2021){Rosenthal}, {Fulton}, {Hirsch},
  {Isaacson}, {Howard}, {Dedrick}, {Sherstyuk}, {Blunt}, {Petigura}, {Knutson},
  {Behmard}, {Chontos}, {Crepp}, {Crossfield}, {Dalba}, {Fischer}, {Henry},
  {Kane}, {Kosiarek}, {Marcy}, {Rubenzahl}, {Weiss}, \&
  {Wright}}]{2021ApJS..255....8R}
{Rosenthal}, L.~J., {Fulton}, B.~J., {Hirsch}, L.~A., {et~al.} 2021, \apjs,
  255, 8

\bibitem[{{Safronov}(1972)}]{1972epcf.book.....S}
{Safronov}, V.~S. 1972, {Evolution of the protoplanetary cloud and formation of
  the earth and planets.}

\bibitem[{{Sahlmann} {et~al.}(2011){Sahlmann}, {S{\'e}gransan}, {Queloz},
  {Udry}, {Santos}, {Marmier}, {Mayor}, {Naef}, {Pepe}, \&
  {Zucker}}]{2011A&A...525A..95S}
{Sahlmann}, J., {S{\'e}gransan}, D., {Queloz}, D., {et~al.} 2011, \aap, 525,
  A95

\bibitem[{{Schib} {et~al.}(2023){Schib}, {Mordasini}, \&
  {Helled}}]{2023A&A...669A..31S}
{Schib}, O., {Mordasini}, C., \& {Helled}, R. 2023, \aap, 669, A31

\bibitem[{{Sozzetti} {et~al.}(2006){Sozzetti}, {Udry}, {Zucker}, {Torres},
  {Beuzit}, {Latham}, {Mayor}, {Mazeh}, {Naef}, {Perrier}, {Queloz}, \&
  {Sivan}}]{2006A&A...449..417S}
{Sozzetti}, A., {Udry}, S., {Zucker}, S., {et~al.} 2006, \aap, 449, 417

\bibitem[{{Stefanik} {et~al.}(1994){Stefanik}, {Latham}, {Scarfe}, {Mazeh},
  {Davis}, \& {Torres}}]{1994AAS...184.4307S}
{Stefanik}, R.~P., {Latham}, D.~W., {Scarfe}, C.~D., {et~al.} 1994, in American
  Astronomical Society Meeting Abstracts, Vol. 184, American Astronomical
  Society Meeting Abstracts \#184, 43.07

\bibitem[{{Teague} {et~al.}(2018){Teague}, {Bae}, {Bergin}, {Birnstiel}, \&
  {Foreman-Mackey}}]{2018ApJ...860L..12T}
{Teague}, R., {Bae}, J., {Bergin}, E.~A., {Birnstiel}, T., \& {Foreman-Mackey},
  D. 2018, \apjl, 860, L12

\bibitem[{{Tinney} {et~al.}(2001){Tinney}, {Butler}, {Marcy}, {Jones}, {Penny},
  {Vogt}, {Apps}, \& {Henry}}]{2001ApJ...551..507T}
{Tinney}, C.~G., {Butler}, R.~P., {Marcy}, G.~W., {et~al.} 2001, \apj, 551, 507

\bibitem[{{Trifonov} {et~al.}(2020){Trifonov}, {Tal-Or}, {Zechmeister},
  {Kaminski}, {Zucker}, \& {Mazeh}}]{2020A&A...636A..74T}
{Trifonov}, T., {Tal-Or}, L., {Zechmeister}, M., {et~al.} 2020, \aap, 636, A74

\bibitem[{{Udry} {et~al.}(2002){Udry}, {Mayor}, {Naef}, {Pepe}, {Queloz},
  {Santos}, \& {Burnet}}]{2002A&A...390..267U}
{Udry}, S., {Mayor}, M., {Naef}, D., {et~al.} 2002, \aap, 390, 267

\bibitem[{{Udry} {et~al.}(2000){Udry}, {Mayor}, {Naef}, {Pepe}, {Queloz},
  {Santos}, {Burnet}, {Confino}, \& {Melo}}]{2000A&A...356..590U}
{Udry}, S., {Mayor}, M., {Naef}, D., {et~al.} 2000, \aap, 356, 590

\bibitem[{{van Leeuwen}(2007)}]{2007A&A...474..653V}
{van Leeuwen}, F. 2007, \aap, 474, 653

\bibitem[{{Venner} {et~al.}(2021){Venner}, {Vanderburg}, \&
  {Pearce}}]{2021AJ....162...12V}
{Venner}, A., {Vanderburg}, A., \& {Pearce}, L.~A. 2021, \aj, 162, 12

\bibitem[{{Vogt}(1987)}]{1987PASP...99.1214V}
{Vogt}, S.~S. 1987, \pasp, 99, 1214

\bibitem[{{Whitworth}(2018)}]{2018haex.bookE..95W}
{Whitworth}, A.~P. 2018, in Handbook of Exoplanets, ed. H.~J. {Deeg} \& J.~A.
  {Belmonte}, 95

\bibitem[{{Wilson} {et~al.}(2016){Wilson}, {H{\'e}brard}, {Santos}, {Sahlmann},
  {Montagnier}, {Astudillo-Defru}, {Boisse}, {Bouchy}, {Rey}, {Arnold},
  {Bonfils}, {Bourrier}, {Courcol}, {Deleuil}, {Delfosse}, {D{\'\i}az},
  {Ehrenreich}, {Forveille}, {Moutou}, {Pepe}, {Santerne}, {S{\'e}gransan}, \&
  {Udry}}]{2016A&A...588A.144W}
{Wilson}, P.~A., {H{\'e}brard}, G., {Santos}, N.~C., {et~al.} 2016, \aap, 588,
  A144

\bibitem[{{Winn}(2022)}]{2022AJ....164..196W}
{Winn}, J.~N. 2022, \aj, 164, 196

\bibitem[{{Wittenmyer} {et~al.}(2009){Wittenmyer}, {Endl}, {Cochran},
  {Ram{\'\i}rez}, {Reffert}, {MacQueen}, \& {Shetrone}}]{2009AJ....137.3529W}
{Wittenmyer}, R.~A., {Endl}, M., {Cochran}, W.~D., {et~al.} 2009, \aj, 137,
  3529

\bibitem[{{Wittenmyer} {et~al.}(2012){Wittenmyer}, {Horner}, {Tuomi}, {Salter},
  {Tinney}, {Butler}, {Jones}, {O'Toole}, {Bailey}, {Carter}, {Jenkins},
  {Zhang}, {Vogt}, \& {Rivera}}]{2012ApJ...753..169W}
{Wittenmyer}, R.~A., {Horner}, J., {Tuomi}, M., {et~al.} 2012, \apj, 753, 169

\bibitem[{{Zucker} \& {Mazeh}(2001)}]{2001ApJ...562..549Z}
{Zucker}, S. \& {Mazeh}, T. 2001, \apj, 562, 549

\end{thebibliography}


\onecolumn

\begin{appendix}

\section{Tables}
\label{app:tables}

\begin{landscape}

\begin{table}

\renewcommand{\arraystretch}{1.3}

\centering
\caption{Stellar parameters.}
\label{tab:stellar_parameters}

\resizebox{\paperheight-5cm}{!}{


\begin{tabular}{@{}lrlcccccccccc@{}}
\toprule
\multicolumn{1}{c}{{Target}} &
  \multicolumn{1}{c}{{Gaia DR3 ID}} &
  \multicolumn{1}{c}{{Spectral}} &
  {V$_{\mathrm{mag}}$} &
  {G$_{\mathrm{mag}}$} &
  {Distance} &
  {$M_{\star}$} &
  {$R_{\star}$} &
  {$L_{\star}$} &
  {$\rho_{\star}$} &
  {$\log g$} &
  {$T_{\mathrm{eff}}$} &
  {$[{\mathrm{Fe/H}}]$} \\
\multicolumn{1}{c}{} &
  \multicolumn{1}{c}{} &
  \multicolumn{1}{c}{Type} &
  [mag] &
  [mag] &
  [pc] &
  [\Msun] &
  [R$_{\odot}$] &
  [L$_{\odot}$] &
  [cgs] &
  [cgs] &
  [K] &
  [dex] \\ \midrule
HD89707 &
  3751763647996317056 &
  G2V &
  7.19 &
  7.03 &
  $34.87\pm0.15$ &
  $0.99^{+0.14}_{-0.11}$ &
  $1.14^{+0.19}_{-0.18}$ &
  $1.47^{+0.61}_{-0.50}$ &
  $0.98^{+0.54}_{-0.34}$ &
  $4.34\pm0.12$ &
  $5950\pm240$ &
  $-0.14\pm0.29$ \\
HD132406 &
  1594127865540229888 &
  G0V &
   &
  8.29 &
  $70.544\pm0.095$ &
  $1.14^{+0.13}_{-0.15}$ &
  $1.319\pm0.033$ &
  $2.42^{+0.46}_{-0.35}$ &
  $0.70\pm0.12$ &
  $4.253^{+0.060}_{-0.073}$ &
  $6270^{+330}_{-280}$ &
  $-0.17^{+0.24}_{-0.30}$ \\
HD175167 &
  6421118739093252224 &
  G5IV/V &
  8.0 &
  7.84 &
  $71.22^{+0.11}_{-0.12}$ &
  $1.20^{+0.21}_{-0.18}$ &
  $1.77^{+0.20}_{-0.23}$ &
  $4.1^{+1.4}_{-1.1}$ &
  $0.308^{+0.150}_{-0.090}$ &
  $4.03^{+0.12}_{-0.11}$ &
  $6210\pm340$ &
  $-0.21^{+0.26}_{-0.28}$ \\
HD82460 &
  824461960796102528 &
  G0 &
  8.37 &
  8.25 &
  $50.81\pm0.19$ &
  $0.957^{+0.100}_{-0.091}$ &
  $1.033^{+0.095}_{-0.100}$ &
  $1.23^{+0.32}_{-0.29}$ &
  $1.24^{+0.37}_{-0.28}$ &
  $4.397^{+0.073}_{-0.079}$ &
  $5980\pm220$ &
  $-0.25\pm0.26$ \\
HD81040 &
  637329067477530368 &
  G0V &
   &
  7.57 &
  $34.407\pm0.050$ &
  $0.900^{+0.098}_{-0.083}$ &
  $0.963\pm0.098$ &
  $1.16^{+0.33}_{-0.28}$ &
  $1.44^{+0.46}_{-0.34}$ &
  $4.432^{+0.076}_{-0.082}$ &
  $6100\pm220$ &
  $-0.52^{+0.28}_{-0.30}$ \\
HD92320 &
  855523714036230016 &
  F2 &
  8.38 &
  8.23 &
  $45.58\pm0.11$ &
  $0.900\pm0.073$ &
  $0.924\pm0.041$ &
  $0.96^{+0.15}_{-0.13}$ &
  $1.61^{+0.24}_{-0.22}$ &
  $4.463^{+0.045}_{-0.050}$ &
  $5950^{+190}_{-180}$ &
  $-0.38^{+0.22}_{-0.24}$ \\
HD77065 &
  685029558383335168 &
  K2 &
   &
  8.53 &
  $32.985\pm0.076$ &
  $0.791^{+0.044}_{-0.041}$ &
  $0.759\pm0.022$ &
  $0.378^{+0.050}_{-0.039}$ &
  $2.55^{+0.23}_{-0.21}$ &
  $4.576\pm0.029$ &
  $5200^{+160}_{-140}$ &
  $-0.15^{+0.17}_{-0.18}$ \\
HD68638A &
  1035000055055287680 &
  G8V &
   &
  7.3 &
  $32.513\pm0.065$ &
  $1.00^{+0.12}_{-0.11}$ &
  $1.10\pm0.11$ &
  $1.60^{+0.44}_{-0.38}$ &
  $1.06^{+0.34}_{-0.26}$ &
  $4.360^{+0.078}_{-0.085}$ &
  $6180\pm240$ &
  $-0.33^{+0.25}_{-0.27}$ \\
CD-4610046 &
  5999024986946599808 &
  G0 &
  10.62 &
  10.54 &
  $102.76\pm0.40$ &
  $0.860\pm0.049$ &
  $0.825^{+0.015}_{-0.014}$ &
  $0.518^{+0.048}_{-0.031}$ &
  $2.16\pm0.16$ &
  $4.540^{+0.027}_{-0.029}$ &
  $5391^{+140}_{-97}$ &
  $-0.04^{+0.14}_{-0.16}$ \\
HD52756 &
  5563001178343925376 &
  K2IV &
  8.47 &
  8.22 &
  $32.372\pm0.022$ &
  $0.835^{+0.051}_{-0.048}$ &
  $0.814\pm0.016$ &
  $0.546^{+0.064}_{-0.060}$ &
  $2.19\pm0.18$ &
  $4.540^{+0.030}_{-0.032}$ &
  $5500\pm170$ &
  $-0.21^{+0.17}_{-0.18}$ \\
HD40503 &
  2884087104955208064 &
  K2/3V &
  9.21 &
  8.97 &
  $39.205\pm0.026$ &
  $0.797\pm0.044$ &
  $0.762\pm0.020$ &
  $0.387^{+0.054}_{-0.044}$ &
  $2.54\pm0.21$ &
  $4.576\pm0.029$ &
  $5210^{+180}_{-160}$ &
  $-0.14^{+0.17}_{-0.18}$ \\
HD30246 &
  3309006602007842048 &
  G5 &
  8.28 &
  8.14 &
  $48.79\pm0.19$ &
  $0.956^{+0.074}_{-0.087}$ &
  $0.980\pm0.021$ &
  $1.10^{+0.14}_{-0.10}$ &
  $1.43^{+0.15}_{-0.17}$ &
  $4.437^{+0.038}_{-0.049}$ &
  $5970^{+200}_{-160}$ &
  $-0.24^{+0.21}_{-0.25}$ \\
HIP66074 &
  1712614124767394816 &
  M0V &
   &
  9.75 &
  $35.399^{+0.015}_{-0.014}$ &
  $0.737^{+0.036}_{-0.033}$ &
  $0.700\pm0.015$ &
  $0.189^{+0.033}_{-0.022}$ &
  $3.03^{+0.23}_{-0.21}$ &
  $4.615\pm0.025$ &
  $4550^{+200}_{-150}$ &
  $0.17^{+0.20}_{-0.22}$ \\
BD-170063 &
  2367734656180397952 &
  K4Vk: &
  9.72 &
  9.21 &
  $34.516\pm0.025$ &
  $0.778^{+0.038}_{-0.036}$ &
  $0.736^{+0.017}_{-0.018}$ &
  $0.256^{+0.031}_{-0.027}$ &
  $2.75\pm0.21$ &
  $4.595\pm0.025$ &
  $4780^{+140}_{-120}$ &
  $0.13^{+0.17}_{-0.16}$ \\
BD+291539 &
  873616860770228352 &
  G5 &
   &
  9.12 &
  $62.32^{+0.15}_{-0.14}$ &
  $0.920^{+0.063}_{-0.068}$ &
  $0.915\pm0.016$ &
  $0.799^{+0.100}_{-0.082}$ &
  $1.70\pm0.16$ &
  $4.480^{+0.034}_{-0.039}$ &
  $5710^{+200}_{-170}$ &
  $-0.10^{+0.18}_{-0.22}$ \\
HD166356 &
  2161507648230817792 &
  K0 &
  7.54 &
  7.37 &
  $61.65\pm0.20$ &
  $1.39^{+0.18}_{-0.24}$ &
  $1.83\pm0.11$ &
  $5.2^{+1.7}_{-1.3}$ &
  $0.317^{+0.089}_{-0.075}$ &
  $4.054^{+0.084}_{-0.100}$ &
  $6470^{+520}_{-470}$ &
  $-0.08^{+0.30}_{-0.35}$ \\
HD162020 &
  5957920668132624256 &
  K3V &
  9.12 &
  8.77 &
  $31.385\pm0.060$ &
  $0.797^{+0.045}_{-0.040}$ &
  $0.770\pm0.017$ &
  $0.413^{+0.056}_{-0.050}$ &
  $2.46^{+0.20}_{-0.18}$ &
  $4.567\pm0.028$ &
  $5270^{+190}_{-180}$ &
  $-0.18^{+0.17}_{-0.19}$ \\
HD164427 &
  6647630950597964544 &
  G0+V &
  6.89 &
  6.73 &
  $38.41^{+0.39}_{-0.37}$ &
  $1.19^{+0.13}_{-0.15}$ &
  $1.404\pm0.037$ &
  $2.66^{+0.48}_{-0.34}$ &
  $0.601^{+0.110}_{-0.097}$ &
  $4.216^{+0.062}_{-0.069}$ &
  $6220^{+310}_{-240}$ &
  $-0.06^{+0.22}_{-0.25}$ \\
HD48679 &
  1142214430312151424 &
  G0 &
  8.85 &
  8.69 &
  $66.98\pm0.37$ &
  $1.034\pm0.097$ &
  $1.127\pm0.036$ &
  $1.40^{+0.19}_{-0.17}$ &
  $1.02^{+0.15}_{-0.14}$ &
  $4.349\pm0.052$ &
  $5910^{+210}_{-200}$ &
  $0.01^{+0.20}_{-0.23}$ \\
HD112758 &
  3626268998574790656 &
  G9V &
  7.59 &
  7.33 &
  $20.07^{+0.16}_{-0.15}$ &
  $0.815^{+0.077}_{-0.066}$ &
  $0.800^{+0.082}_{-0.075}$ &
  $0.54^{+0.18}_{-0.14}$ &
  $2.27^{+0.59}_{-0.49}$ &
  $4.548^{+0.059}_{-0.068}$ &
  $5530\pm220$ &
  $-0.32\pm0.24$ \\
HD3277 &
  4994200964065634432 &
  G8V &
  7.44 &
  7.27 &
  $29.28^{+0.23}_{-0.21}$ &
  $0.942^{+0.110}_{-0.096}$ &
  $1.02\pm0.13$ &
  $1.22^{+0.40}_{-0.35}$ &
  $1.27^{+0.51}_{-0.37}$ &
  $4.403^{+0.091}_{-0.099}$ &
  $5980\pm220$ &
  $-0.29\pm0.27$ \\
HD114762 &
  3937211745905473024 &
  F9 &
   &
  7.15 &
  $38.17\pm0.16$ &
  $0.98^{+0.17}_{-0.15}$ &
  $1.18^{+0.16}_{-0.18}$ &
  $1.67^{+0.57}_{-0.51}$ &
  $0.87^{+0.46}_{-0.28}$ &
  $4.30\pm0.12$ &
  $6040\pm240$ &
  $-0.26^{+0.48}_{-0.66}$ \\
HD140913 &
  1224551770875466496 &
  G0V &
  8.05 &
  7.92 &
  $48.794\pm0.088$ &
  $0.987^{+0.081}_{-0.094}$ &
  $1.034\pm0.021$ &
  $1.32^{+0.15}_{-0.11}$ &
  $1.26\pm0.15$ &
  $4.403^{+0.042}_{-0.050}$ &
  $6090^{+180}_{-140}$ &
  $-0.26^{+0.21}_{-0.23}$ \\
HD151528 &
  4133650458966620672 &
  G8IV-V &
  7.61 &
  7.45 &
  $37.01\pm0.18$ &
  $1.02^{+0.14}_{-0.11}$ &
  $1.16^{+0.16}_{-0.17}$ &
  $1.60^{+0.61}_{-0.50}$ &
  $0.93^{+0.46}_{-0.29}$ &
  $4.32\pm0.11$ &
  $6010\pm250$ &
  $-0.13^{+0.25}_{-0.27}$ \\
HD17289 &
  4724313637321332864 &
  G0V &
  7.43 &
  7.3 &
  $45.76^{+1.10}_{-0.99}$ &
  $1.11^{+0.19}_{-0.16}$ &
  $1.37^{+0.21}_{-0.23}$ &
  $2.66^{+1.20}_{-0.98}$ &
  $0.63^{+0.39}_{-0.21}$ &
  $4.23^{+0.13}_{-0.12}$ &
  $6300^{+390}_{-370}$ &
  $-0.29\pm0.28$ \\
HD148284 &
  1318110830190386048 &
  G0 &
  9.02 &
  8.84 &
  $122.04\pm0.51$ &
  $1.39^{+0.17}_{-0.23}$ &
  $1.875^{+0.062}_{-0.069}$ &
  $4.92^{+1.70}_{-0.99}$ &
  $0.295^{+0.069}_{-0.064}$ &
  $4.033^{+0.076}_{-0.095}$ &
  $6270^{+580}_{-410}$ &
  $0.02^{+0.23}_{-0.27}$ \\
HD132032 &
  1181993180456516864 &
  G5 &
  8.11 &
  7.97 &
  $55.89\pm0.21$ &
  $1.055^{+0.097}_{-0.120}$ &
  $1.143\pm0.026$ &
  $1.79^{+0.27}_{-0.21}$ &
  $1.00^{+0.13}_{-0.14}$ &
  $4.347^{+0.047}_{-0.062}$ &
  $6250^{+250}_{-220}$ &
  $-0.27^{+0.23}_{-0.26}$ \\
HD17155 &
  4753355209745022208 &
  K4V &
  9.04 &
  8.71 &
  $28.082\pm0.072$ &
  $0.744^{+0.037}_{-0.034}$ &
  $0.708\pm0.017$ &
  $0.272^{+0.034}_{-0.030}$ &
  $2.95^{+0.24}_{-0.20}$ &
  $4.610\pm0.025$ &
  $4950^{+150}_{-140}$ &
  $-0.13\pm0.16$ \\
HR810 &
  4745373133284418816 &
  F8V &
  5.4 &
  5.26 &
  $17.357\pm0.011$ &
  $1.073^{+0.094}_{-0.120}$ &
  $1.140\pm0.027$ &
  $1.88^{+0.25}_{-0.16}$ &
  $1.02^{+0.13}_{-0.14}$ &
  $4.355^{+0.044}_{-0.060}$ &
  $6330^{+230}_{-170}$ &
  $-0.30^{+0.26}_{-0.30}$ \\
HD5433 &
  2778298280881817984 &
  G5 &
  8.65 &
  8.53 &
  $63.44\pm0.43$ &
  $0.984^{+0.100}_{-0.092}$ &
  $1.062^{+0.080}_{-0.076}$ &
  $1.26^{+0.29}_{-0.24}$ &
  $1.17^{+0.28}_{-0.23}$ &
  $4.382^{+0.065}_{-0.070}$ &
  $5930\pm220$ &
  $-0.13\pm0.26$ \\
HD91669 &
  3750881083756656128 &
  K0/1III &
  9.7 &
  9.49 &
  $71.81\pm0.26$ &
  $0.954^{+0.075}_{-0.065}$ &
  $0.963^{+0.100}_{-0.067}$ &
  $0.76^{+0.22}_{-0.13}$ &
  $1.50^{+0.34}_{-0.37}$ &
  $4.449^{+0.060}_{-0.082}$ &
  $5500^{+160}_{-140}$ &
  $0.16^{+0.19}_{-0.20}$ \\
BD-004475 &
  2651390587219807744 &
  G0 &
   &
  8.79 &
  $43.11^{+0.18}_{-0.19}$ &
  $0.835\pm0.052$ &
  $0.815\pm0.021$ &
  $0.557^{+0.068}_{-0.062}$ &
  $2.18^{+0.20}_{-0.19}$ &
  $4.539^{+0.031}_{-0.033}$ &
  $5520\pm170$ &
  $-0.24^{+0.19}_{-0.20}$ \\ \bottomrule
\end{tabular}%

}

\end{table}

\end{landscape}

\begin{table}[]
\centering

\caption{Radial velocity data used for our analysis for each target and their sources.}

\label{tab:rv_data}

\resizebox{0.41\textwidth}{!}{%
\begin{tabular}{@{}llrrrl@{}}
\toprule
Target     & Instrument & N$_{obs}$ & $\langle\sigma_{RV}\rangle$ & Timespan$^{\dagger}$ & Source                     \\
     &  &  & [\ms] & [days] &                      \\ \midrule
BD+291539  & SOPHIE+    & 16        & 3.2                         & 1095     & (1) \\ \midrule
BD-004475  & SOPHIE     & 6         & 11.2                        & 4053     & (2) \\
           & SOPHIE+    & 10        & 3.0                         &          & (2) \\ \midrule
BD-170063  & CORALIE14  & 14        & 5.7                         & 7185     & (0) \\
           & HARPS03    & 28        & 1.7                         &          & (3) \\
           & HIRES      & 12        & 1.6                         &          & (4) \\ \midrule
CD-4610046 & CORALIE14  & 14        & 9.8                         & 427      & (5) \\ \midrule
HD112758   & CORALIE14  & 9         & 4.0                         & 14820    & (0) \\
           & CORAVEL    & 70        & 366.0                       &          & (6) \\ \midrule
HD114762   & COUDE      & 27        & 33.6                        & 11183    & (7) \\
           & HAMILTON   & 57        & 17.5                        &          & (8) \\
           & HIRES      & 24        & 1.7                         &          & (9) \\ \midrule
HD132032   & SOPHIE     & 18        & 4.4                         & 735      & (10) \\ \midrule
HD132406   & ELODIE     & 17        & 12.0                        & 1078     & (11) \\
           & SOPHIE     & 4         & 3.8                         &          & (2) \\ \midrule
HD140913   & CORAVEL    & 53        & 422.3                       & 9825     & (6) \\
           & HIRES      & 5         & 1.5                         &          & (4) \\ \midrule
HD148284   & HIRES      & 27        & 1.6                         & 3323     & (4) \\ \midrule
HD151528   & CORALIE14  & 3         & 4.4                         & 8107     & (0) \\
           & CORALIE98  & 11        & 8.1                         &          & (0) \\ \midrule
HD162020   & CORALIE07  & 31        & 4.9                         & 8759     & (0) \\
           & CORALIE14  & 22        & 6.3                         &          & (0) \\
           & CORALIE98  & 52        & 9.6                         &          & (0) \\ \midrule
HD164427   & CORALIE07  & 8         & 3.4                         & 12353    & (0) \\
           & CORALIE14  & 4         & 4.5                         &          & (0) \\
           & CORALIE98  & 18        & 8.6                         &          & (0) \\
           & CORAVEL    & 5         & 330.0                       &          & (6) \\
           & UCLES      & 27        & 6.2                         &          & (12) \\ \midrule
HD166356   & SOPHIE     & 8         & 3.8                         & 43       & (2) \\ \midrule
HD17155    & CORALIE07  & 9         & 5.0                         & 7648     & (0) \\
           & CORALIE14  & 5         & 5.4                         &          & (0) \\
           & CORALIE98  & 6         & 8.8                         &          & (0) \\ \midrule
HD17289    & CORALIE07  & 13        & 4.1                         & 5428     & (0) \\
           & CORALIE98  & 26        & 9.3                         &          & (0) \\ \midrule
HD175167   & MIKE       & 13        & 4.2                         & 1828     & (13) \\ \midrule
HD30246    & SOPHIE     & 55        & 5.6                         & 3668     & (2) \\ \midrule
HD3277     & CORALIE07  & 14        & 3.5                         & 8493     & (0) \\
           & CORALIE14  & 5         & 3.1                         &          & (0) \\
           & CORALIE98  & 11        & 10.7                        &          & (0) \\ \midrule
HD48679    & SOPHIE     & 21        & 4.0                         & 846      & (2) \\ \midrule
HD52756    & CORALIE07  & 8         & 7.6                         & 4442     & (0) \\
           & CORALIE98  & 27        & 5.5                         &          & (0) \\ \midrule
HD5433     & SOPHIE     & 9         & 10.2                        & 4472     & (2) \\
           & SOPHIE+    & 11        & 3.7                         &          & (2) \\ \midrule
HD68638A   & ELODIE     & 27        & 13.3                        & 2187     & (11) \\ \midrule
HD77065    & SOPHIE     & 23        & 4.4                         & 1021     & (2) \\ \midrule
HD81040    & ELODIE     & 23        & 14.0                        & 5744     & (11) \\
           & HIRES      & 3         & 11.7                        &          & (14) \\
           & SOPHIE     & 8         & 2.2                         &          & (2) \\ \midrule
HD82460    & SOPHIE     & 14        & 4.2                         & 1487     & (2) \\ \midrule
HD89707    & CORALIE07  & 14        & 4.0                         & 14951    & (0) \\
           & CORALIE14  & 4         & 4.9                         &          & (0) \\
           & CORALIE98  & 32        & 6.1                         &          & (0) \\
           & CORAVEL    & 64        & 400.5                       &          & (6) \\ \midrule
HD91669    & TULL       & 18        & 10.0                        & 1504     & (15) \\ \midrule
HD92320    & SOPHIE     & 16        & 3.7                         & 1437     & (2) \\ \midrule
HIP66074   & HIRES      & 11        & 1.7                         & 1398     & (4) \\ \midrule
HR810      & CES        & 91        & 16.7                        & 4861    & (16) \\
           & CORALIE98  & 38        & 4.9                         &          & (0) \\
           & HARPS03    & 9         & 1.0                         &          & (17) \\ \bottomrule
\end{tabular}%
}

\vspace{0.5em} 
\small
\textbf{Note:} $^{\dagger}$ Total timespan considering all instruments. 
\textbf{Sources:} (0) this work; (1) \cite{2019A&A...631A.125K}; (2) SOPHIE archive: \url{http://atlas.obs-hp.fr/sophie/}; (3) \cite{2009A&A...496..513M}; (4) \cite{2017AJ....153..208B}; (5) \cite{2023A&A...674A..10H}; (6) DACE: \url{https://dace.unige.ch/radialVelocities/}; (7) \cite{1991ApJ...380L..35C}; (8) \cite{2006ApJ...646..505B}; (9) \cite{2021ApJS..255....8R}; (10) \cite{2016A&A...588A.144W}; (11) ELODIE archive: \url{http://atlas.obs-hp.fr/elodie/}; (12) \cite{2001ApJ...551..507T}; (13) \cite{2010ApJ...711.1229A}; (14) \cite{2006A&A...449..417S}; (15) \cite{2009AJ....137.3529W}; (16) \cite{2000A&A...353L..33K}; (17) \cite{2020A&A...636A..74T}

\end{table}


\renewcommand{\arraystretch}{1.5}

\begin{table}[h!]

\centering
\caption{MCMC results from the RV-only fit.}
\resizebox{\paperwidth-4cm}{!}{

\begin{tabular}{lccccccc}
\toprule
{Target} & {K} & {P} & {a} & {$T_p$} & {ecc} & {$\omega$} & {$M_2 \sin I$} \\
         & {[m/s]} & {[days]} & {[au]} & {[days]} &  & {[deg]} & {[\Mjup]} \\
\midrule
HIP66074 & $17.6^{+3.3}_{-1.7}$ & $300.3^{+2.8}_{-3.0}$ & $0.793\pm0.013$ & $56^{+21}_{-23}$ & $0.38^{+0.12}_{-0.11}$ & $-90\pm14$ & $0.440^{+0.056}_{-0.038}$ \\
HR810 & $55.7\pm1.3$ & $307.50\pm0.11$ & $0.878\pm0.022$ & $-2.5^{+8.1}_{-7.4}$ & $0.141\pm0.023$ & $-4.4^{+9.6}_{-8.9}$ & $1.772\pm0.097$ \\
BD-170063 & $173.3^{+1.1}_{-1.0}$ & $655.62\pm0.17$ & $1.358\pm0.022$ & $-139.4\pm1.4$ & $0.5456^{+0.0036}_{-0.0034}$ & $112.43^{+0.95}_{-1.03}$ & $5.25\pm0.17$ \\
HD132406 & $109^{+15}_{-13}$ & $970^{+19}_{-34}$ & $1.994^{+0.088}_{-0.094}$ & $-28^{+57}_{-104}$ & $0.313^{+0.058}_{-0.068}$ & $-143.1^{+16.5}_{-10.0}$ & $5.49^{+0.85}_{-0.71}$ \\
HD81040 & $165.6^{+7.7}_{-8.6}$ & $1003.6^{+6.7}_{-4.4}$ & $1.895^{+0.062}_{-0.066}$ & $127^{+28}_{-16}$ & $0.540^{+0.036}_{-0.044}$ & $79.2^{+5.2}_{-7.1}$ & $6.38\pm0.51$ \\
HD175167 & $148.3\pm4.1$ & $1243^{+34}_{-10}$ & $2.41^{+0.13}_{-0.14}$ & $1^{+66}_{-27}$ & $0.5122^{+0.0095}_{-0.0105}$ & $-15.6^{+5.0}_{-1.6}$ & $7.65\pm0.87$ \\
HD68638 & $325^{+21}_{-17}$ & $240.75\pm0.41$ & $0.757^{+0.028}_{-0.030}$ & $-28^{+11}_{-11}$ & $0.560^{+0.038}_{-0.035}$ & $-91.2\pm4.9$ & $8.24\pm0.70$ \\
HD114762 & $620.29\pm0.70$ & $83.91709^{+6.7e-4}_{-6.1e-4}$ & $0.373^{+0.019}_{-0.021}$ & $-30.784\pm0.036$ & $0.34413^{+1.01e-3}_{-9.2e-4}$ & $-156.15\pm0.13$ & $12.4^{+1.3}_{-1.4}$ \\
HD162020 & $1810.9^{+1.8}_{-1.7}$ & $8.4282377\pm2.7e-6$ & $0.0751\pm0.0013$ & $4.1877^{+0.0036}_{-0.0033}$ & $0.28151^{+7.6e-4}_{-8.4e-4}$ & $28.73^{+0.17}_{-0.16}$ & $14.96\pm0.53$ \\
HD91669 & $933.1^{+4.0}_{-4.4}$ & $497.60^{+0.85}_{-0.76}$ & $1.157\pm0.024$ & $-110.0^{+6.1}_{-5.2}$ & $0.4488\pm0.0029$ & $161.37\pm0.58$ & $28.8\pm1.2$ \\
BD-004475 & $737.2^{+9.3}_{-15.5}$ & $723.55^{+0.70}_{-0.87}$ & $1.614^{+0.052}_{-0.055}$ & $-31.8^{+2.0}_{-3.0}$ & $0.3849^{+0.0073}_{-0.0068}$ & $-94.1^{+1.6}_{-2.7}$ & $31.4\pm2.2$ \\
HD112758 & $1644.93^{+1.00}_{-0.94}$ & $103.2517\pm0.0048$ & $0.402^{+0.011}_{-0.012}$ & $28.16^{+0.20}_{-0.19}$ & $0.1531^{+0.0018}_{-0.0020}$ & $-23.39\pm0.63$ & $32.7\pm1.9$ \\
HD48679 & $1221.1\pm6.9$ & $1083\pm29$ & $2.085\pm0.075$ & $-333.59\pm0.15$ & $0.8220\pm0.0032$ & $155.62\pm0.30$ & $35.9\pm2.3$ \\
HD140913 & $1846^{+91}_{-76}$ & $147.931^{+0.020}_{-0.022}$ & $0.545^{+0.016}_{-0.017}$ & $-0.2\pm1.1$ & $0.542^{+0.015}_{-0.018}$ & $26.2\pm3.1$ & $40.0^{+3.2}_{-3.0}$ \\
HD148284 & $1022.5\pm1.6$ & $339.301\pm0.031$ & $1.062^{+0.049}_{-0.053}$ & $94.40\pm0.16$ & $0.3898\pm0.0011$ & $35.48\pm0.17$ & $40.2\pm3.8$ \\
HD30246 & $1145.2\pm4.6$ & $989.54\pm0.54$ & $1.914^{+0.052}_{-0.055}$ & $-71.53\pm0.53$ & $0.6588\pm0.0029$ & $-88.90\pm0.49$ & $41.0\pm2.3$ \\
HD77065 & $2818.1^{+1.5}_{-1.4}$ & $119.1142\pm0.0025$ & $0.4381\pm0.0079$ & $-31.3044\pm0.0076$ & $0.69410\pm3.4e-4$ & $105.943\pm0.064$ & $42.0\pm1.5$ \\
HD5433 & $1790^{+150}_{-130}$ & $575.1^{+1.5}_{-1.4}$ & $1.346^{+0.043}_{-0.045}$ & $67.6\pm1.7$ & $0.781\pm0.022$ & $83.6^{+4.1}_{-3.9}$ & $45.4\pm3.4$ \\
HD164427 & $2215.1^{+2.7}_{-2.9}$ & $108.53856^{+3.9e-4}_{-3.5e-4}$ & $0.472^{+0.018}_{-0.019}$ & $-20.645\pm0.021$ & $0.54916^{+7.4e-4}_{-7.9e-4}$ & $-3.205\pm0.090$ & $48.8\pm3.8$ \\
CD-4610046 & $1978\pm13$ & $242.48\pm0.31$ & $0.724\pm0.014$ & $53.01\pm0.55$ & $0.4276^{+0.0042}_{-0.0048}$ & $129.92^{+0.41}_{-0.44}$ & $49.6\pm1.9$ \\
HD17289 & $1396.9^{+5.3}_{-4.8}$ & $561.637\pm0.071$ & $1.379^{+0.069}_{-0.077}$ & $-258.59^{+0.63}_{-0.58}$ & $0.5292\pm0.0023$ & $51.80^{+0.45}_{-0.42}$ & $51.6^{+5.3}_{-5.6}$ \\
HD89707 & $5110\pm210$ & $298.241\pm0.015$ & $0.870^{+0.035}_{-0.038}$ & $75.03\pm0.55$ & $0.9464\pm0.0041$ & $107.9^{+2.4}_{-2.2}$ & $53.7\pm4.6$ \\
HD92320 & $2593.7\pm6.7$ & $145.413\pm0.012$ & $0.522^{+0.014}_{-0.014}$ & $26.89\pm0.32$ & $0.3259\pm0.0013$ & $-22.23\pm0.60$ & $59.1\pm3.2$ \\
HD52756 & $4949.4^{+2.7}_{-3.1}$ & $52.865777^{+6.1e-5}_{-5.6e-5}$ & $0.2595\pm0.0051$ & $-13.9268^{+0.0071}_{-0.0060}$ & $0.67782^{+2.7e-4}_{-3.2e-4}$ & $139.915\pm0.044$ & $59.6\pm2.4$ \\
BD+291539 & $2398.7\pm3.3$ & $175.928\pm0.044$ & $0.598^{+0.014}_{-0.015}$ & $-30.61\pm0.12$ & $0.2771\pm0.0019$ & $-49.83\pm0.26$ & $60.1\pm2.8$ \\
HD166356 & $1560^{+600}_{-390}$ & $275^{+104}_{-80}$ & $0.92^{+0.23}_{-0.19}$ & $70^{+1160}_{-900}$ & $0.129^{+0.119}_{-0.085}$ & $77\pm25$ & $61^{+33}_{-20}$ \\
HD132032 & $1953.1^{+1.9}_{-1.6}$ & $274.45\pm0.18$ & $0.841^{+0.028}_{-0.030}$ & $-53.2\pm1.2$ & $0.0838\pm0.0019$ & $-0.88^{+0.96}_{-0.85}$ & $64.5\pm4.4$ \\
HD3277 & $4071.5^{+1.1}_{-1.2}$ & $46.151369^{+8.1e-5}_{-7.3e-5}$ & $0.2468^{+0.0087}_{-0.0093}$ & $-18.8750\pm0.0099$ & $0.28462\pm2.5e-4$ & $-39.464^{+0.063}_{-0.071}$ & $66.2\pm4.8$ \\
HD151528 & $2884.8^{+4.7}_{-7.4}$ & $211.8478^{+0.0073}_{-0.0067}$ & $0.700^{+0.027}_{-0.030}$ & $97.168\pm0.070$ & $0.5344^{+0.0017}_{-0.0018}$ & $112.285^{+0.103}_{-0.094}$ & $72.4\pm5.9$ \\
HD17155 & $2596^{+25}_{-26}$ & $1425.92^{+0.44}_{-0.49}$ & $2.246\pm0.036$ & $429.2\pm1.0$ & $0.7755^{+0.0039}_{-0.0041}$ & $-82.89^{+0.19}_{-0.22}$ & $74.5\pm2.4$ \\
HD82460 & $4560^{+650}_{-550}$ & $589.44^{+0.43}_{-0.36}$ & $1.356^{+0.044}_{-0.046}$ & $56.6^{+4.8}_{-5.1}$ & $0.904^{+0.020}_{-0.023}$ & $-96.9^{+9.6}_{-8.3}$ & $77.8\pm5.7$ \\
\bottomrule
\end{tabular}

}
\label{tab:rvonlyMCMC}
\end{table}


\begin{landscape}
\renewcommand{\arraystretch}{1.5}

\begin{table}

\centering
\caption{MCMC results from the combined Gaia \& RV fit.}
\resizebox{\paperheight-5cm}{!}{

\begin{tabular}{@{}lcccccccccccc@{}}
\toprule
{Target} &
  {$\varpi$} &
  {$M_1$} &
  {$M_2 \sin I$ $^{\dagger}$} &
  {$M_2$} &
  {K} &
  {$a_r$} &
  {P} &
  {$T_p$ $^{\ddagger}$} &
  {ecc} &
  {$\omega$} &
  {$\Omega$} &
  {Inc} \\
   &
  {[mas]} &
  {[\Msun]} &
  {[\Mjup]} &
  {[\Mjup]} &
  {[m/s]} &
  {[au]} &
  {[days]} &
  {[days]} &
   &
  {[deg]} &
  {[deg]} &
  {[deg]} \\ \midrule
  \multicolumn{3}{l}{\textit{Robust solutions}} &
  \multicolumn{1}{l}{} &
  \multicolumn{1}{l}{} &
  \multicolumn{1}{l}{} &
  \multicolumn{1}{l}{} &
  \multicolumn{1}{l}{} &
  \multicolumn{1}{l}{} &
  \multicolumn{1}{l}{} &
  \multicolumn{1}{l}{} &
  \multicolumn{1}{l}{} &
  \multicolumn{1}{l}{} \\
BD-170063 &
  $28.993\pm0.018$ &
  $0.778\pm0.037$ &
  $5.25\pm0.17$ &
  $5.325\pm0.036$ &
  $173.35^{+0.76}_{-0.69}$ &
  $1.361\pm0.021$ &
  $655.641^{+0.070}_{-0.076}$ &
  $-139.40^{+0.36}_{-0.32}$ &
  $0.5455\pm0.0025$ &
  $112.41\pm0.43$ &
  $127.0^{+5.1}_{-3.6}$ &
  $82.4^{+2.8}_{-2.0}$ \\
HD132406 &
  $14.198\pm0.016$ &
  $1.14\pm0.14$ &
  $5.49^{+0.85}_{-0.71}$ &
  $6.96^{+0.89}_{-0.67}$ &
  $102.7\pm8.2$ &
  $1.942^{+0.080}_{-0.085}$ &
  $920^{+23}_{-12}$ &
  $-189^{+75}_{-45}$ &
  $0.279^{+0.049}_{-0.060}$ &
  $-126.9\pm8.6$ &
  $64^{+11}_{-14}$ &
  $132.3^{+5.4}_{-6.0}$ \\
HD81040 &
  $29.011\pm0.023$ &
  $0.900\pm0.090$ &
  $6.38\pm0.51$ &
  $7.19^{+0.23}_{-0.20}$ &
  $170.8^{+6.0}_{-5.6}$ &
  $1.900^{+0.061}_{-0.066}$ &
  $1004.8\pm2.4$ &
  $126^{+14}_{-15}$ &
  $0.528\pm0.028$ &
  $72.7\pm3.9$ &
  $19.3\pm4.6$ &
  $111.4^{+4.2}_{-4.8}$ \\
HD68638A &
  $30.701\pm0.023$ &
  $1.00\pm0.11$ &
  $8.24\pm0.70$ &
  $35.1\pm1.4$ &
  $300\pm15$ &
  $0.765^{+0.027}_{-0.029}$ &
  $240.59^{+0.25}_{-0.26}$ &
  $-32.0^{+6.4}_{-6.9}$ &
  $0.487^{+0.035}_{-0.038}$ &
  $-91.9\pm5.1$ &
  $143.0\pm5.7$ &
  $166.51\pm0.63$ \\
HD91669 &
  $13.767\pm0.015$ &
  $0.835\pm0.052$ &
  $28.8\pm1.2$ &
  $38.09\pm0.64$ &
  $933.1^{+3.9}_{-4.5}$ &
  $1.175\pm0.023$ &
  $498.73^{+0.67}_{-0.99}$ &
  $-102.0^{+5.0}_{-7.2}$ &
  $0.4485^{+0.0032}_{-0.0029}$ &
  $161.45\pm0.63$ &
  $244.9^{+2.4}_{-2.6}$ &
  $51.2^{+1.3}_{-1.1}$ \\
HD30246 &
  $20.687\pm0.028$ &
  $0.956\pm0.080$ &
  $41.0\pm2.3$ &
  $42.18\pm0.23$ &
  $1143.8\pm4.9$ &
  $1.942^{+0.051}_{-0.054}$ &
  $990.08\pm0.58$ &
  $-72.19^{+0.54}_{-0.57}$ &
  $0.6605\pm0.0030$ &
  $-90.10^{+0.50}_{-0.53}$ &
  $25.1\pm1.4$ &
  $85.0\pm1.2$ \\
BD-004475 &
  $23.103^{+0.015}_{-0.014}$ &
  $1.07\pm0.11$ &
  $31.4\pm2.2$ &
  $50.93^{+0.47}_{-0.71}$ &
  $747^{+16}_{-18}$ &
  $1.639^{+0.051}_{-0.054}$ &
  $723.71\pm0.83$ &
  $-29.4^{+3.0}_{-2.3}$ &
  $0.3803^{+0.0066}_{-0.0079}$ &
  $-91.8\pm2.2$ &
  $137.1\pm1.9$ &
  $139.63^{+0.63}_{-0.79}$ \\
HD77065 &
  $30.553\pm0.018$ &
  $0.791\pm0.043$ &
  $42.0\pm1.5$ &
  $66.73^{+0.79}_{-0.73}$ &
  $2818.3\pm1.5$ &
  $0.4496\pm0.0075$ &
  $119.1141^{+0.0028}_{-0.0025}$ &
  $-31.3047\pm0.0079$ &
  $0.69410\pm3.5e-4$ &
  $105.941^{+0.067}_{-0.060}$ &
  $290.91^{+0.91}_{-0.78}$ &
  $41.52\pm0.55$ \\
CD-4610046 &
  $9.687^{+0.017}_{-0.016}$ &
  $0.860\pm0.048$ &
  $49.6\pm1.9$ &
  $66.8^{+1.2}_{-1.1}$ &
  $1973^{+12}_{-11}$ &
  $0.741\pm0.013$ &
  $242.32\pm0.28$ &
  $53.28\pm0.51$ &
  $0.4255^{+0.0045}_{-0.0041}$ &
  $129.70\pm0.39$ &
  $316.4\pm1.3$ &
  $128.9\pm1.1$ \\
HD52756 &
  $30.820^{+0.016}_{-0.018}$ &
  $0.835\pm0.049$ &
  $59.6\pm2.4$ &
  $66.94^{+0.91}_{-0.86}$ &
  $4948.9\pm3.0$ &
  $0.2660\pm0.0049$ &
  $52.865770^{+6.8e-5}_{-6.2e-5}$ &
  $-13.9278\pm0.0068$ &
  $0.67786^{+2.8e-4}_{-3.2e-4}$ &
  $139.900\pm0.044$ &
  $27.6^{+1.8}_{-2.6}$ &
  $69.2\pm1.9$ \\
HD92320 &
  $22.131\pm0.017$ &
  $0.900\pm0.073$ &
  $59.1\pm3.2$ &
  $67.36\pm0.57$ &
  $2595.1\pm7.0$ &
  $0.535^{+0.013}_{-0.014}$ &
  $145.418^{+0.013}_{-0.012}$ &
  $26.92\pm0.33$ &
  $0.3259\pm0.0014$ &
  $-22.46^{+0.59}_{-0.62}$ &
  $76.70\pm0.98$ &
  $113.1\pm1.1$ \\
HD132032 &
  $17.599\pm0.026$ &
  $1.06\pm0.11$ &
  $64.5\pm4.4$ &
  $69.55^{+0.66}_{-0.63}$ &
  $1953.6\pm1.8$ &
  $0.859^{+0.027}_{-0.029}$ &
  $274.51\pm0.19$ &
  $-53.0^{+1.1}_{-1.2}$ &
  $0.0830\pm0.0019$ &
  $-1.00^{+0.91}_{-0.86}$ &
  $262.0^{+2.2}_{-2.1}$ &
  $75.1^{+2.1}_{-1.9}$ \\
BD+291539 &
  $16.118\pm0.014$ &
  $0.920\pm0.065$ &
  $60.1\pm2.8$ &
  $70.1^{+1.3}_{-1.2}$ &
  $2398.7\pm3.3$ &
  $0.612\pm0.014$ &
  $175.926^{+0.043}_{-0.045}$ &
  $-30.60\pm0.12$ &
  $0.2772\pm0.0019$ &
  $-49.81^{+0.29}_{-0.27}$ &
  $160.7^{+3.2}_{-3.6}$ &
  $115.9\pm2.0$ \\
HD17155 &
  $35.267\pm0.012$ &
  $0.744\pm0.035$ &
  $74.5\pm2.4$ &
  $85.99^{+0.40}_{-0.46}$ &
  $2581^{+28}_{-31}$ &
  $2.322\pm0.033$ &
  $1422.53^{+0.43}_{-0.53}$ &
  $408.64^{+1.12}_{-0.98}$ &
  $0.7649^{+0.0039}_{-0.0045}$ &
  $-80.78^{+0.49}_{-0.56}$ &
  $182.14\pm0.28$ &
  $70.44\pm0.30$ \\
HD140913 &
  $20.461\pm0.017$ &
  $0.987\pm0.087$ &
  $40.0^{+3.2}_{-3.0}$ &
  $93.3^{+1.7}_{-1.6}$ &
  $2023\pm73$ &
  $0.561^{+0.015}_{-0.016}$ &
  $147.912^{+0.022}_{-0.023}$ &
  $-1.67\pm0.90$ &
  $0.5223\pm0.0081$ &
  $23.7^{+2.1}_{-1.8}$ &
  $311.0\pm1.6$ &
  $30.3\pm1.3$ \\
HD148284 &
  $8.652^{+0.028}_{-0.069}$ &
  $1.39\pm0.20$ &
  $40.2\pm3.8$ &
  $103.6^{+2.0}_{-2.3}$ &
  $1022.8^{+1.9}_{-1.7}$ &
  $1.087^{+0.047}_{-0.051}$ &
  $339.307^{+0.094}_{-0.036}$ &
  $94.35^{+0.19}_{-0.31}$ &
  $0.3897^{+0.0013}_{-0.0012}$ &
  $35.42^{+0.20}_{-0.87}$ &
  $324.1^{+2.2}_{-2.9}$ &
  $156.00^{+0.45}_{-0.56}$ \\
HD48679 &
  $14.903^{+0.014}_{-0.017}$ &
  $1.034\pm0.096$ &
  $35.9\pm2.3$ &
  $108.88^{+0.91}_{-1.50}$ &
  $1234.1^{+7.9}_{-12.0}$ &
  $2.255^{+0.074}_{-0.091}$ &
  $1172.6^{+8.8}_{-91.9}$ &
  $-333.34^{+0.19}_{-0.24}$ &
  $0.8311^{+0.0013}_{-0.0069}$ &
  $156.29^{+0.29}_{-0.41}$ &
  $328.5^{+7.0}_{-1.1}$ &
  $20.87\pm0.18$ \\
HD166356 &
  $16.208\pm0.017$ &
  $1.39\pm0.21$ &
  $61^{+33}_{-20}$ &
  $110.9\pm3.4$ &
  $2760^{+140}_{-130}$ &
  $0.914^{+0.041}_{-0.045}$ &
  $260.69^{+0.20}_{-0.22}$ &
  $-83.9^{+5.5}_{-5.1}$ &
  $0.412^{+0.035}_{-0.033}$ &
  $108.1^{+4.1}_{-3.8}$ &
  $182.3\pm1.7$ &
  $111.1\pm1.5$ \\
HD151528 &
  $27.150^{+0.024}_{-0.022}$ &
  $1.02\pm0.13$ &
  $72.4\pm5.9$ &
  $145.7^{+1.0}_{-1.1}$ &
  $2883.2\pm3.0$ &
  $0.730^{+0.025}_{-0.027}$ &
  $211.8506\pm0.0044$ &
  $97.204^{+0.062}_{-0.049}$ &
  $0.5338^{+0.0011}_{-0.0012}$ &
  $112.348^{+0.076}_{-0.088}$ &
  $220.72\pm0.29$ &
  $147.20^{+0.24}_{-0.26}$ \\
HD114762 &
  $25.335^{+0.040}_{-0.036}$ &
  $0.98\pm0.16$ &
  $12.4^{+1.3}_{-1.4}$ &
  $218.3^{+3.5}_{-2.2}$ &
  $620.38\pm0.65$ &
  $0.397^{+0.017}_{-0.019}$ &
  $83.91720\pm6.1e-4$ &
  $-30.781^{+0.039}_{-0.030}$ &
  $0.34405\pm9.0e-4$ &
  $-156.15^{+0.15}_{-0.13}$ &
  $156.41^{+0.58}_{-0.70}$ &
  $3.699^{+0.035}_{-0.052}$ \\
HD112758 &
  $47.679\pm0.020$ &
  $0.815\pm0.071$ &
  $32.7\pm1.9$ &
  $257.28\pm0.93$ &
  $1644.6\pm1.6$ &
  $0.4392\pm0.0099$ &
  $103.2495\pm0.0049$ &
  $27.617^{+0.085}_{-0.107}$ &
  $0.1577^{+0.0017}_{-0.0015}$ &
  $-25.14^{+0.19}_{-0.28}$ &
  $155.20^{+0.20}_{-0.15}$ &
  $8.716\pm0.029$ \\
HD164427 &
  $25.550^{+0.021}_{-0.019}$ &
  $1.19\pm0.14$ &
  $48.8\pm3.8$ &
  $355.5^{+2.6}_{-2.9}$ &
  $2216.2^{+2.8}_{-3.1}$ &
  $0.513^{+0.015}_{-0.016}$ &
  $108.53855\pm3.3e-4$ &
  $-20.642\pm0.017$ &
  $0.54944\pm7.3e-4$ &
  $-3.187^{+0.078}_{-0.070}$ &
  $337.69^{+0.41}_{-0.49}$ &
  $9.340^{+0.066}_{-0.058}$ \\
HD162020 &
  $31.708^{+0.012}_{-0.014}$ &
  $0.797\pm0.042$ &
  $14.96\pm0.53$ &
  $410.8^{+5.8}_{-5.3}$ &
  $1811.2^{+1.3}_{-1.6}$ &
  $0.0859\pm0.0010$ &
  $8.4282388^{+1.4e-6}_{-2.6e-6}$ &
  $4.1874^{+0.0026}_{-0.0023}$ &
  $0.28126\pm5.7e-4$ &
  $28.70^{+0.13}_{-0.12}$ &
  $288.93^{+0.67}_{-0.73}$ &
  $177.273^{+0.030}_{-0.027}$ \\
  \midrule
\multicolumn{3}{l}{\textit{Challenging cases*}} &
  \multicolumn{1}{l}{} &
  \multicolumn{1}{l}{} &
  \multicolumn{1}{l}{} &
  \multicolumn{1}{l}{} &
  \multicolumn{1}{l}{} &
  \multicolumn{1}{l}{} &
  \multicolumn{1}{l}{} &
  \multicolumn{1}{l}{} &
  \multicolumn{1}{l}{} &
  \multicolumn{1}{l}{} \\
  
HIP66074 &
  $28.2199\pm0.0099$ &
  $0.737\pm0.034$ &
  $0.440^{+0.056}_{-0.038}$ &
  $2.46\pm0.42$ &
  $15.96^{+1.17}_{-0.95}$ &
  $0.790\pm0.012$ &
  $298.3^{+1.1}_{-1.2}$ &
  $46.8\pm7.3$ &
  $0.243\pm0.059$ &
  $-80^{+12}_{-10}$ &
  $338^{+11}_{-15}$ &
  $9.8^{+2.0}_{-1.5}$ \\
HD175167 &
  $14.109\pm0.020$ &
  $1.20\pm0.19$ &
  $7.65\pm0.87$ &
  $15.7^{+1.9}_{-1.7}$ &
  $178^{+24}_{-18}$ &
  $2.33^{+0.12}_{-0.13}$ &
  $1177^{+27}_{-25}$ &
  $-186\pm77$ &
  $0.514\pm0.034$ &
  $-29.3^{+8.9}_{-11.2}$ &
  $65.4^{+7.8}_{-6.1}$ &
  $35.8^{+2.2}_{-2.3}$ \\
HD5433 &
  $15.740^{+0.022}_{-0.026}$ &
  $0.985\pm0.096$ &
  $45.4\pm3.4$ &
  $69.19^{+0.91}_{-0.83}$ &
  $1657^{+72}_{-69}$ &
  $1.376^{+0.040}_{-0.044}$ &
  $574.92\pm0.67$ &
  $67.56\pm0.80$ &
  $0.757\pm0.013$ &
  $88.1\pm2.4$ &
  $74.9^{+2.4}_{-2.6}$ &
  $41.34\pm0.77$ \\
HD82460 &
  $19.995\pm0.024$ &
  $0.957\pm0.096$ &
  $77.8\pm5.7$ &
  $67.2^{+1.2}_{-1.0}$ &
  $2280^{+113}_{-93}$ &
  $1.383^{+0.042}_{-0.045}$ &
  $587.6\pm2.4$ &
  $69.1\pm3.5$ &
  $0.760^{+0.024}_{-0.022}$ &
  $-81.4\pm1.3$ &
  $173.5\pm1.3$ &
  $67.0^{+1.2}_{-1.0}$ \\
HD89707 &
  $29.150\pm0.028$ &
  $0.99\pm0.12$ &
  $53.7\pm4.6$ &
  $105.0^{+1.4}_{-1.6}$ &
  $10170^{+410}_{-840}$ &
  $0.899^{+0.033}_{-0.036}$ &
  $298.228^{+0.022}_{-0.017}$ &
  $73.91^{+1.04}_{-0.49}$ &
  $0.97951^{+3.6e-4}_{-7.1e-4}$ &
  $128.5^{+2.3}_{-6.9}$ &
  $169.2^{+9.1}_{-3.2}$ &
  $42.8^{+1.6}_{-3.5}$ \\
HD3277 &
  $35.258^{+0.019}_{-0.018}$ &
  $0.94\pm0.10$ &
  $66.2\pm4.8$ &
  $489.4^{+2.5}_{-2.1}$ &
  $4071.34\pm0.95$ &
  $0.2823^{+0.0067}_{-0.0071}$ &
  $46.151350^{+8.2e-5}_{-6.2e-5}$ &
  $-18.8742^{+0.0108}_{-0.0079}$ &
  $0.28453^{+3.2e-4}_{-2.6e-4}$ &
  $-39.454^{+0.069}_{-0.061}$ &
  $272.57^{+0.19}_{-0.18}$ &
  $169.810^{+0.040}_{-0.035}$ \\
HD17289 &
  $19.872\pm0.015$ &
  $1.11\pm0.17$ &
  $51.6^{+5.3}_{-5.6}$ &
  $540.6^{+2.1}_{-2.0}$ &
  $1385.6^{+6.6}_{-7.7}$ &
  $1.567^{+0.054}_{-0.058}$ &
  $561.877^{+0.100}_{-0.085}$ &
  $-257.44\pm0.63$ &
  $0.5152\pm0.0037$ &
  $51.85^{+0.60}_{-0.64}$ &
  $4.70^{+0.51}_{-0.57}$ &
  $172.917\pm0.044$ \\ \bottomrule
\end{tabular}%

}

\vspace{0.5em} 
\small
\textbf{Note:} Targets are sorted by the true mass of the companion. $a_r$ is the relative semi major axis. $^{\dagger}$ The minimum mass was obtained from the RV only fit. $^{\ddagger}$ The time of periastron is taken in relation to the Gaia DR3 epoch of 2016.0 (BJD=2457389.0). \textbf{*} Caution should be exercised when considering the solutions for these systems, as they may be unreliable due to discrepancies between the Gaia and RV solutions.

\label{tab:jointMCMC}
\end{table}

\end{landscape}




\section{RV orbital solutions phase-folded plots} 
\label{app:rvplots}

    \onecolumn
    	{
    	\centering
    \captionof{figure}{Radial velocity curves of the joint solution, residuals, and phase pholds for the companions detected around each star. Additionally, at the bottom we show the Z-score (the difference in standard deviations) between the Gaia only orbital solution and the combined Gaia-RV orbital solution.}
    \begin{longtable}[h!]{c c}
    \label{fig:app-phasefolded}
    
    \endfirsthead
    \multicolumn{2}{c}{\small \textbf{Fig. \thetable\ } continued.}\\
    \endhead

        \includegraphics[width=0.45\linewidth]{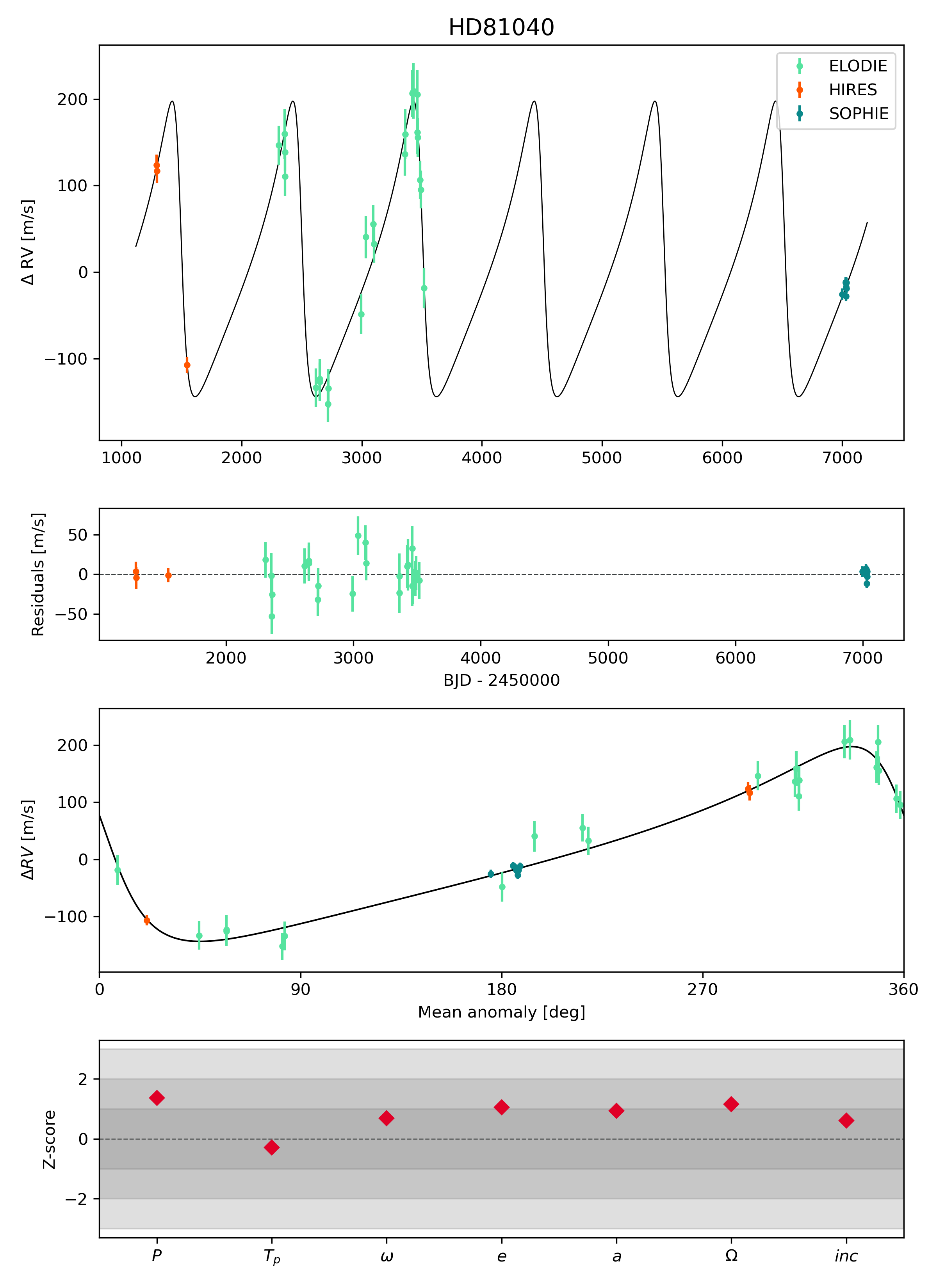}&
        \includegraphics[width=0.45\linewidth]{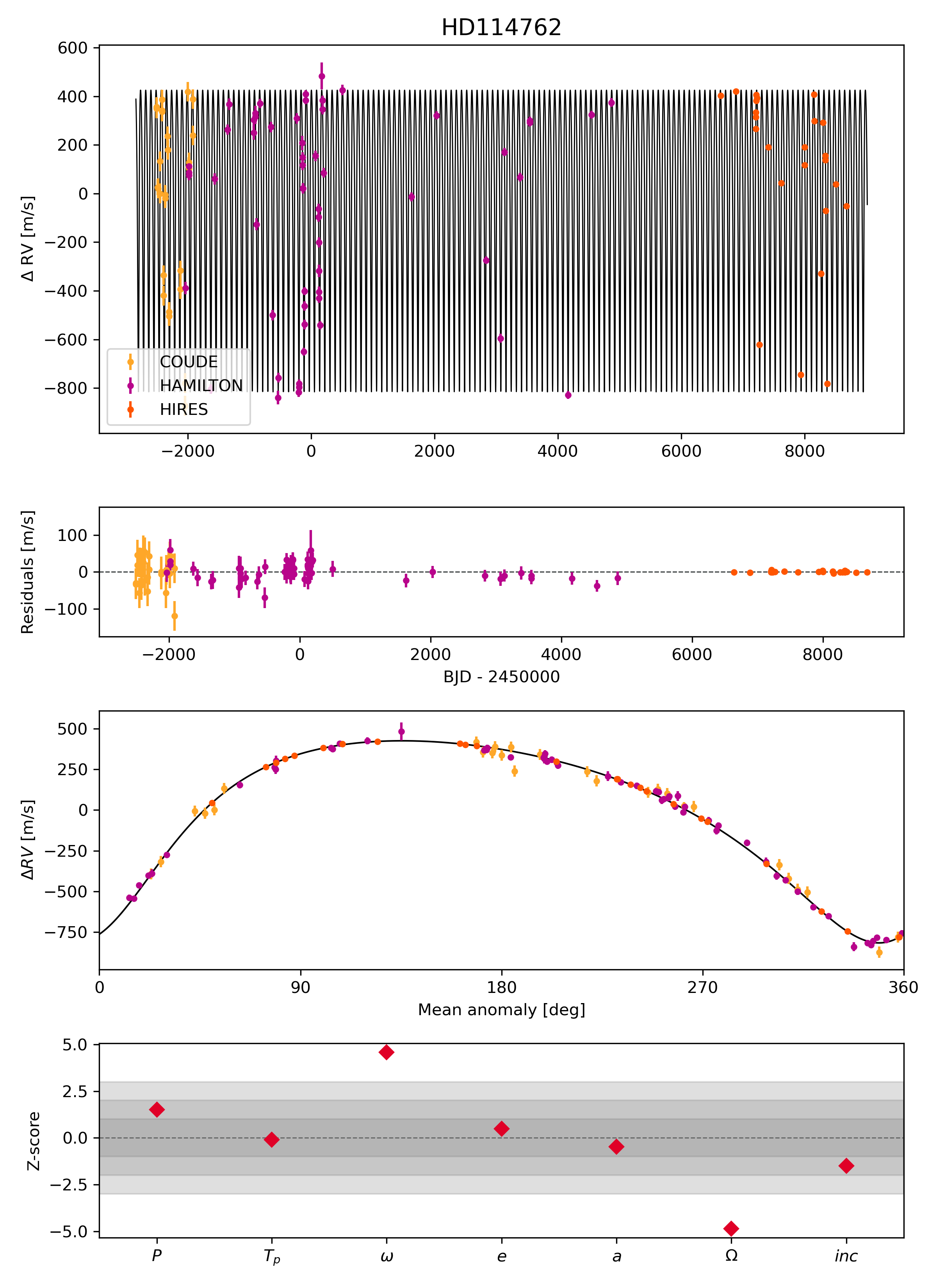}\\
        
        \includegraphics[width=0.45\linewidth]{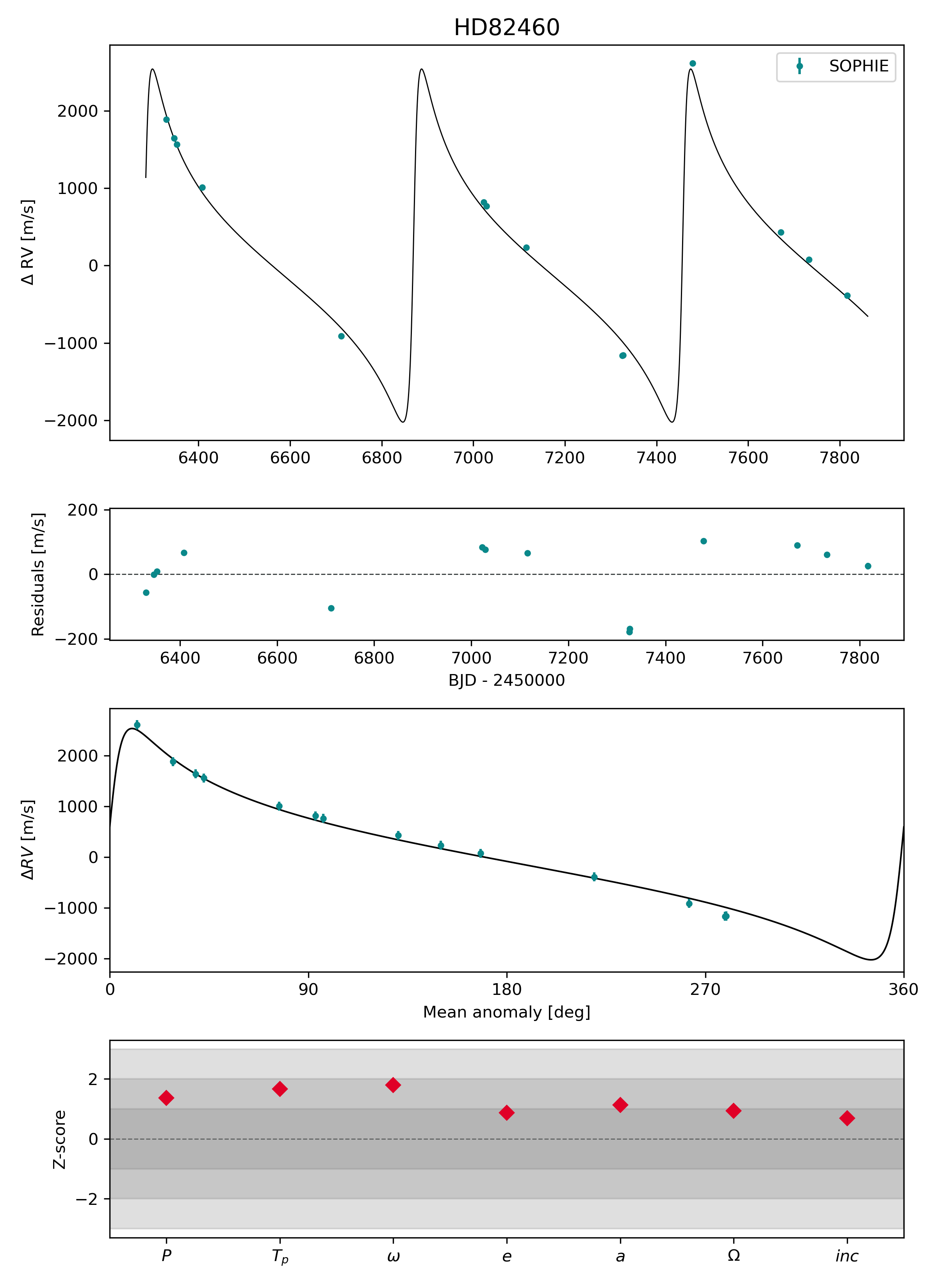}&
        \includegraphics[width=0.45\linewidth]{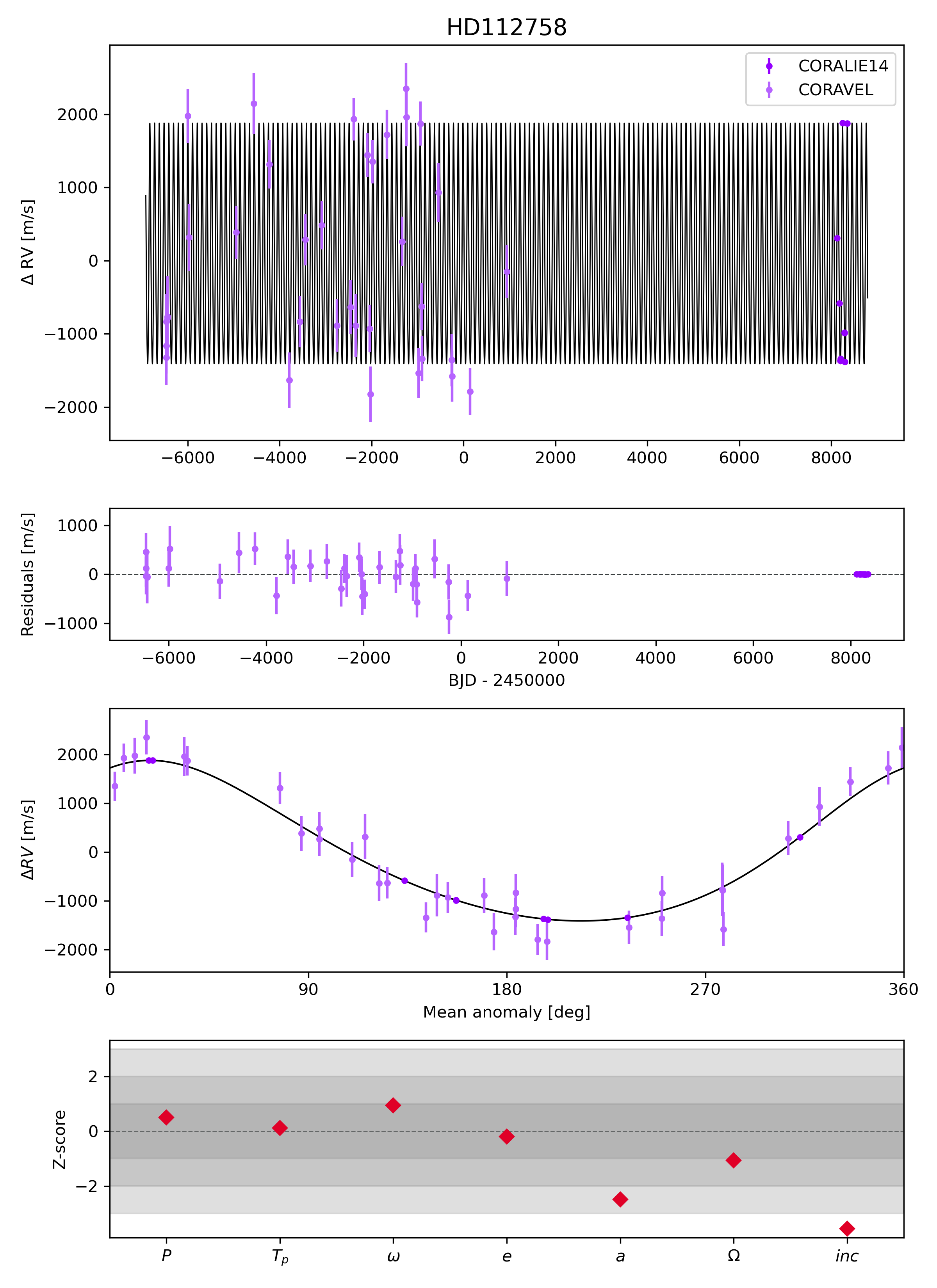}\\
        
        \includegraphics[width=0.45\linewidth]{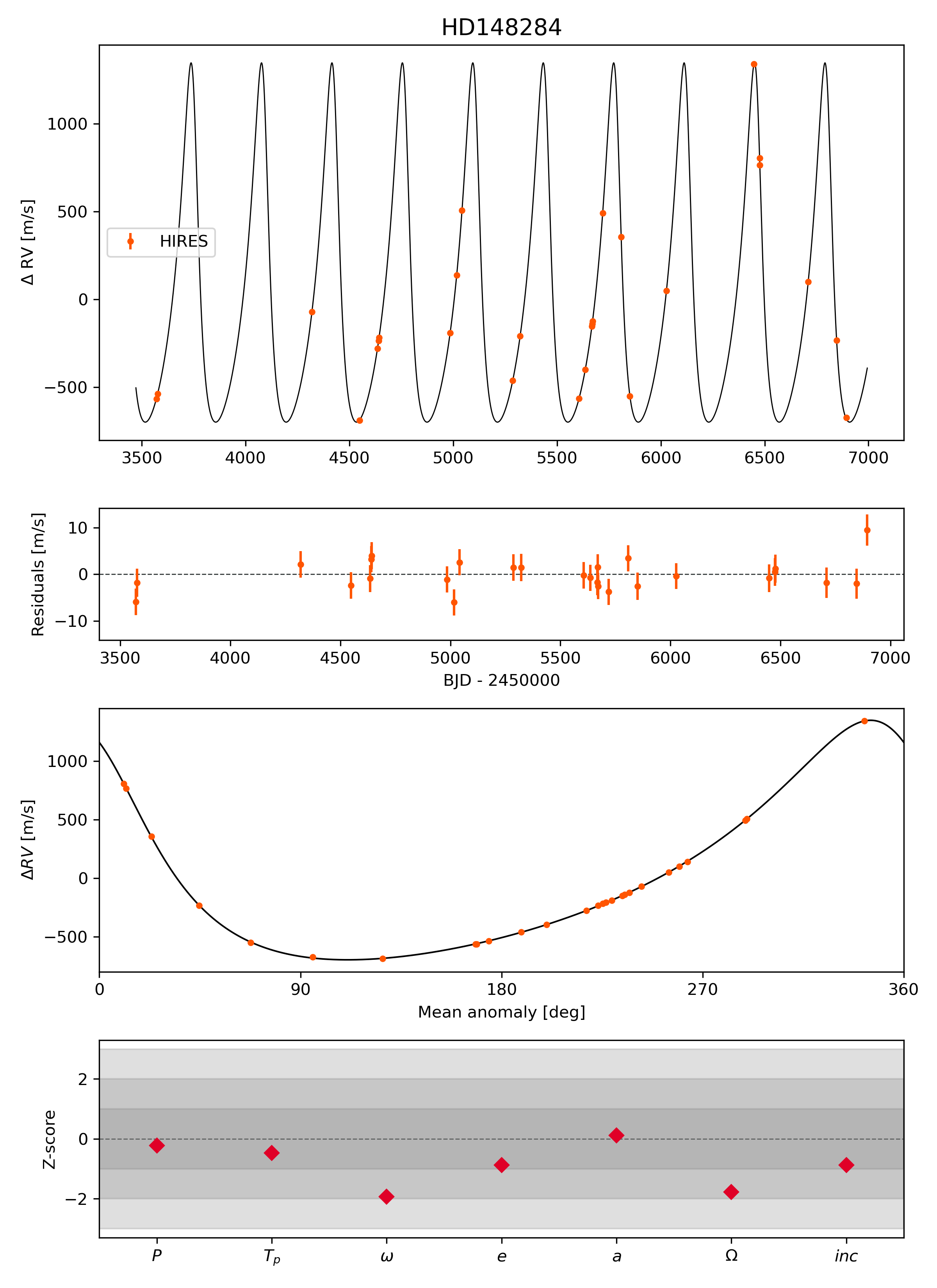}&
        \includegraphics[width=0.45\linewidth]{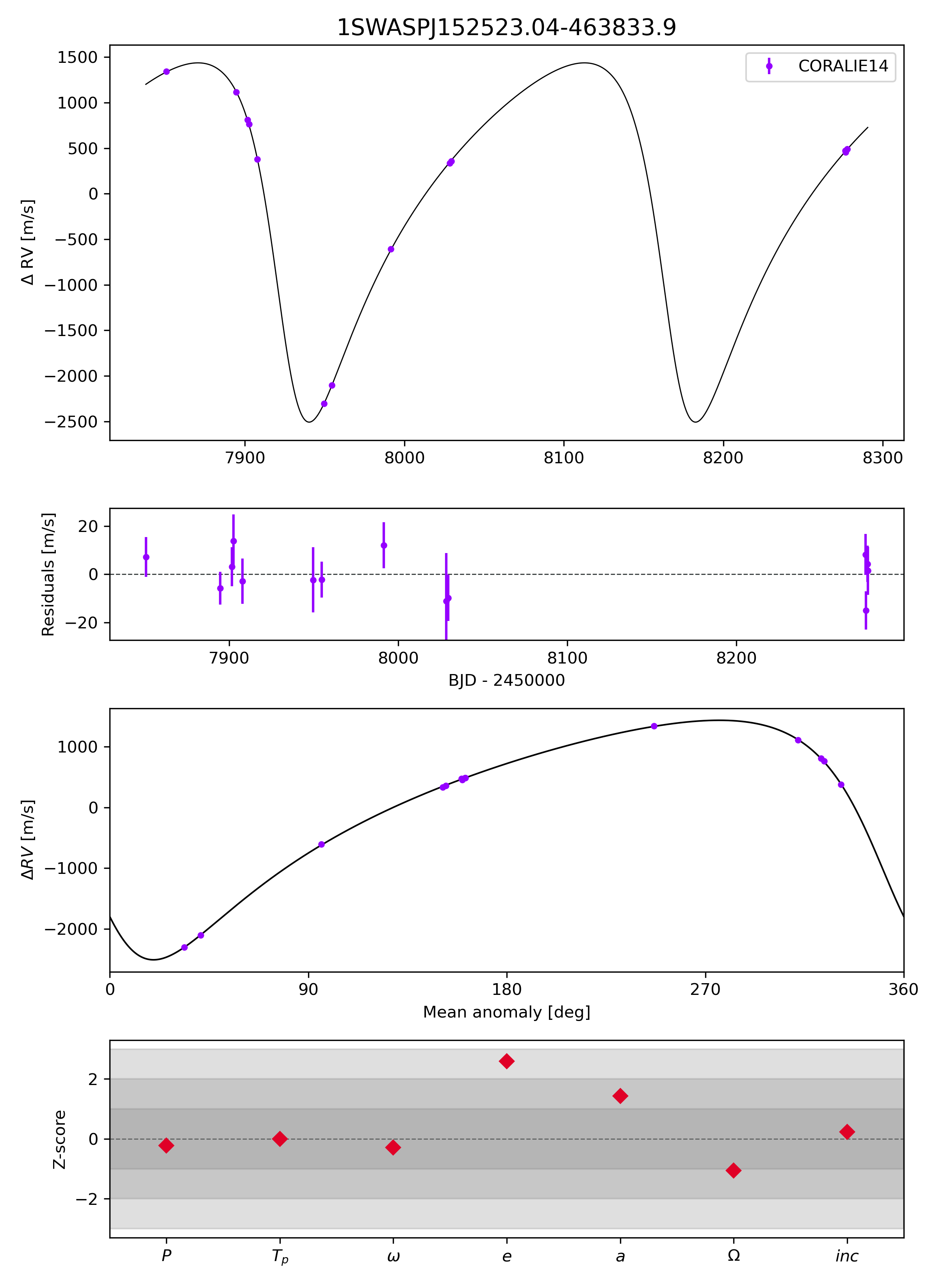}\\
        
        \includegraphics[width=0.45\linewidth]{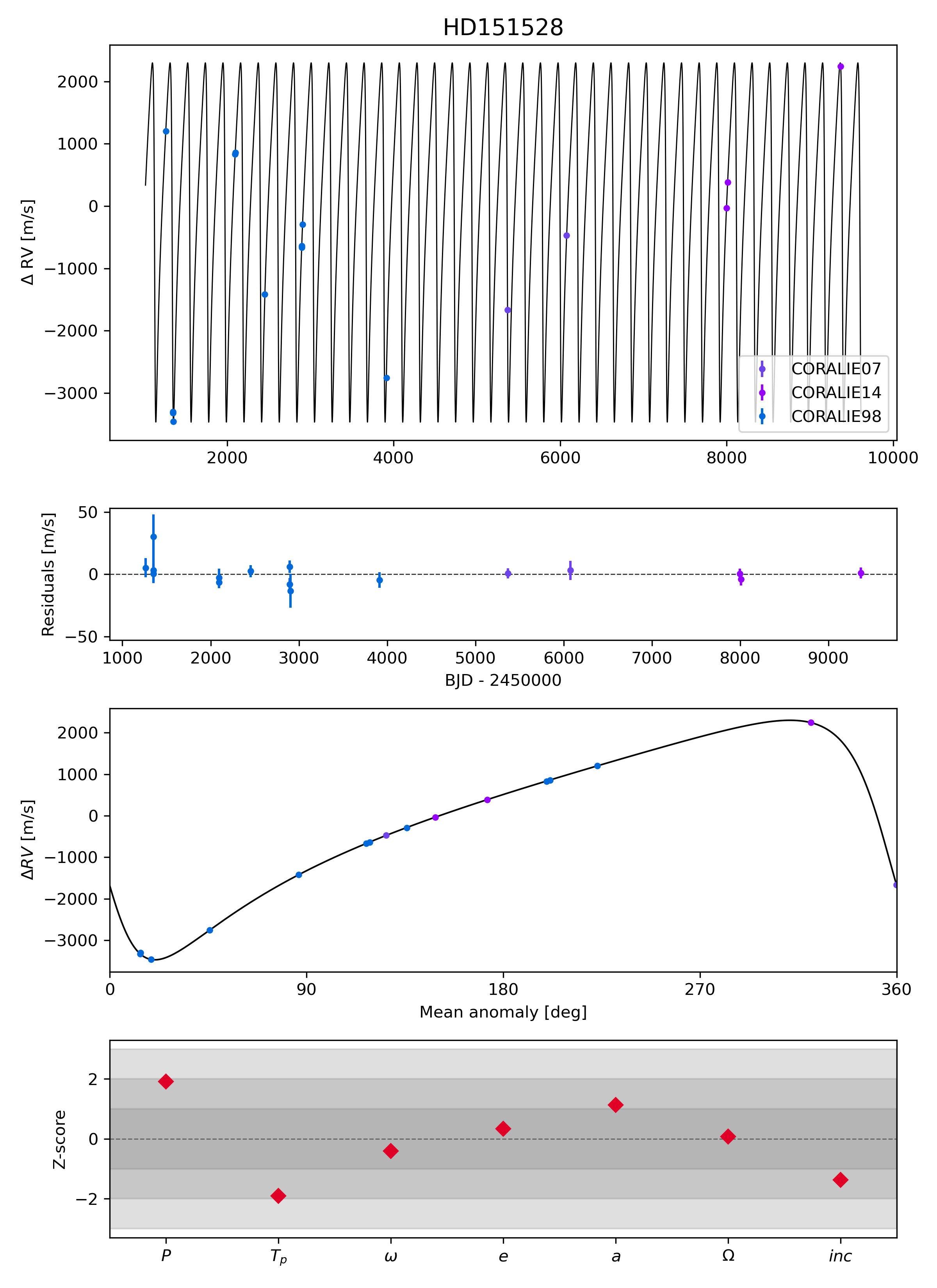}&
        \includegraphics[width=0.45\linewidth]{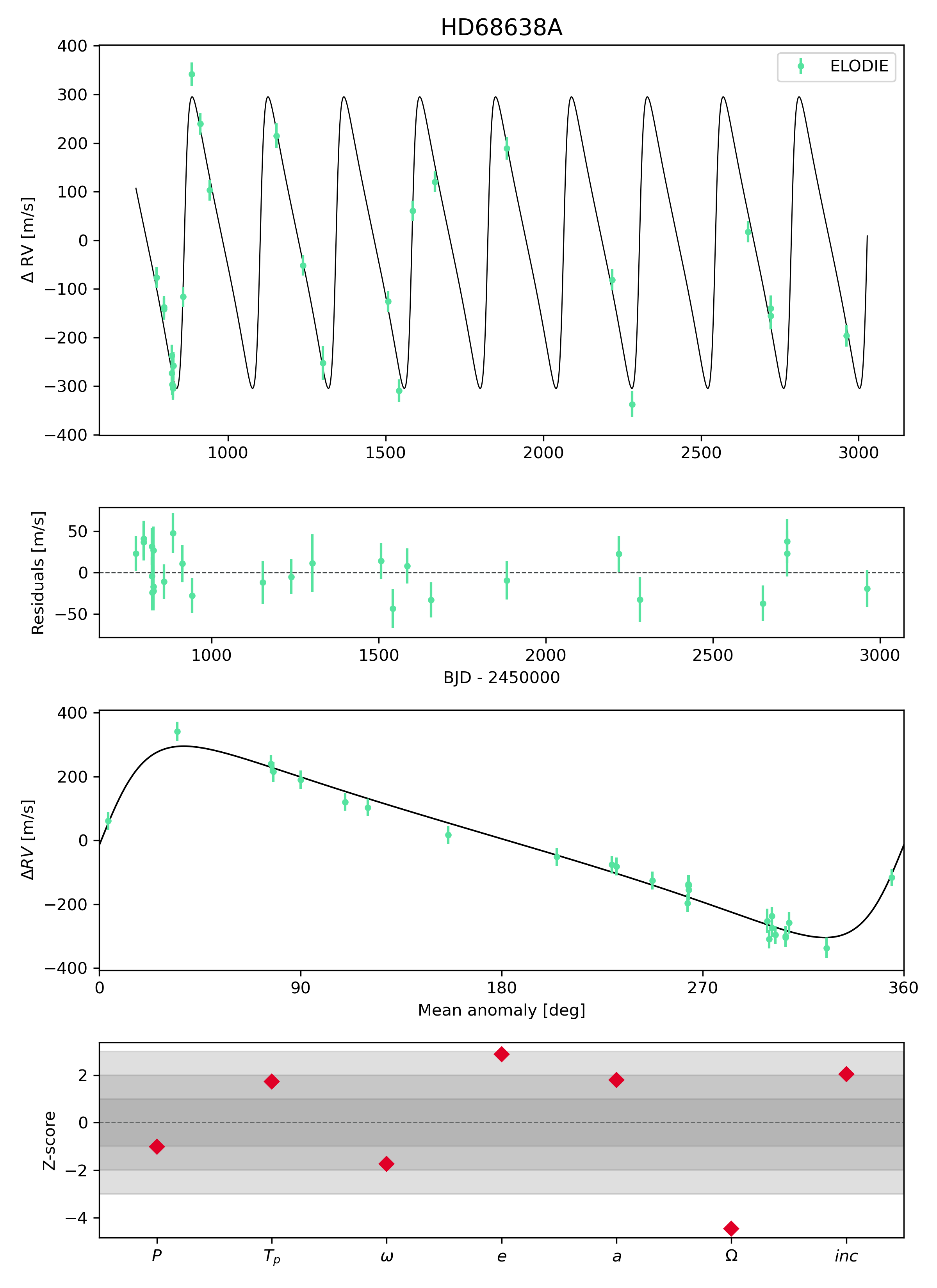}\\

        \includegraphics[width=0.45\linewidth]{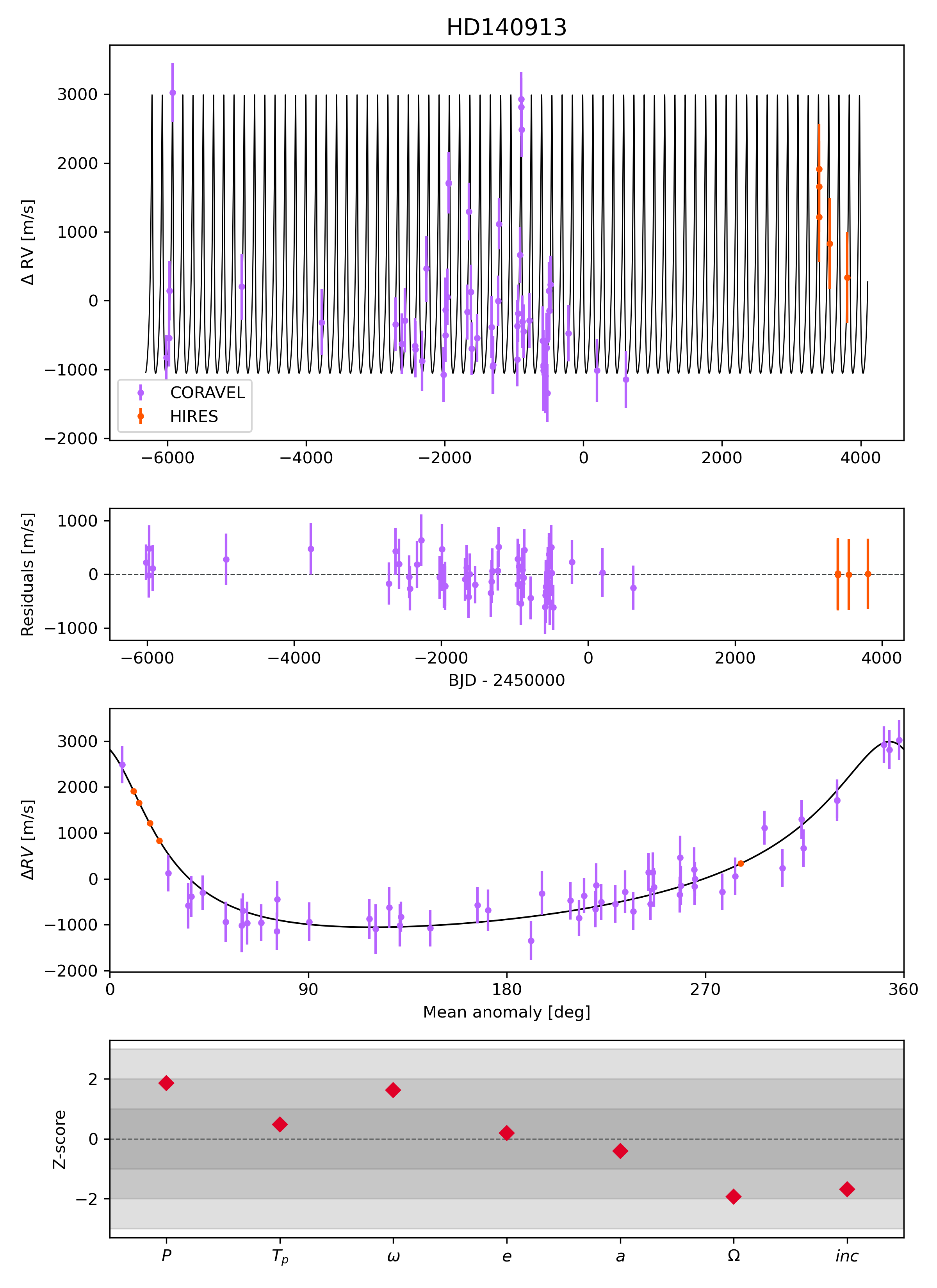}&
        \includegraphics[width=0.45\linewidth]{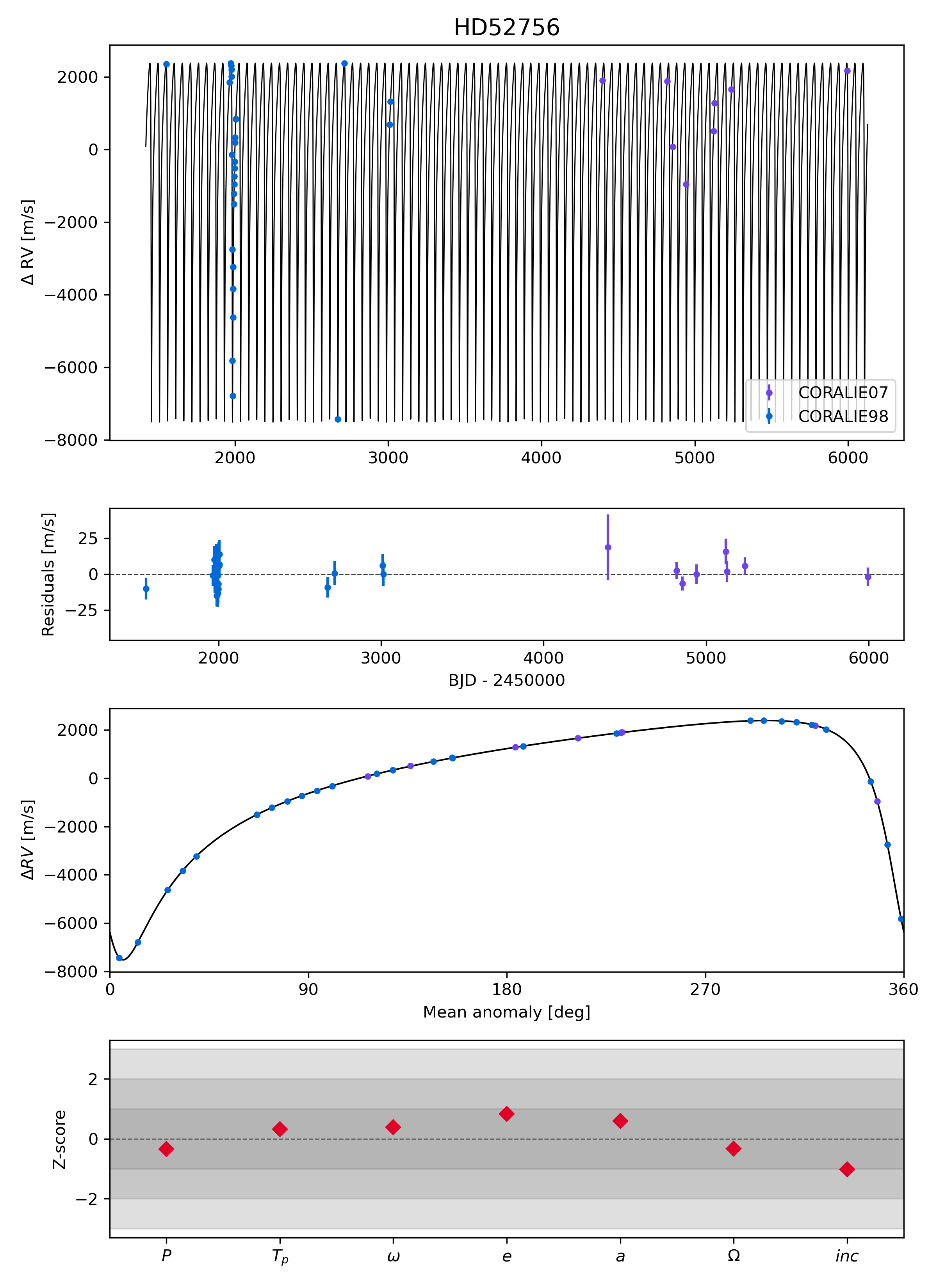}\\

        \includegraphics[width=0.45\linewidth]{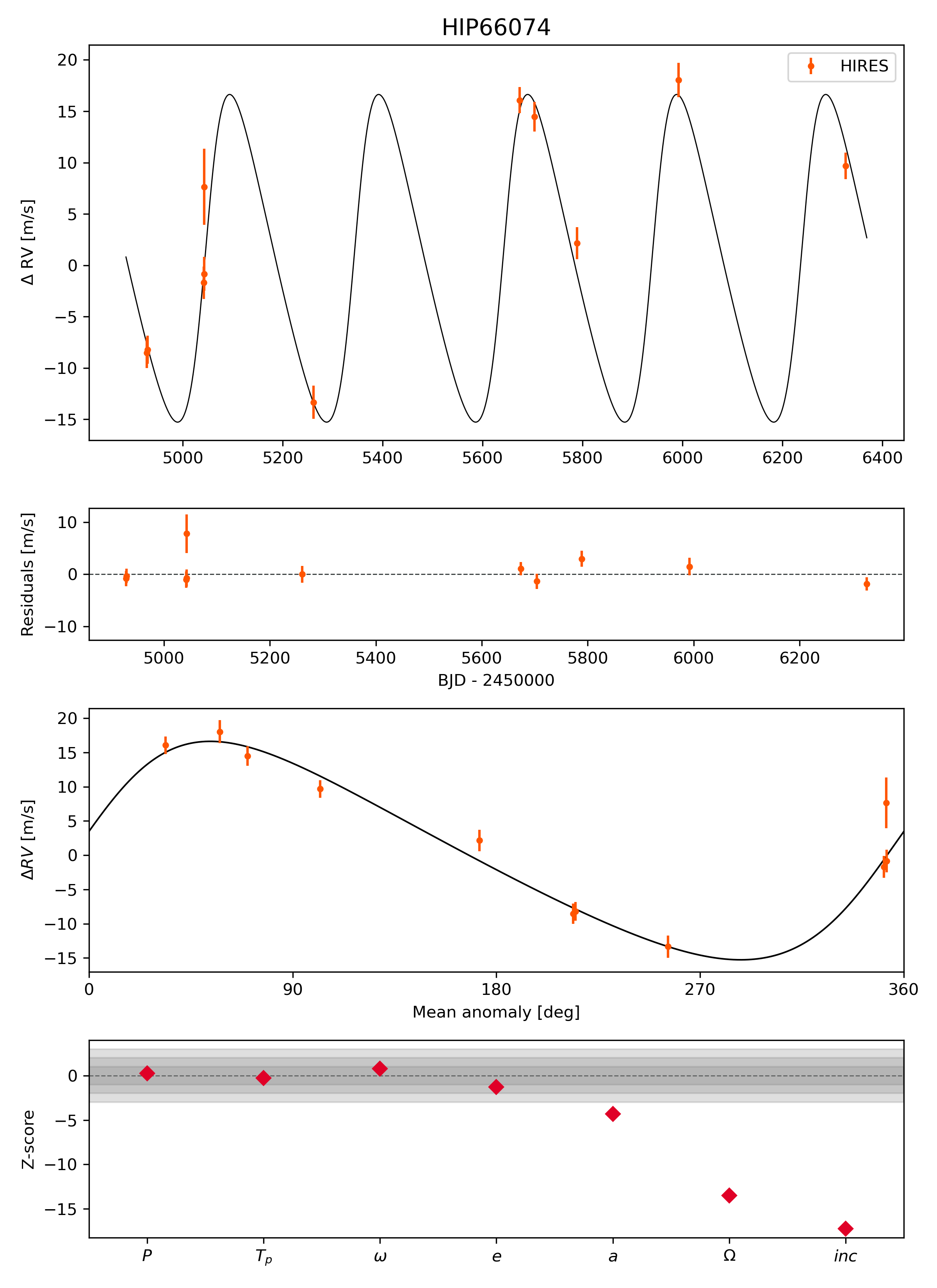}&
        \includegraphics[width=0.45\linewidth]{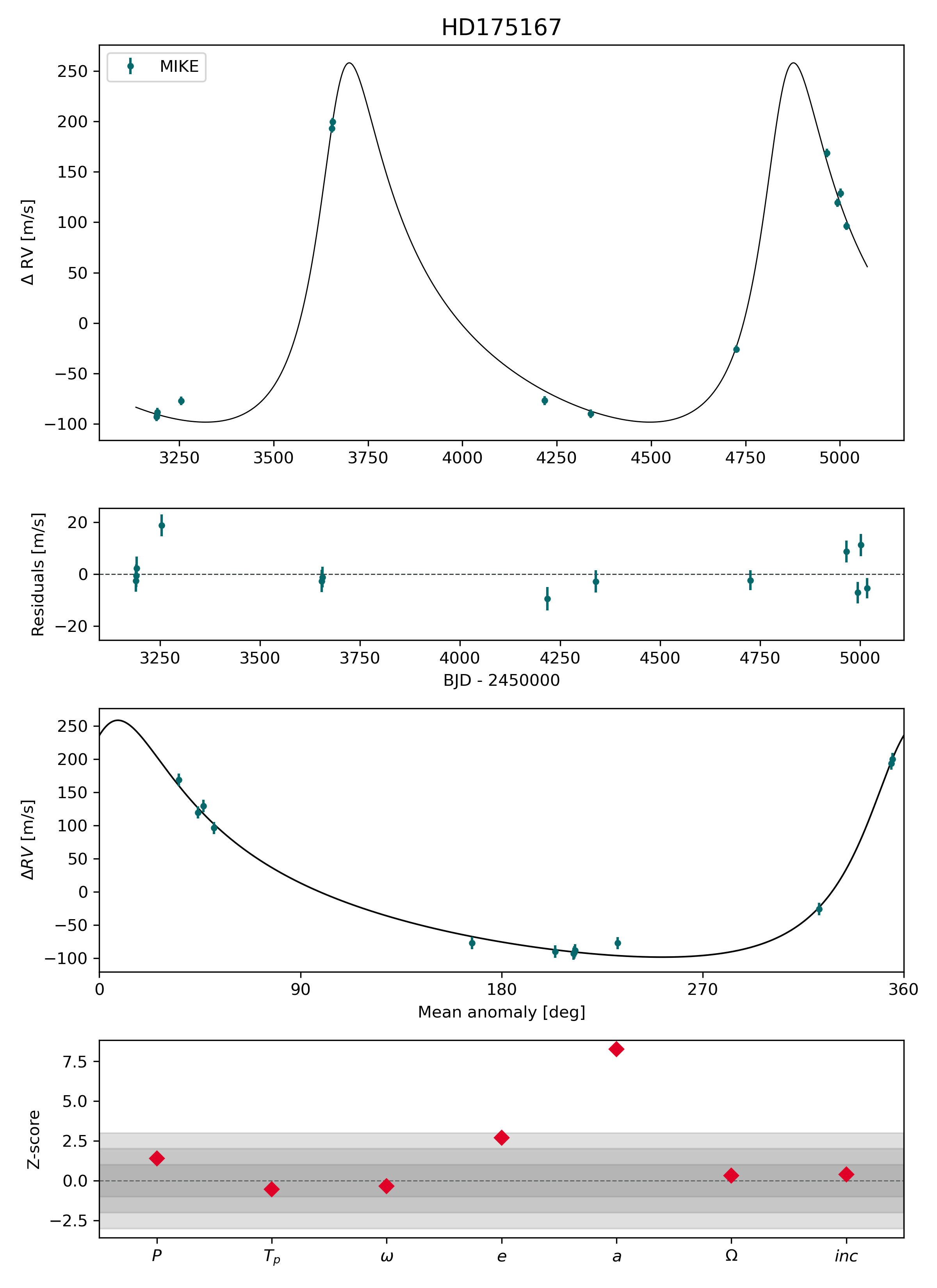}\\
        
        \includegraphics[width=0.45\linewidth]{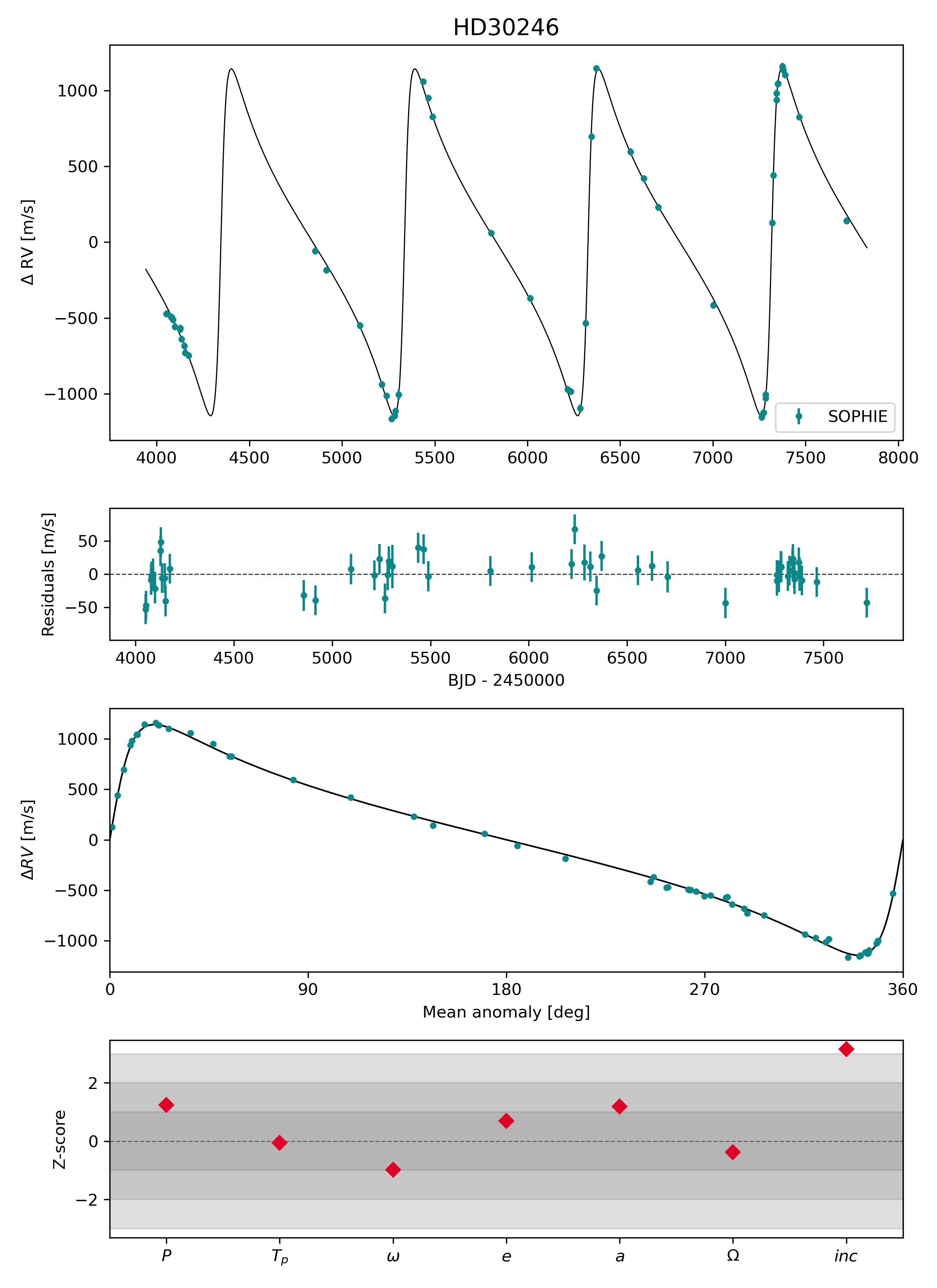}&
        \includegraphics[width=0.45\linewidth]{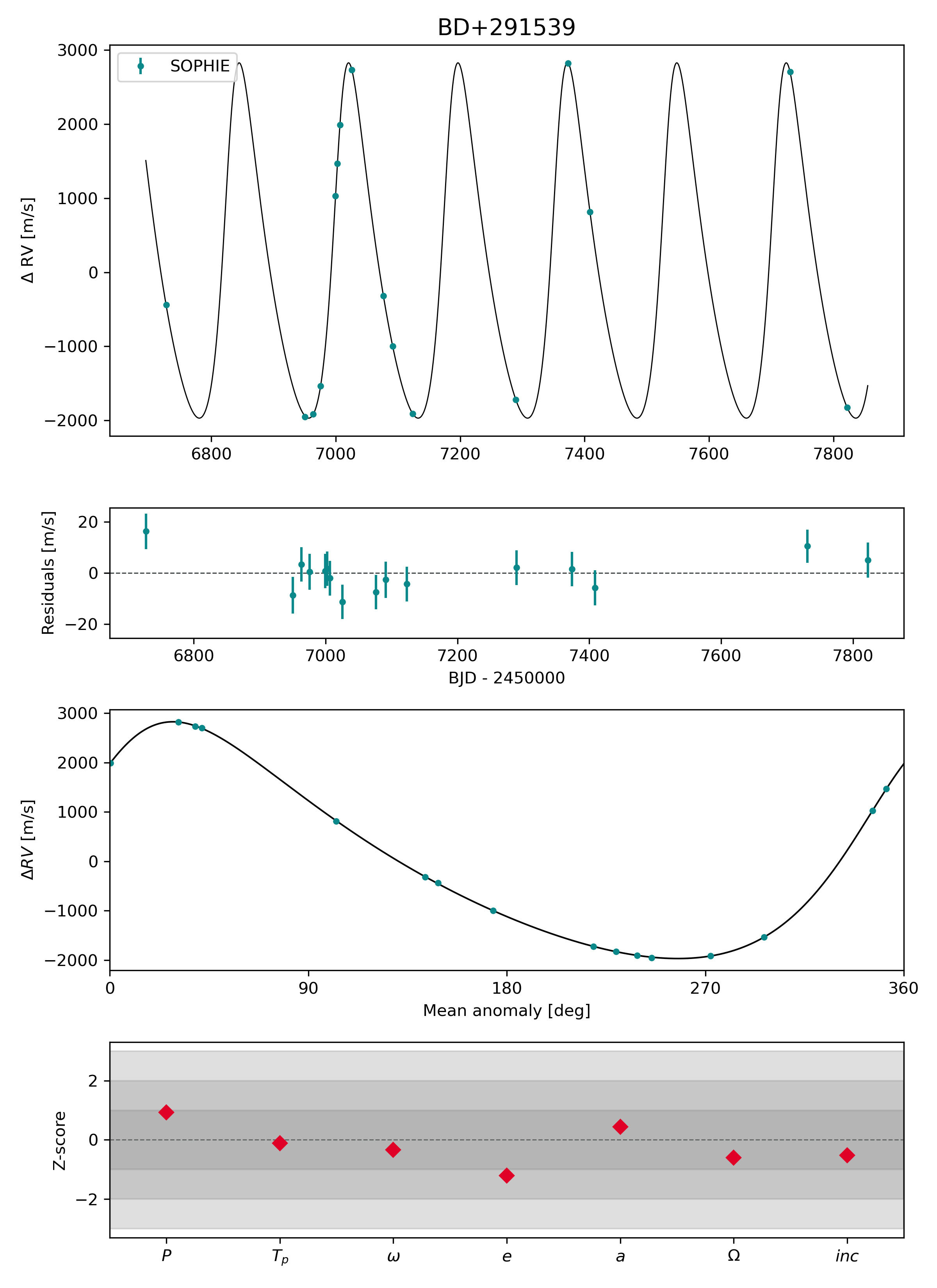}\\
        
        \includegraphics[width=0.45\linewidth]{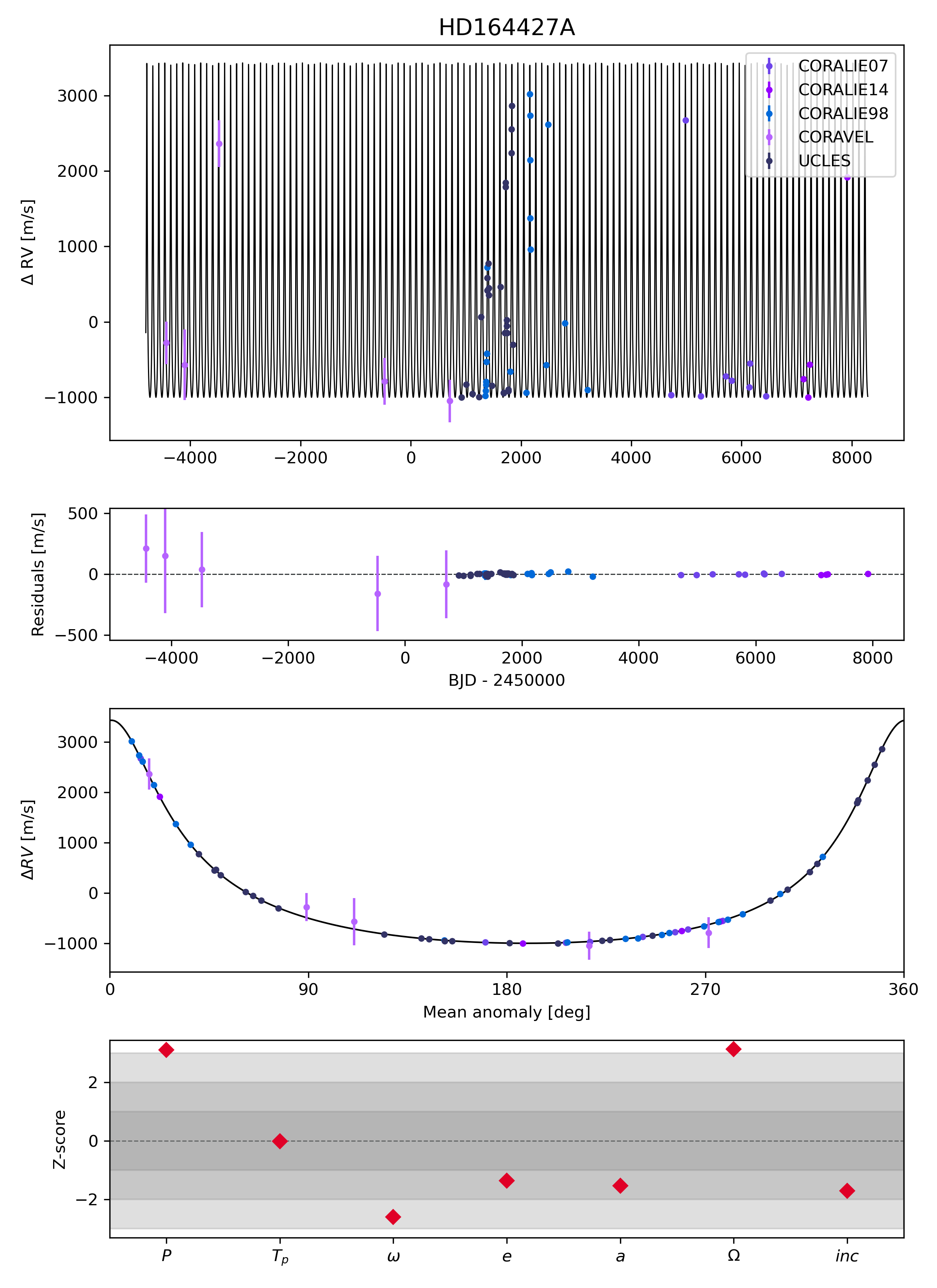}&
        \includegraphics[width=0.45\linewidth]{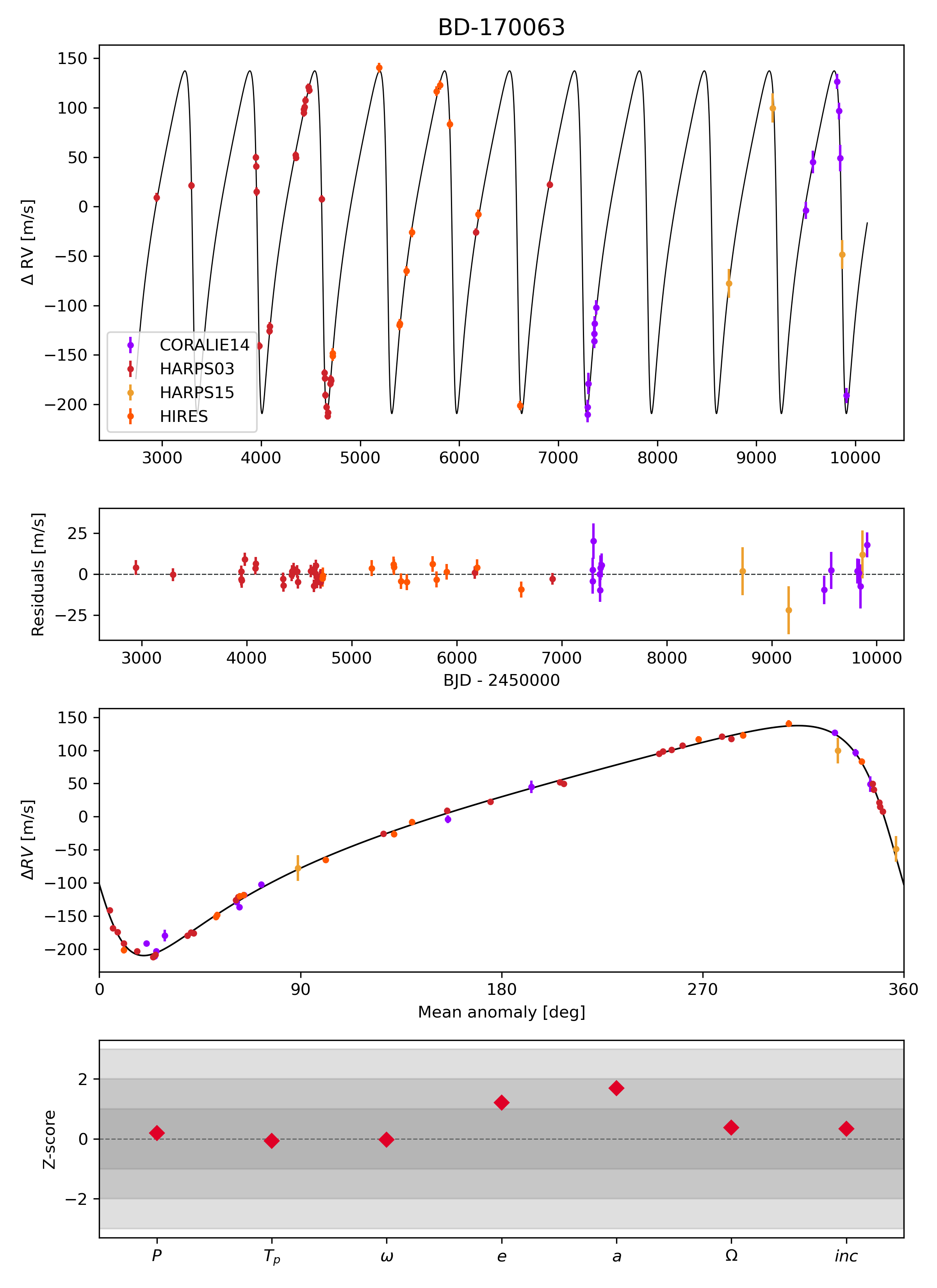}\\
        
        \includegraphics[width=0.45\linewidth]{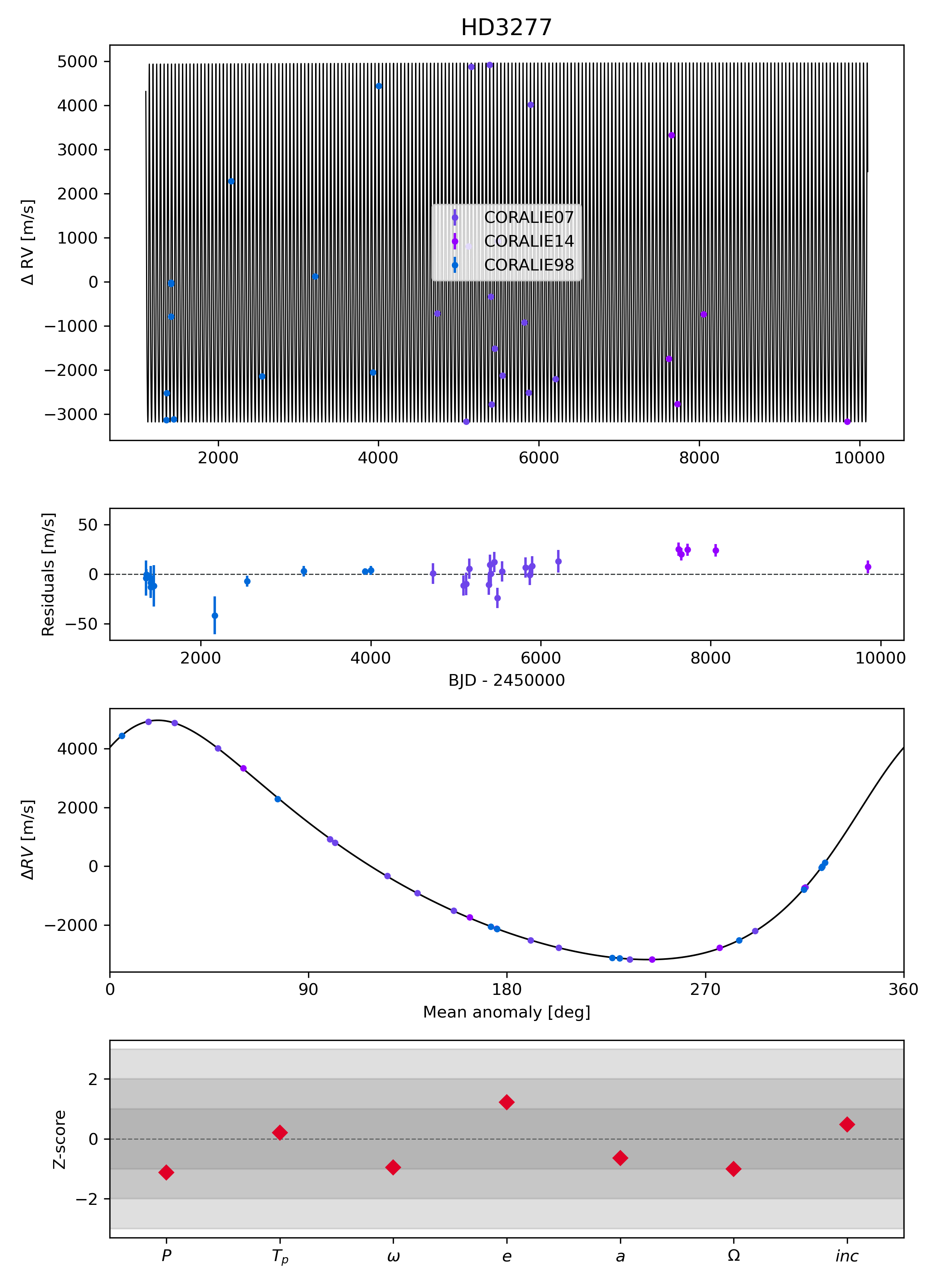}&
        \includegraphics[width=0.45\linewidth]{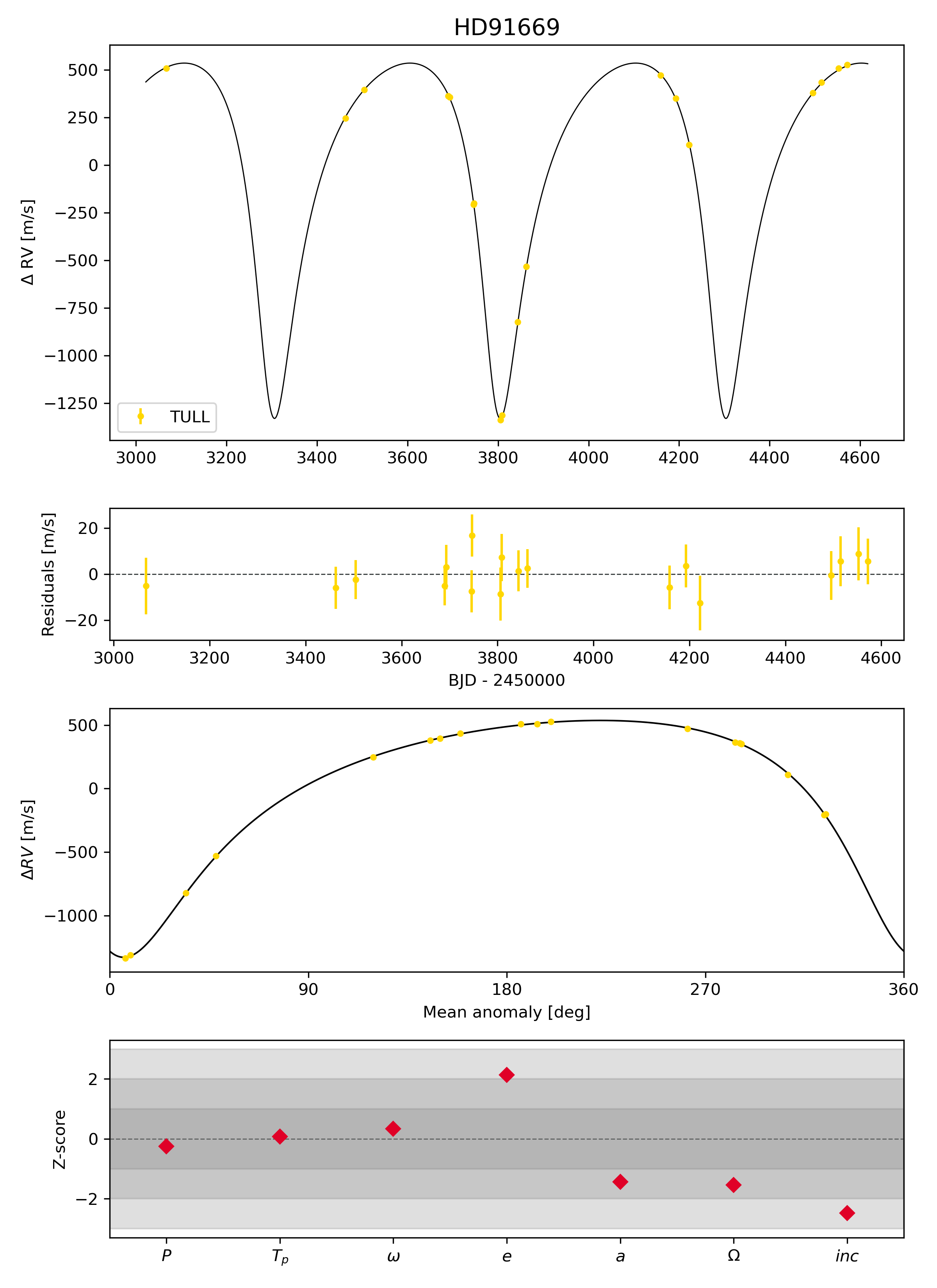}\\
        
        \includegraphics[width=0.45\linewidth]{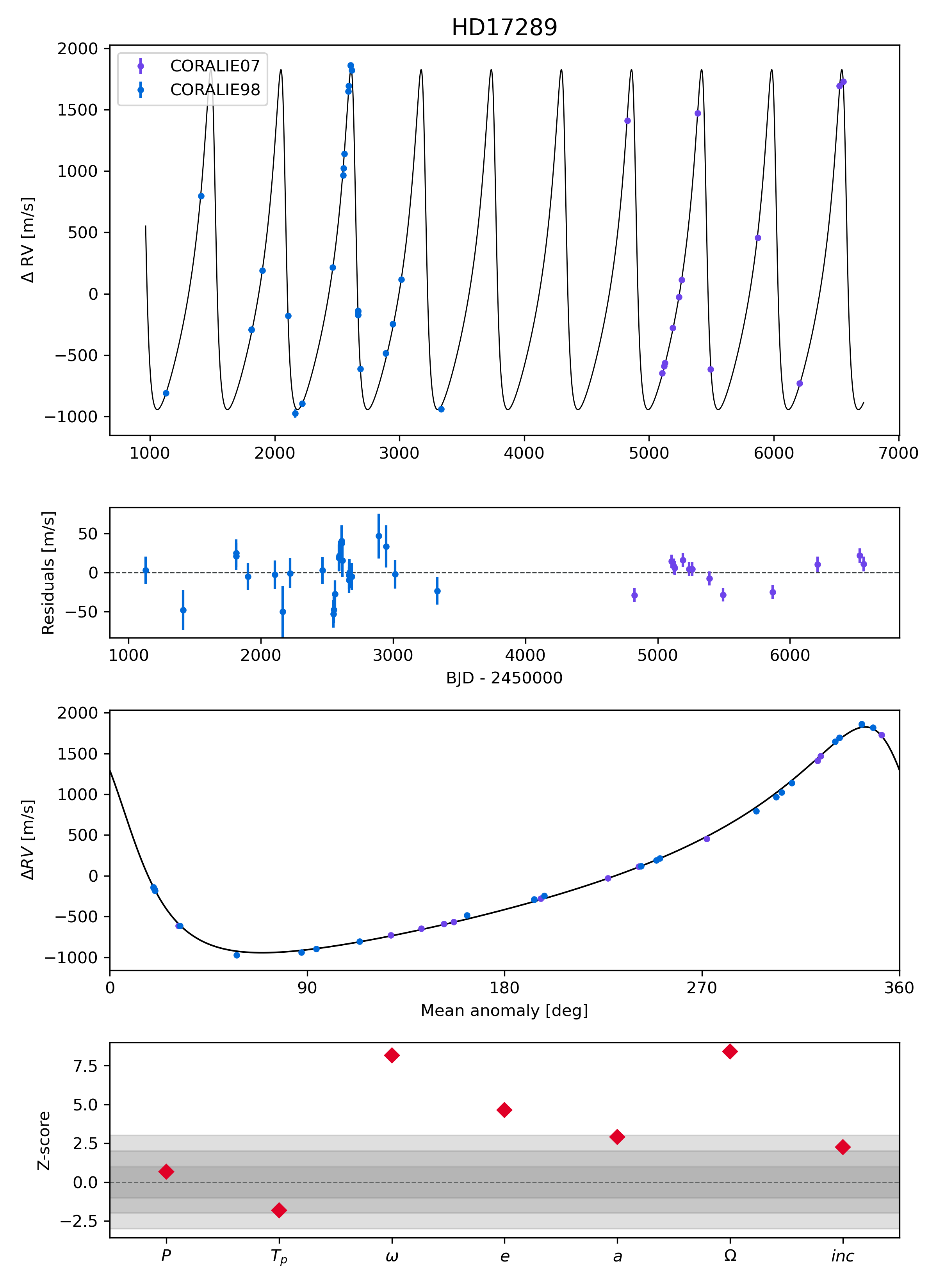}&
        \includegraphics[width=0.45\linewidth]{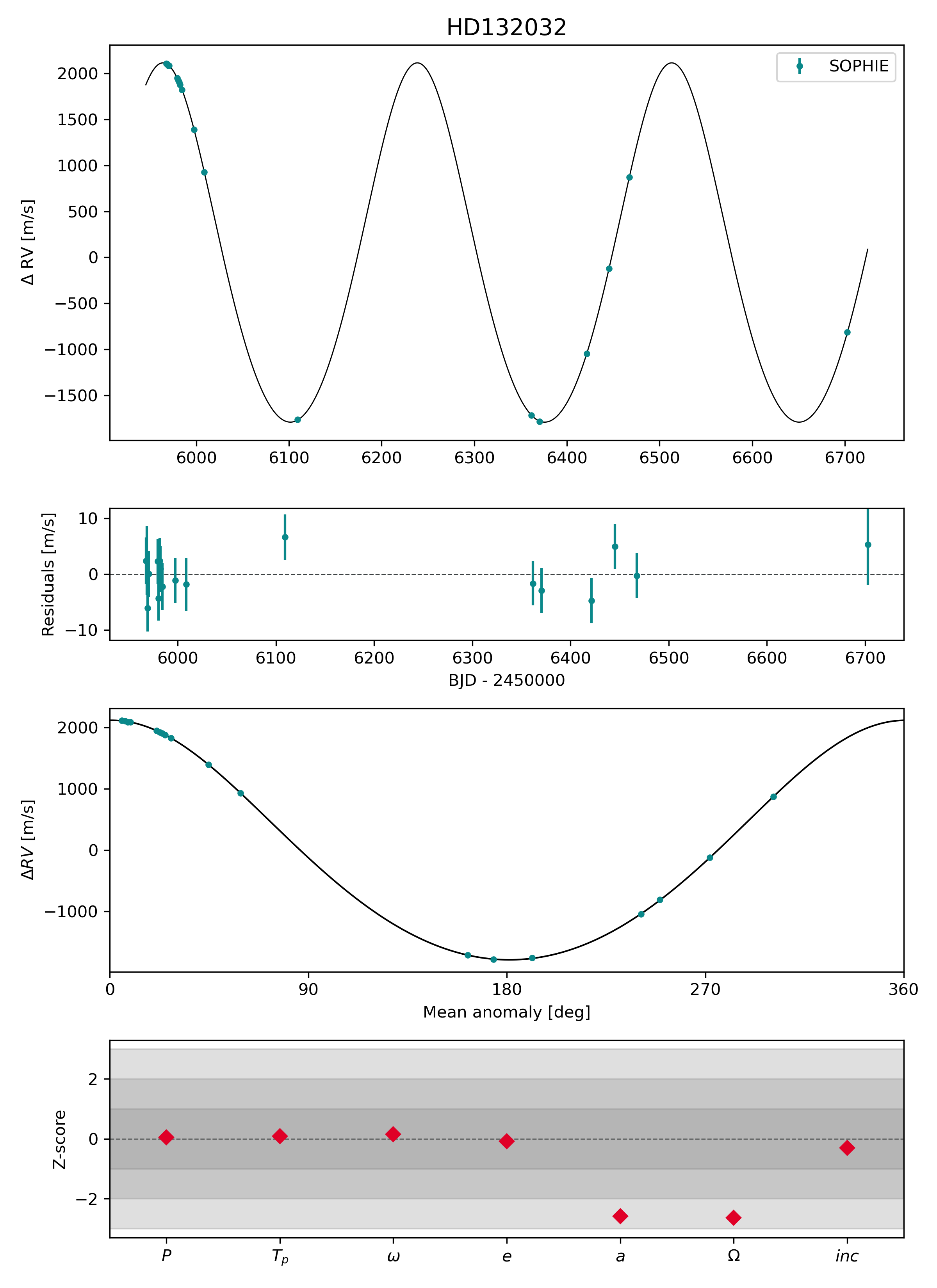}\\
        
        \includegraphics[width=0.45\linewidth]{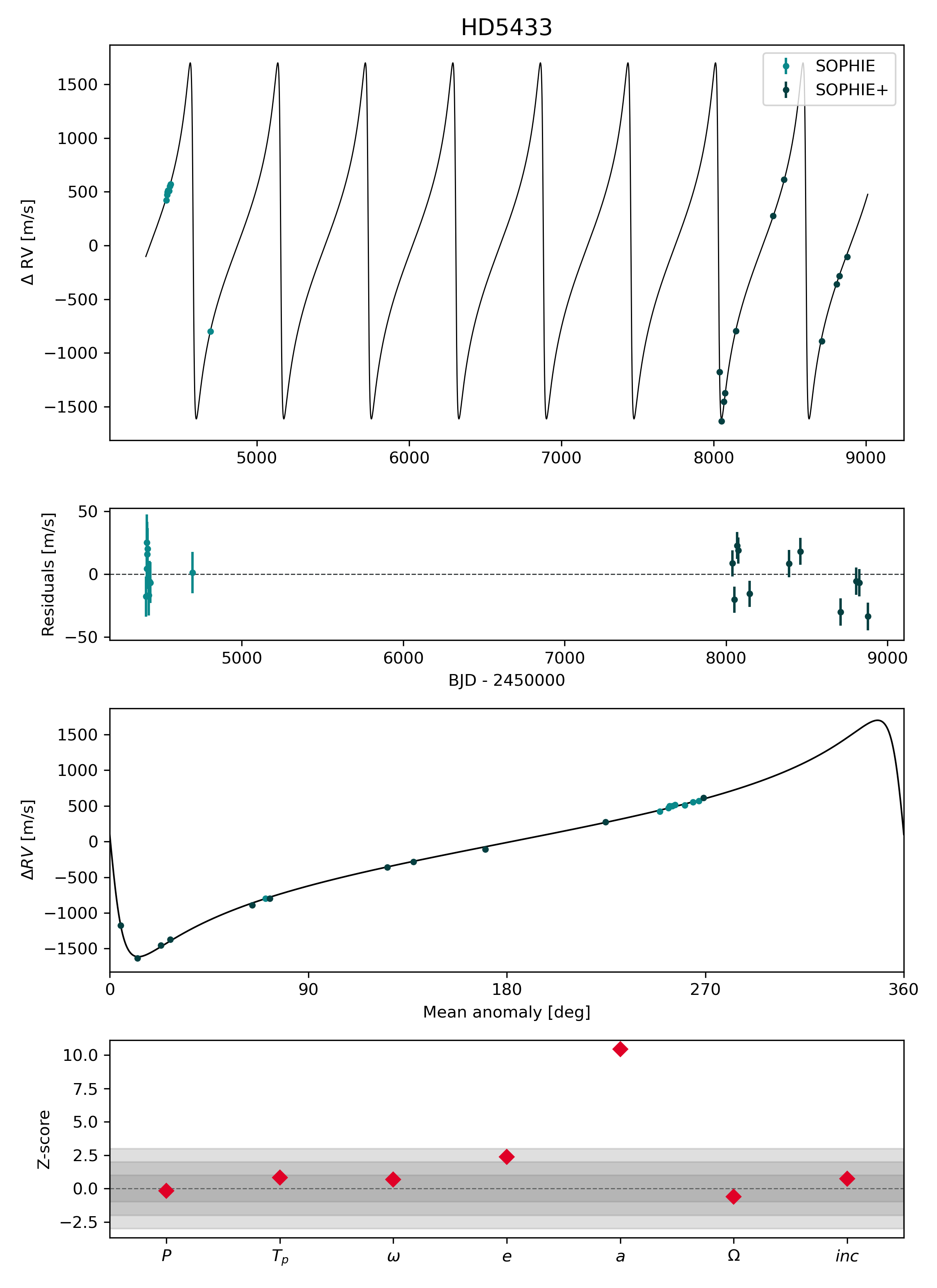}&
        \includegraphics[width=0.45\linewidth]{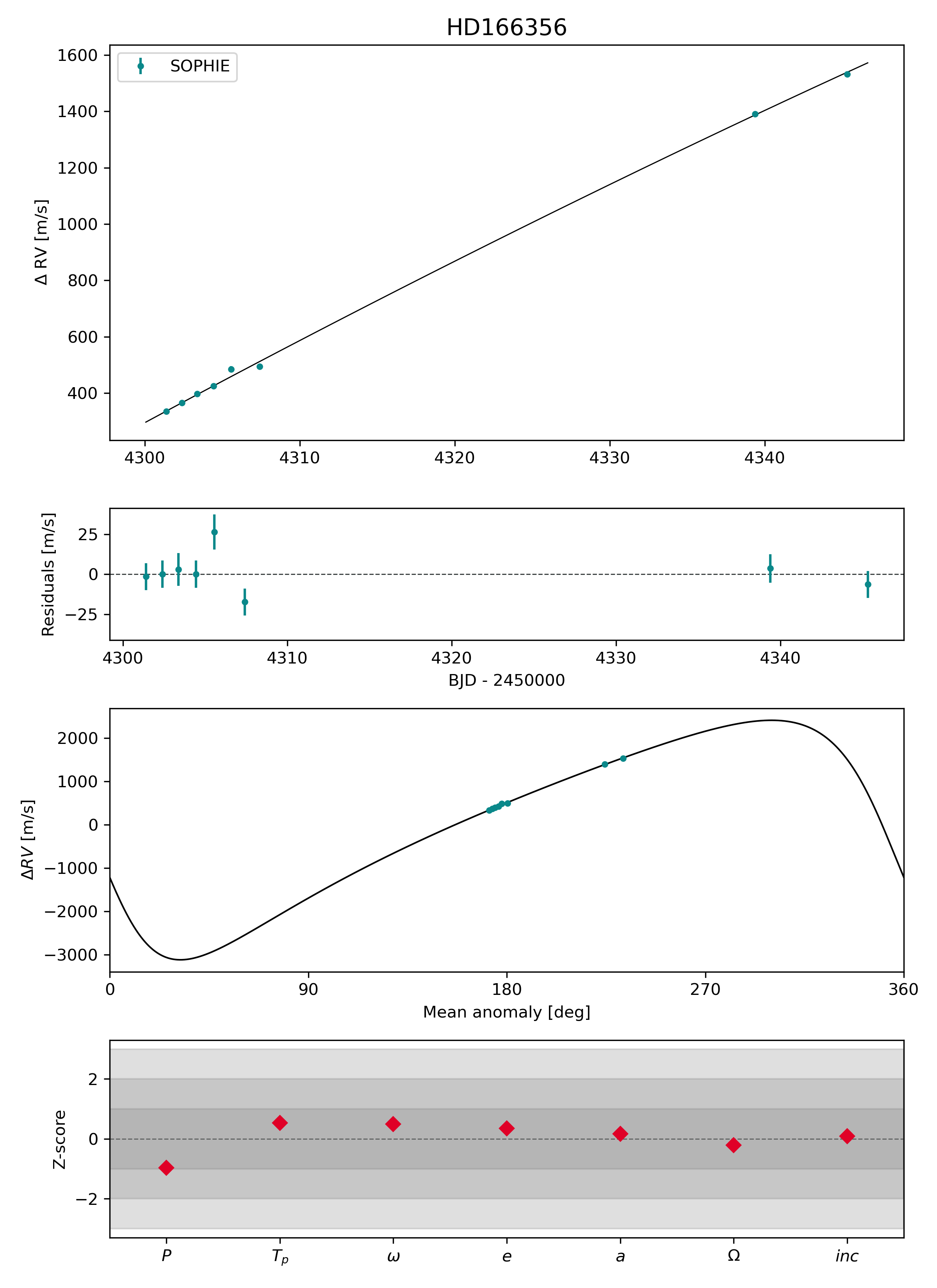}\\
        
        \includegraphics[width=0.45\linewidth]{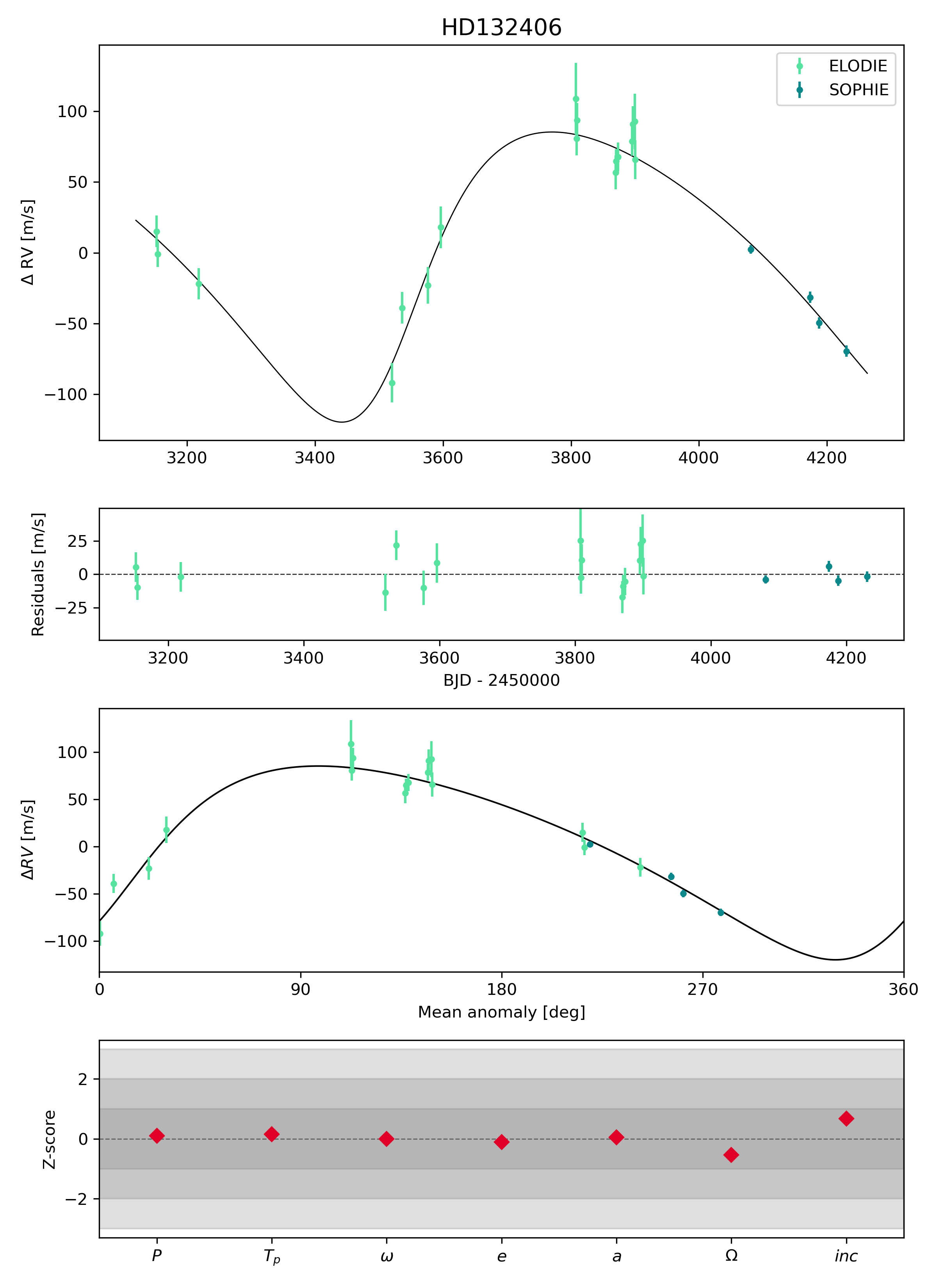}&
        \includegraphics[width=0.45\linewidth]{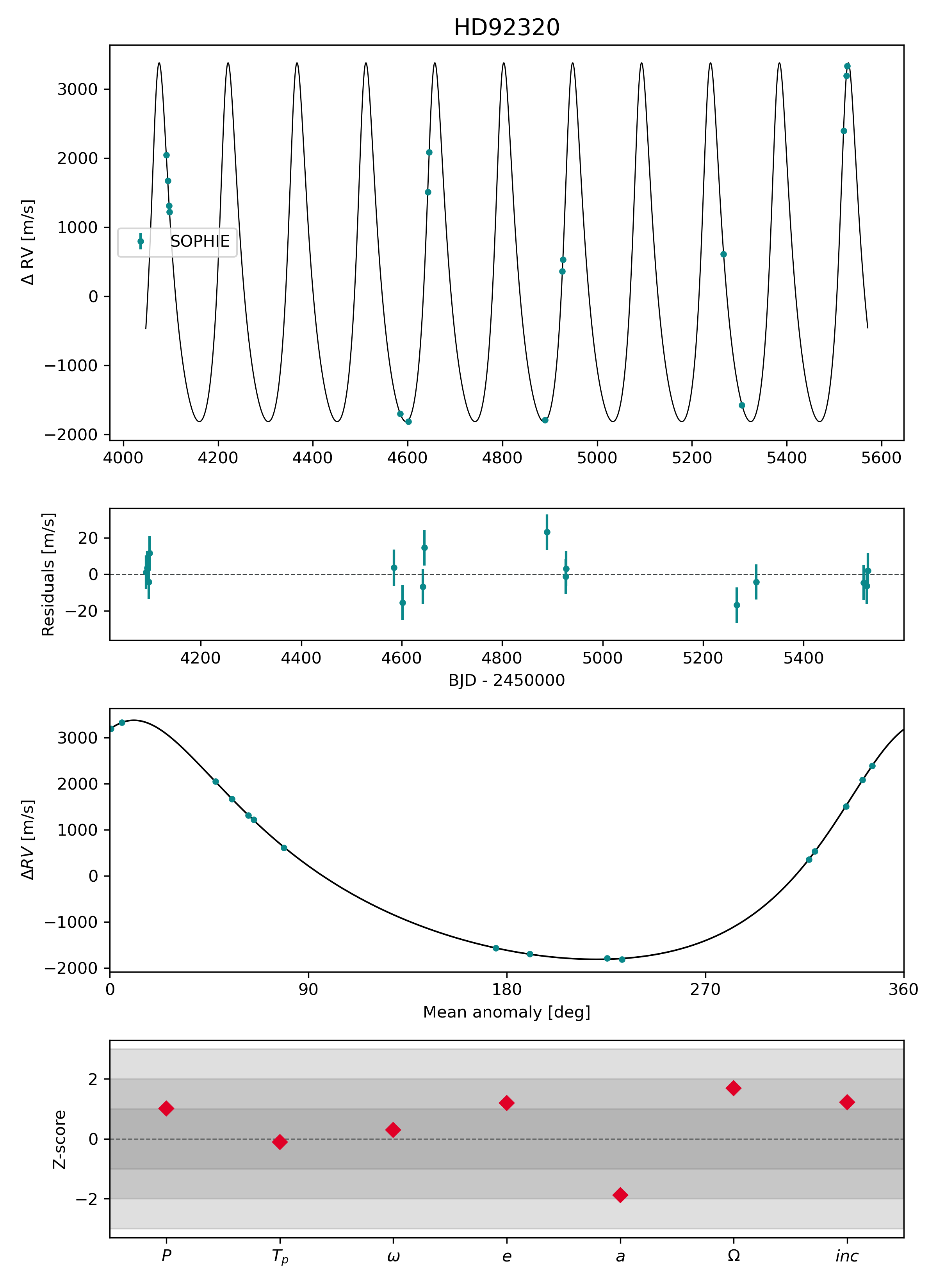}\\
        
        \includegraphics[width=0.45\linewidth]{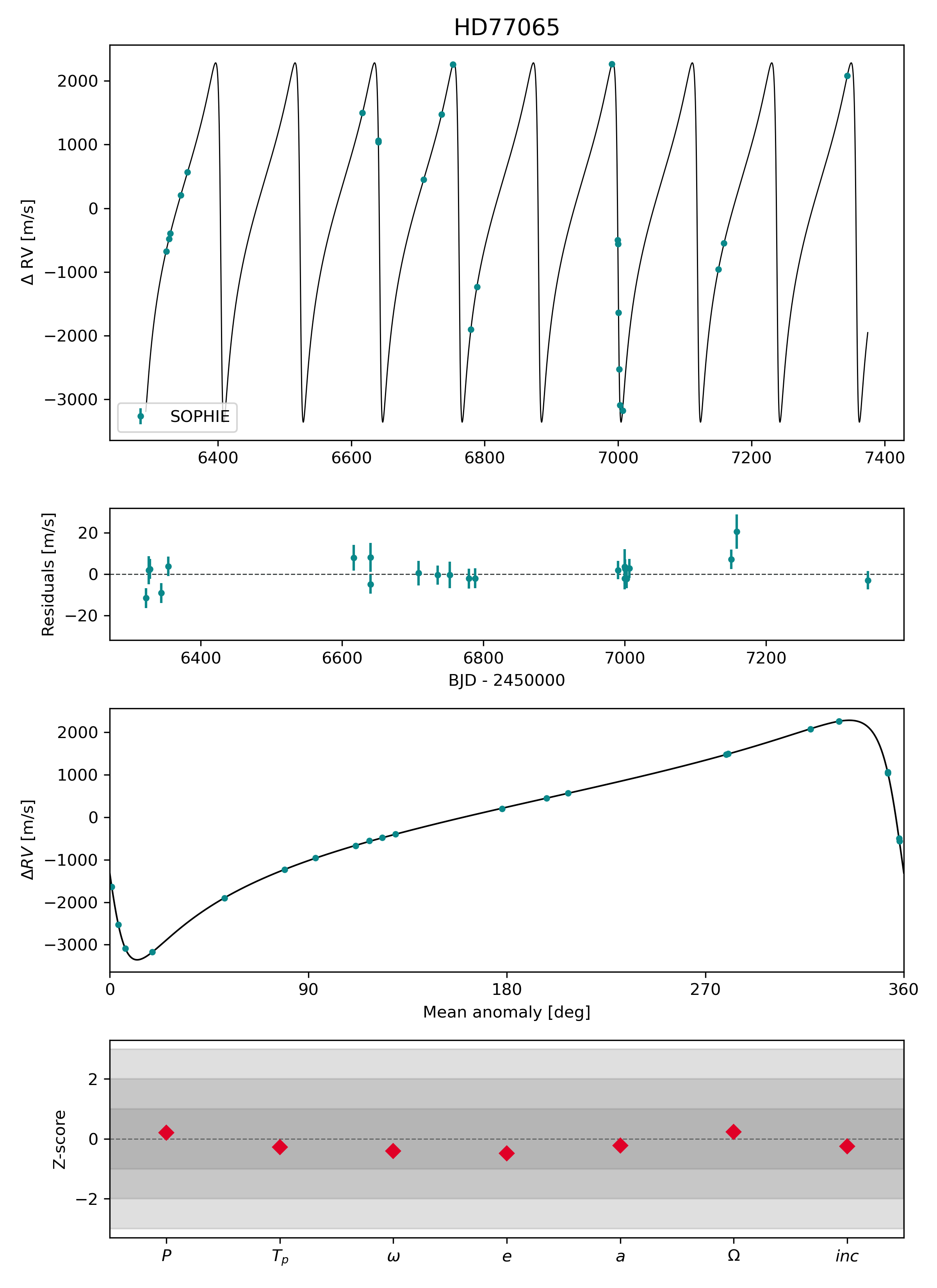}&
        \includegraphics[width=0.45\linewidth]{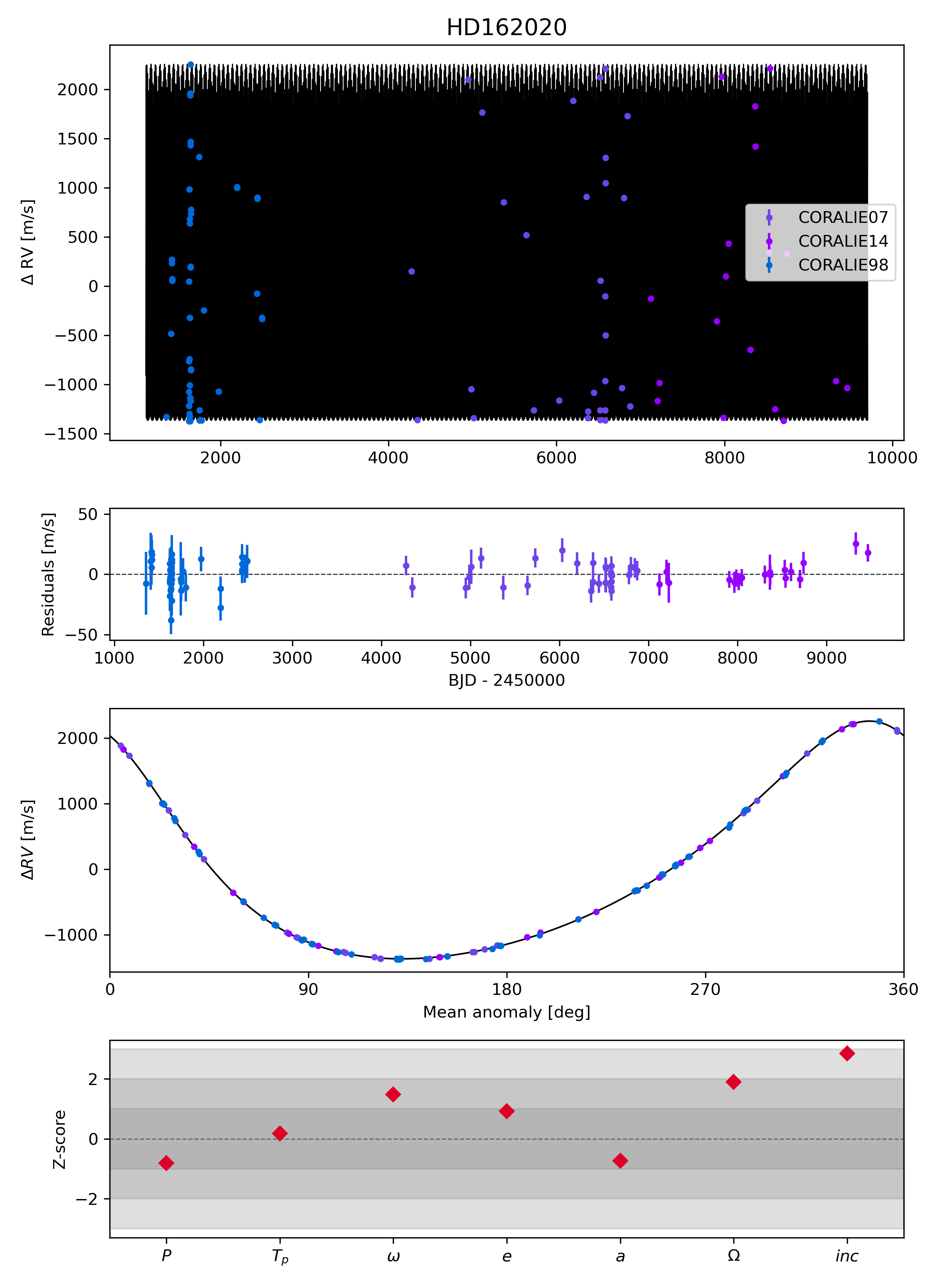}\\
        
        \includegraphics[width=0.45\linewidth]{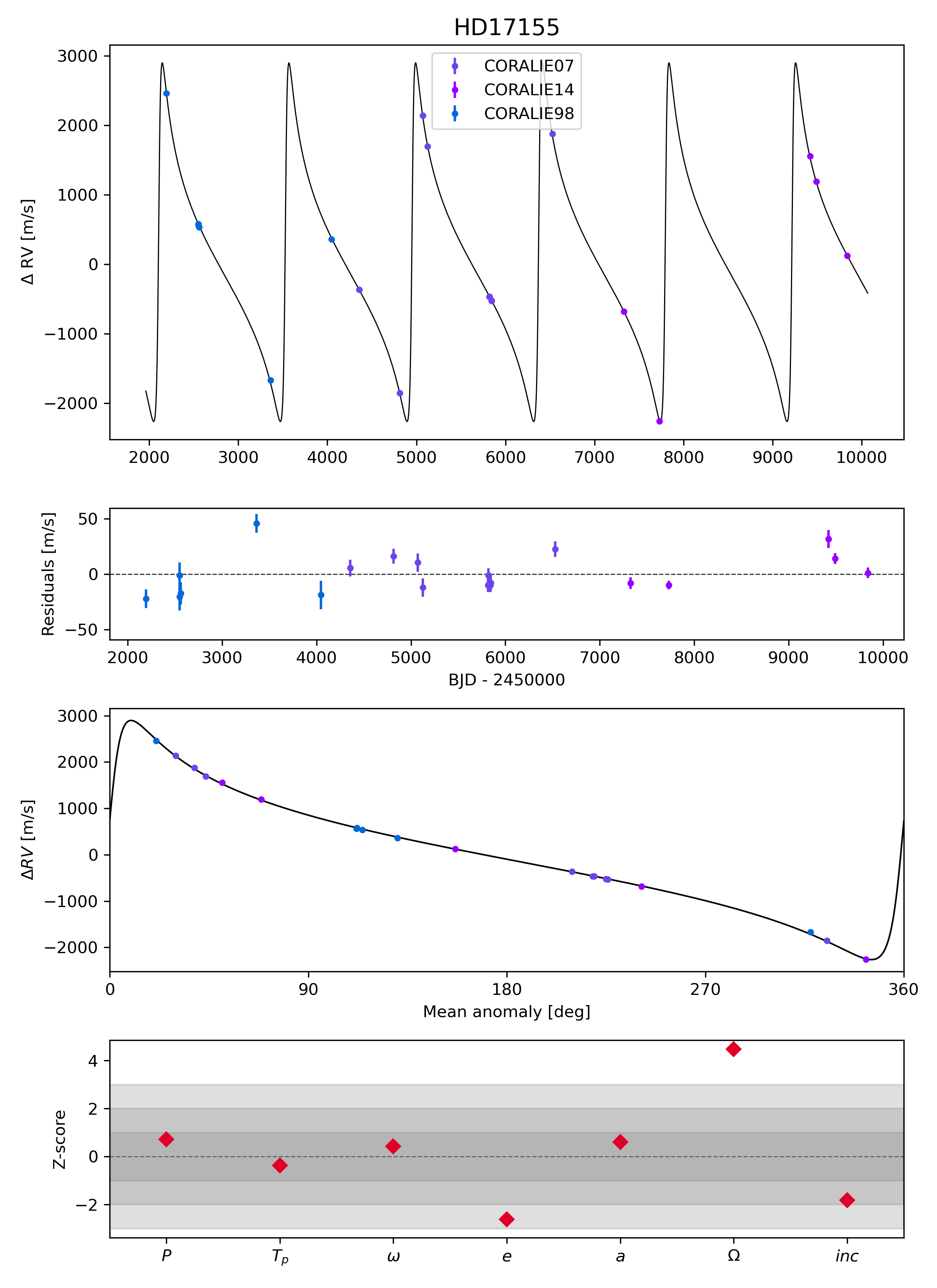}&
        \includegraphics[width=0.45\linewidth]{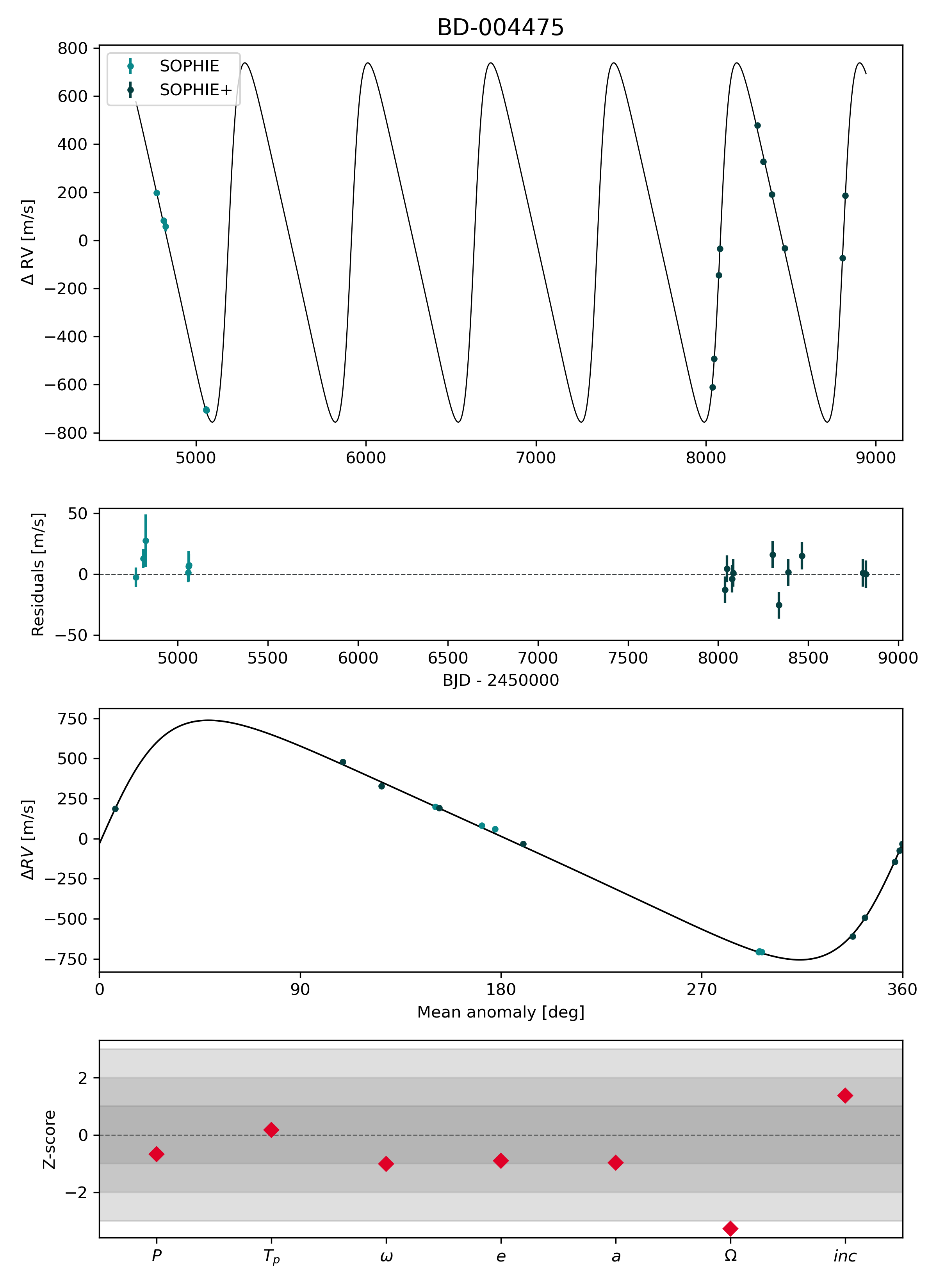}\\
        
        \includegraphics[width=0.45\linewidth]{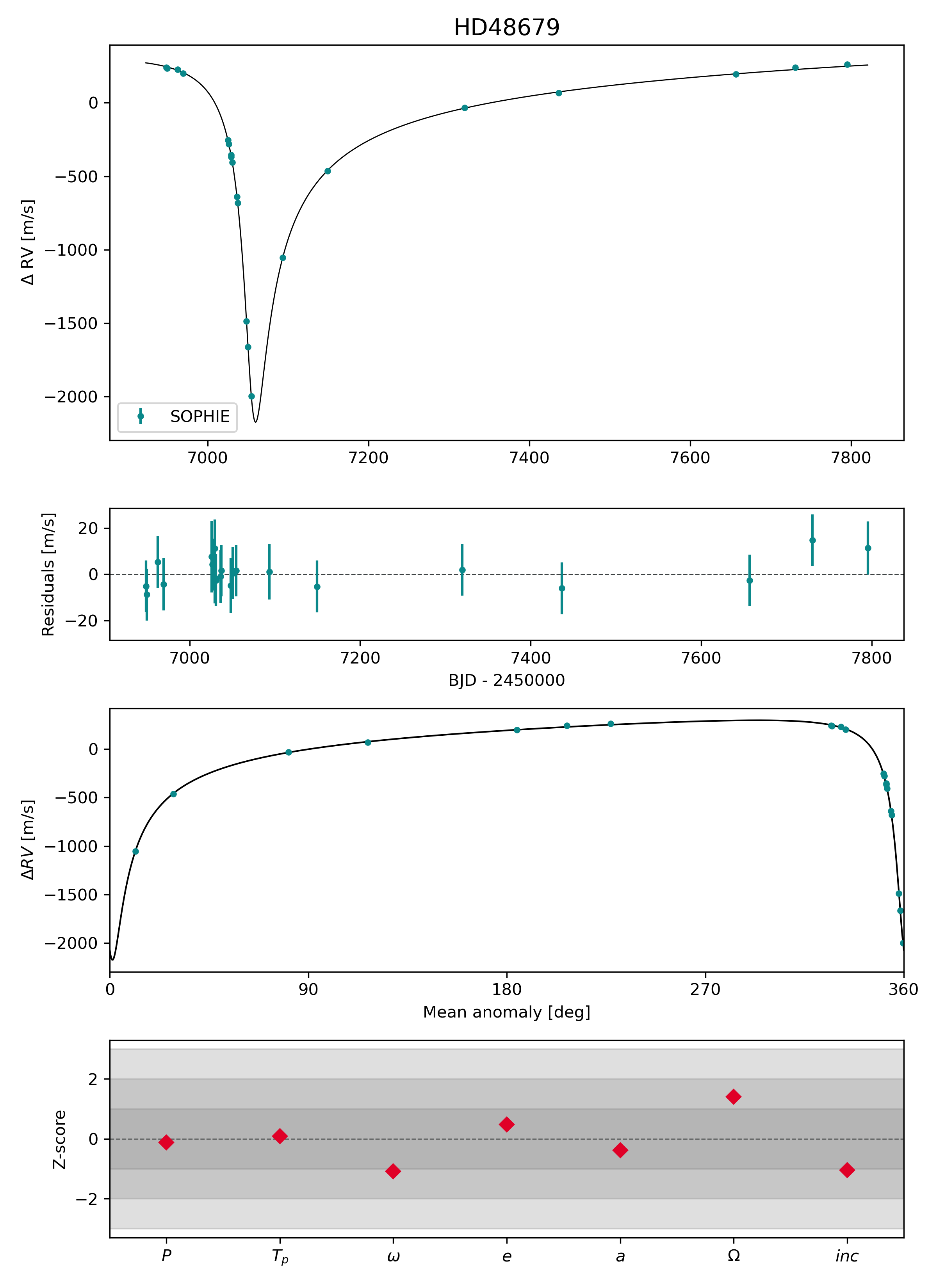}&
        \includegraphics[width=0.45\linewidth]{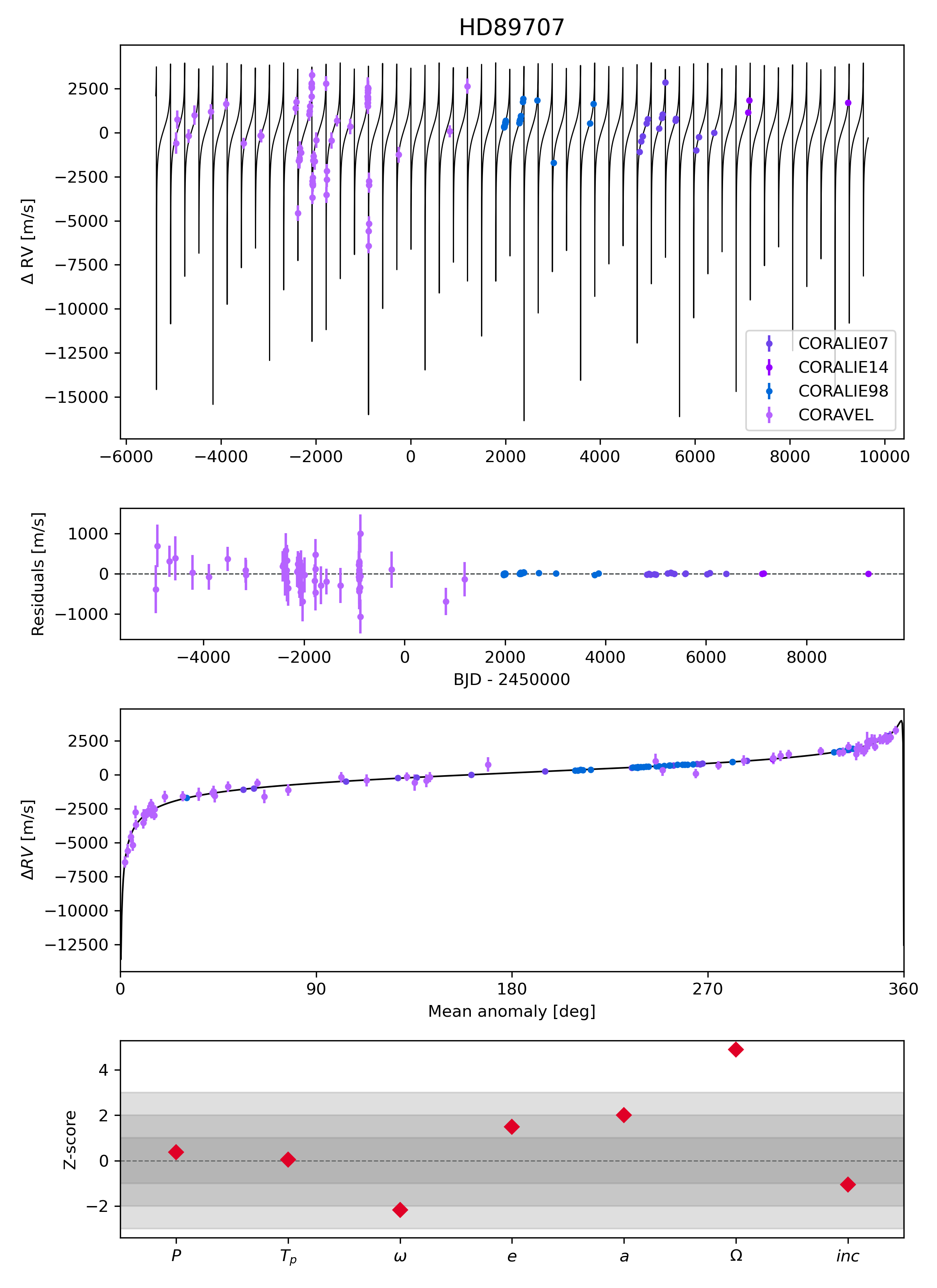}\\

\end{longtable}
    	
    	}
    \clearpage
    \twocolumn

\end{appendix}


\end{document}